\documentclass[%
 reprint,
 twocolumn,
 amsmath,amssymb,
aip,jcp,
]{revtex4-2}

\usepackage{natbib}
\usepackage{graphicx}
\usepackage{dcolumn}
\usepackage{bm}
\usepackage{xcolor}

\begin{document}

\title{Perspective: Simulations of nonradiative processes in semiconductor nanocrystals}

\author{Dipti Jasrasaria}
\email{djasrasaria@berkeley.edu}
\address{Department of Chemistry, University of California, Berkeley, California 94720, USA}

\author{Daniel Weinberg}
\email{d\_weinberg@berkeley.edu}
\affiliation{Department of Chemistry, University of California, Berkeley, California 94720, USA}
\affiliation{Materials Sciences Division, Lawrence Berkeley National Laboratory, Berkeley, California 94720, USA}

\author{John P. Philbin}
\email{jphilbin@g.harvard.edu}
\affiliation{Harvard John A. Paulson School of Engineering and Applied Sciences, Harvard University, Cambridge, MA 02138, USA}

\author{Eran Rabani}
\email{eran.rabani@berkeley.edu}
\affiliation{Department of Chemistry, University of California, Berkeley, California 94720, USA}
\affiliation{Materials Sciences Division, Lawrence Berkeley National Laboratory, Berkeley, California 94720, USA}
\affiliation{The Raymond and Beverly Sackler Center of Computational Molecular and Materials Science, Tel Aviv University, Tel Aviv 69978, Israel}

\date{\today}

\begin{abstract}
The description of carrier dynamics in spatially confined semiconductor nanocrystals (NCs), which have enhanced electron-hole and exciton-phonon interactions, is a great challenge for modern computational science. These NCs typically contain thousands of atoms and tens of thousands of valence electrons with a discrete spectra at low excitation energies, similar to atoms and molecules, that converges to the continuum bulk limit at higher energies. Computational methods developed for molecules are limited to very small nanoclusters, and methods for bulk systems with periodic boundary conditions are not suitable due to the lack of translational symmetry in NCs. This perspective focuses on our recent efforts in developing a unified atomistic model based on the semiempirical pseudopotential approach, which is parametrized by first-principle calculations and validated against experimental measurements, to describe two of the main nonradiative relaxation processes of quantum confined excitons: exciton cooling and Auger recombination. We focus on the description of both electron-hole and exciton-phonon interactions in our approach and discuss the role of size, shape, and interfacing on the electronic properties and dynamics for II-VI and III-V semiconductor NCs.
\end{abstract}

\maketitle

\section {Introduction}
\label{sec:intro}
Colloidal semiconductor nanocrystals (NCs) offer an idealized test bed to explore the behavior of excitons and multiexcitons from the discrete, molecular limit to the continous, bulk limit.\cite{Alivisatos1996,Scholes2006NatMater,Klimov2014,EfrosBrus2021,PGS2021,Kagan2021} At low excitation energies, NCs have discrete spectra due to quantum confinement effects, which resemble those of atoms and molecules, while at higher excitation energies, due to increasingly large densities of states, their spectra converge to the bulk continuum limit. Understanding the interplay of degeneracy, size, shape, and material composition on NC electronic structure has been the subject of numerous studies over the past several decades.\cite{Efros1982,Murray1993,Wang1994,Dabbousi1997,Efros2000,Scholes2006NatMater,Rabani2010,Boles2016,Weiss2021}

From a theoretical perspective, the description of excitons and multiexcitons in semiconductor NCs poses several challenges. Because NCs contain several hundreds to thousands of atoms and valence electrons, quantum chemistry techniques that were developed to study the excited states of molecules are far too computationally expensive to be applicable to NCs. Thus, early work focused on the development of continuum approaches starting from the bulk limit, within a family of effective mass models, to describe the quantum confinement of excitons and dielectric screening.\cite{Efros1982,Rossetti1983} The most popular single-parabolic band approximation provides a \textit{qualitative} description of the optical properties of NCs, but it does not account for non-parabolic effects and valence-band degeneracies that are important in NCs. A more \textit{quantitative} description based on the multi-band effective mass model revealed rich behavior and provided accurate predictions of the exciton fine structure and band-edge exciton splittings as well as their dependence on the size, shape, and crystal structure of the NC.\cite{Ekimov1993,Norris1994,Norris1996a,Efros2000} In addition, the inclusion of many-body exchange interactions of electrons and holes resulted in optically forbidden dark excitons and explained the non-monotonic temperature dependence of the radiative lifetime in NCs.\cite{Nirmal1994,Nirmal1995,Norris1996b,Efros1996}

Despite the significant progress made based on the effective mass model, the lack of an underlying atomistic description has limited the application primarily to the description of optical properties, which are less sensitive to the atomistic detail of the NC, particularly in the strong confinement limit ($R < a_{\text{B}}$, where $R$ is the NC radius and $a_{\text{B}}$ is the bulk exciton Bohr radius). To account for inhomogeneities in the NC structure, semiempirical pseudopotential models, which became popular in the 1970s and 1980s to describe the electronic and optical properties of bulk semiconductors and surfaces,\cite{Cohen1966} were employed and further developed to study excitons in a variety of semiconductor NCs,\cite{Friesner1991,Wang1994,Wang1996,Rabani1999b} demonstrating remarkable success in postdicting and predicting the exciton fine structure\cite{Norris1996b} as well as the roles of defects,\cite{Califano2005,Jasrasaria2020} stress, and strain on the electronic structure.\cite{Wang1999, Mattila1999}

\begin{figure*}[ht]
\includegraphics[width=\textwidth]{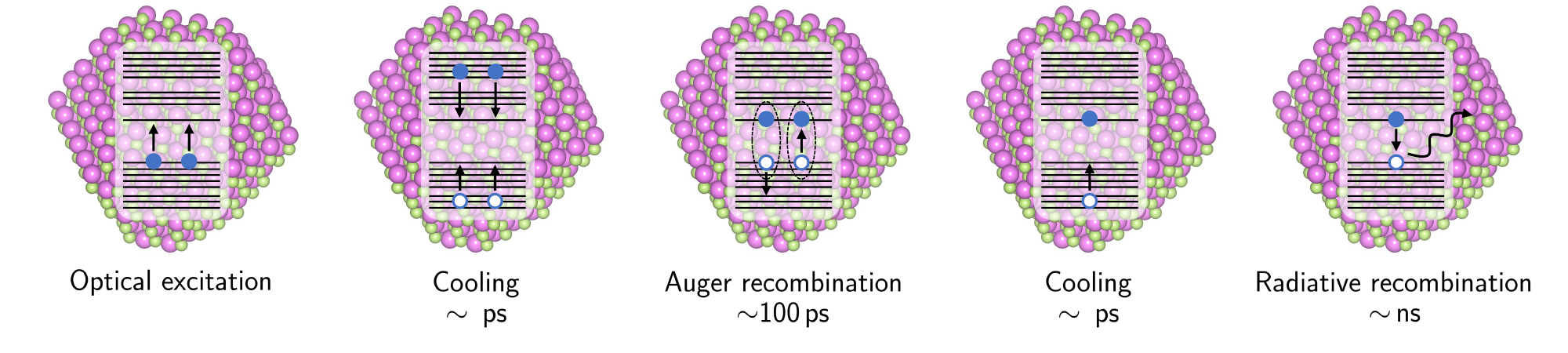}
\caption{\label{fig:AllProcesses}Photoexcitation of a nanocrystal can create multiple electron-hole pairs, which quickly relax to the band edge in a process called cooling. From the band edge, the multi-excitonic state can undergo Auger recombination, which nonradiatively annihilates one exciton and forms a hot electron-hole pair. The hot carriers then cool back to the band edge, from which they can radiatively recombine.}
\end{figure*}

In recent years, growing interest in the dynamics of excitons inspired by novel experimental observations,\cite{GS1999,Klimov2000,GS2005,Oron2007,Pandey2008,Jones2009,Sukhovatkin2009,McArthur2010,Ulbricht2011,Knowles2011,Bae2013,Qin2014,Kambhampati2015,Wu2016,Li2017c,Kaledin2018,Li2019} has shifted the focus for theory to address issues related to the transients of these nonequilibrium species.\cite{Oleg2009} Understanding the radiative and nonradiative decay channels depicted in Fig.~\ref{fig:AllProcesses} as well as the dephasing and energy transfer mechanisms of confined excitons, which are dictated by the exciton-phonon and exciton-exciton couplings, is key to the rational design of NC-based technologies with reduced thermal losses and increased quantum yields. Two central decay channels are the main focus the current perspective.

The first decay channel is that of nonradiative relaxation of hot excitons, or cooling, illustrated in Fig.~\ref{fig:AllProcesses}. The study of this process is motivated by conflicting results for the relaxation times of hot excitons to the band edge in confined structures relative to those of excitons in bulk.\cite{Gfroerer1996,Haiping1996,Heitz1997,Sosnowski1998,Mukai1998,GS1999,Klimov1999,Klimov2000a,Harbold2005,GS2005} Due to the discrete nature of the excitonic levels in confined NCs, exciton cooling \textit{via} phonon emission, especially near the band edge, has been thought to require multi-phonon processes and would, therefore, be inefficient, a phenomenon known as the phonon bottleneck.\cite{Nozik2001} One mechanism for breaking the phonon bottleneck that allows for fast cooling is the Auger process.\cite{Efros2003} In many NCs, holes relax rapidly to the band edge \textit{via} phonon emission because valence band degeneracies and a larger hole effective mass lead to a higher density of hole states with smaller energy spacings that are on the order of the phonon frequencies. An electron, then, can relax to the band edge by nonradiatively transferring its energy to a hole \textit{via} an Auger-like process, and the re-excited hole can quickly relax back to the valence band edge. The Auger-assisted cooling mechanism~\cite{Wang2003} has been supported by experimental observations~\cite{Klimov1999,Klimov2000a,Hendry2006} but, as far as we know, the exciton cooling mechanism has not been confirmed or validated by atomistic calculations,\cite{Kilina2009} mainly because of the significant computational challenges of describing excitons and their coupling to phonons in systems containing thousands of atoms and valence electrons.

The second decay channel involves the nonradiative decay of multiexcitonic states and is motivated by the observation of a ``universal volume scaling law" for Auger recombination (AR) lifetimes in NC.\cite{Klimov2000,Robel2009} At high photo-carrier densities, which are typical of most optoelectronic devices, all semiconductor materials suffer from enhanced exciton-exciton annihilation that occurs primarily \textit{via} AR processes, shown in Fig.~\ref{fig:AllProcesses}, in which one exciton recombines by transferring its energy to another exciton.\cite{Efros2003} This nonradiative process leads to reduced photoluminescence quantum yields and decreases maximum device efficiencies. Thus, understanding the properties of multiexcitonic states and their decay channels is central to improving and further developing many light-induced NC applications. From a theoretical/computational perspective, calculating AR lifetimes within Fermi's golden rule requires a description of the initial biexcitonic (or higher multiexcitonic) state and all possible final electron-hole pair states, a challenging task for NCs of experimentally relevant sizes. Thus, previous theoretical works have relied on effective mass continuum models, which ignore electron-hole correlations in the biexcitonic state,\cite{Li2019a} resulting in a much steeper scaling of the AR lifetimes with the NC volume.\cite{Chepic1990,Vaxenburg2015,Vaxenburg2016} The discrepancy between theory and experiments had been a mystery for several decades. 

In this perspective, we summarize our recent efforts to develop a unified model that address both problems. In Sec.~\ref{sec:methods} we describe the atomistic approach we have adopted to calculate quasiparticle excitations and neutral excitations in semiconductor NCs. First principles approaches, such as time-dependent density functional theory (DFT)~\cite{Chelikowsky2003,Shulenberger2021,Song2022} or many-body perturbation approximations,\cite{Degoli2009} are limited to describing excitons in relatively small clusters, typically those with fewer than 100 atoms, due to their steep computational scaling.\cite{Friesner2005,Voznyy2016} To make meaningful contact with experimental results on NCs that contain thousands of atoms and tens of thousands of electrons, we rely on the semiempirical pseudopotential model~\cite{Wang1994,Wang1996,Rabani1999b,Williamson2000} to describe quasiparticle excitations. We use a converged real-space grid method to represent the single-particle states combined with the filter diagonalization method~\cite{Wall1995,Toledo2002} to compute the single-particle states near the band edge and at higher excitation energies. We then use a subset of converged quasiparticle eigenstates to solve the Bethe-Salpeter equation~\cite{Rohlfing2000} within the static screening approximation to account for electron-hole correlations in neutral optical excitations.\cite{Philbin2018} Sec.~\ref{sec:methods} also provides validation of the approach for the quasiparticle and optical gaps and the exciton binding energies for II-VI and III-V semiconductor NCs in both the strongly ($R<a_{\text{B}}$) and weakly ($R>a_{\text{B}}$) confined regimes.

In Sec.~\ref{sec:exph} we present and assess the accuracy of our approach for determining the exciton-phonon couplings in semiconductors NCs, and we analyze the contributions of acoustic, optical, and surface modes to the overall magnitude of the exciton-phonon couplings. The standard model Hamiltonian that describes a manifold of excitonic states and phonons that are coupled to first order in the atomic displacements is given by:\cite{Giustino2017}
\begin{align}
H = &\sum_{n}E_{n}\left|\psi_{n}\right\rangle \left\langle \psi_{n}\right|+\sum_{\alpha}\hbar\omega_{\alpha}b_{\alpha}^{\dagger}b_{\alpha} \nonumber \\ 
&+\sum_{\alpha nm}V_{n,m}^{\alpha}\left|\psi_{n}\right\rangle \left\langle \psi_{m}\right|q_{\alpha}\,,\label{eq:hamiltonian}
\end{align}
where $\left|\psi_{n}\right\rangle $ describes exciton $n$ with energy $E_{n}$, and $b_{\alpha}^{\dagger}$ and $b_{\alpha}$ are the Boson creation and annihilation operators, respectively, of phonon mode $\alpha$ with frequency $\omega_{\alpha}$ and displacement $q_{\alpha}=\sqrt{\frac{\hbar}{2\omega_{\alpha}}}\left(b_{\alpha}^{\dagger}+b_{\alpha}\right)$. Describing the nonequilibrium dynamics of excitons requires knowledge of the excitonic transition energies $E_{n}$, the phonon modes and their corresponding frequencies, $\omega_\alpha$, and the exciton-phonon couplings, $V_{n,m}^{\alpha}$. Sec.~\ref{sec:exph} provides the details for obtaining both the phonon modes using an atomistic force field~\cite{Zhou2013} and the exciton-phonon couplings directly from the atomistic pseudopotential model described in Sec.~\ref{sec:methods}. We compare the predictions for the reorganization energies (\textit{i.e.}, polaron shifts) computed from $V_{n,n}^{\alpha}$ to experimentally measured Stokes shifts and demonstrate that \textit{acoustic} modes that are delocalized across the NC contribute more significantly than optical modes to the reorganization energy in all NC systems and sizes. Excitons in smaller NCs are more strongly coupled to modes localized near the surface of the NC, while excitons in larger NCs are more strongly coupled to modes in the interior of the NC. The assessment of the exciton-phonon couplings is essential for addressing the dynamics and mechanism for exciton cooling. This topic is further discussed in Sec.~\ref{sec:conclusions}.

Next, in Sec.~\ref{sec:auger} we turn to the ``universal volume scaling law'' for AR lifetimes and present our recent developments for calculating AR lifetimes in NCs that have thousands to tens of thousands of electrons.\cite{Philbin2018} We demonstrate that the inclusion of electron-hole correlations in the initial biexcitonic state (but not in the final electron-hole state) is imperative to capturing the experimentally observed scaling of AR lifetimes with the size and shape of the NC. In addition, we find that electron-hole correlations are essential for obtaining quantitatively accurate lifetimes and that neglecting such correlations can result in AR lifetimes that are orders of magnitude too long. We demonstrate the strength of our approach for 0D spherical quantum dots, 1D nanorods, and 2D nanoplatelets of varying diameters and lengths. To perform these calculations and compute AR lifetimes, we developed a low-scaling approach~\cite{Philbin2020a} based on the stochastic resolution of identity,\cite{Takeshita2017,Dou2019} which is briefly summarized in Sec.~\ref{sec:auger}.

The role of interfaces on reorganization energies and AR lifetimes is the central topic of Sec.~\ref{sec:interfaces}. We focus on core-shell quantum dots and elucidate the shell thickness and band alignment dependencies for quasi-type II CdSe/CdS and type I CdSe/ZnS systems. The introduction of interfaces in these heterostructures allows for wave function engineering that affects electron-hole correlations, exciton-phonon couplings, and exciton-exciton interactions, which impacts both the magnitudes of reorganization energies and AR lifetimes. These insights serve as a starting point for realizing NC systems that readily control both exciton-phonon and exciton-exciton interactions, enabling unique, emergent phenomena, such as room-temperature superfluorescence, fast exciton transport, and near-unity photoluminescence quantum yields. Finally, in Sec.~\ref{sec:conclusions} we summarize the main conclusions and provide an outlook for future directions.


\section{Model Hamiltonian}
\label{sec:methods}
The diversity of dynamic processes in NCs requires a comprehensive model that captures a wide spectrum of physics. The finite size of NCs modifies the electronic structure relative to the bulk material. The continuous conduction and valence bands of the bulk are split into discrete states for finite crystals, and the quantum confinement of carriers gives rise to NCs' hallmark size-dependent optical properties. To properly describe these optical properties, a model must go beyond the ground state to describe the excited electronic configurations. While these excited states are generally well understood in the bulk, quantum confinement complicates our understanding by significantly enhancing the electron-hole interactions.\cite{Williamson2000} The small size of NCs compared to the exciton Bohr radius forces the electron and hole closer to each other than they would be in bulk, increasing the strength of their Coulomb interactions. Additionally, dielectric screening is reduced at the nanoscale as quantum confinement widens the band gap and increases the energy required to polarize the medium. This effect leads to a size-dependent reduction in screening, further contributing to size-dependent modifications of excited states in NCs. These enhanced interactions must be properly considered in order to describe the correlations between electrons and holes and in order to achieve agreement with experimental measurements. 

\begin{figure*}[ht]
\includegraphics[width=\textwidth]{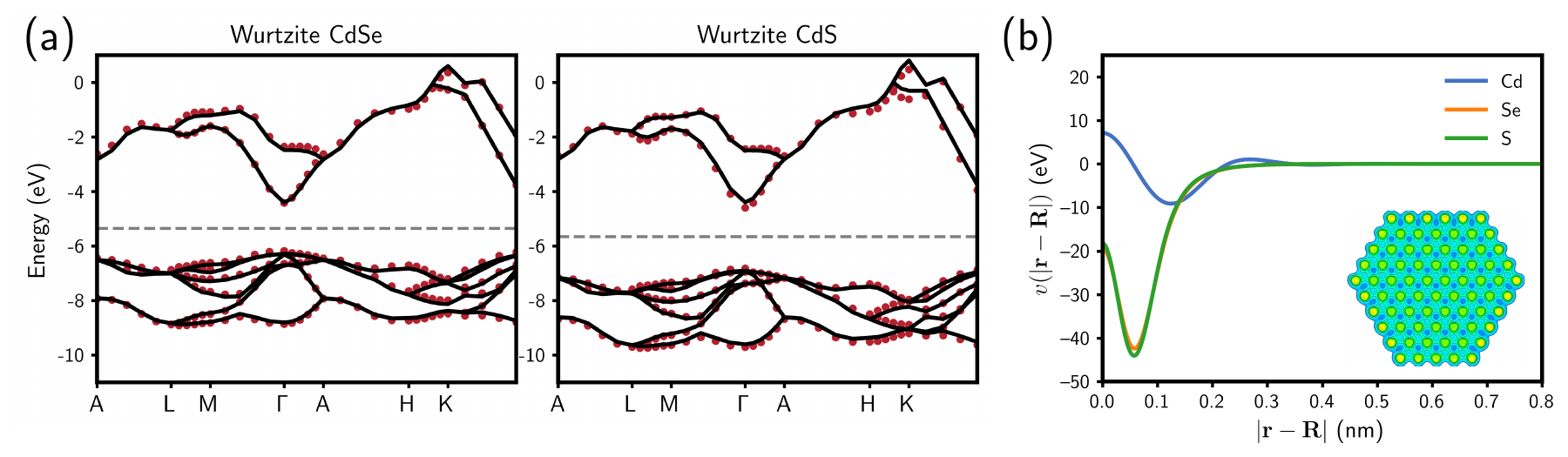}
\caption{\label{fig:fitting}(a) The bulk band structures of wurtzite CdSe (right) and CdS (left) obtained from the pseudopotential Hamiltonian (red points) are compared to literature values\cite{Bergstresser1967} (black lines). The resulting band structures show excellent agreement both around the band gap and across the entire Brilouin zone. (b) The corresponding real-space pseudopotentials for Cd, Se, and S. The inset illustrates a cross-section of the pseudoptential for a wurtzite 3.9~nm CdSe NC as constructed from these atom-centered functions.}
\end{figure*}

Experimentally relevant NCs are highly crystalline, and, in the interior of the structure, they closely resemble the corresponding bulk materials. The atomic configuration aligns closely with the bulk crystalline lattice across the majority of unit cells, suggesting that a description based on bulk bands would be a valid starting point. However, NCs possess additional features that distinguish them from bulk. The NC surface truncates the lattice symmetry, which gives rise to quantum confinement. Core-shell structures also form a nanoscale heterojunction that can introduce significant amounts of strain into the crystal structure.\cite{CaoBanin200, ReissSmall2009} Both these internal interfaces and surfaces cause deformations from crystallinity on the scale of individual atoms, so accurate modeling of NCs must include this atomistic detail. For example, localized trap states at surfaces or interfaces due to atomic defects are ubiquitous in experimental studies of NCs, where they are observed to rapidly quench photoluminescence and result in significantly lower quantum yields.\cite{Wuister2004, Guzelturk2021} An atomistic description of the NC structure allows for the introduction of site-specific defects or alloying to understand their roles in trap formation and to determine the dynamics of trapping in NC systems.\cite{Jasrasaria2020,Enright2022} In addition to the static deformation of the crystal lattice, the effects of lattice fluctuations (\textit{i.e.}, phonons) play a key role in the physics of NCs and must be properly incorporated.\cite{Jasrasaria2021} Finally, in order to make meaningful contact with experimental measurements on NCs that contain thousands of atoms and tens of thousands of electrons, computational evaluation of the model must scale moderately with system size in comparison to first principles approaches. Because NC systems have important size dependent properties, such as optical gaps,\cite{Ekimov1985} radiative lifetimes, and AR lifetimes,\cite{Philbin2020a} and the scaling of these properties with system size is often an important question, the ability to access experimentally relevant sizes with volumes ranging across multiple orders of magnitude is crucial. 

\begin{figure*}[ht]
\includegraphics[width=0.85\textwidth]{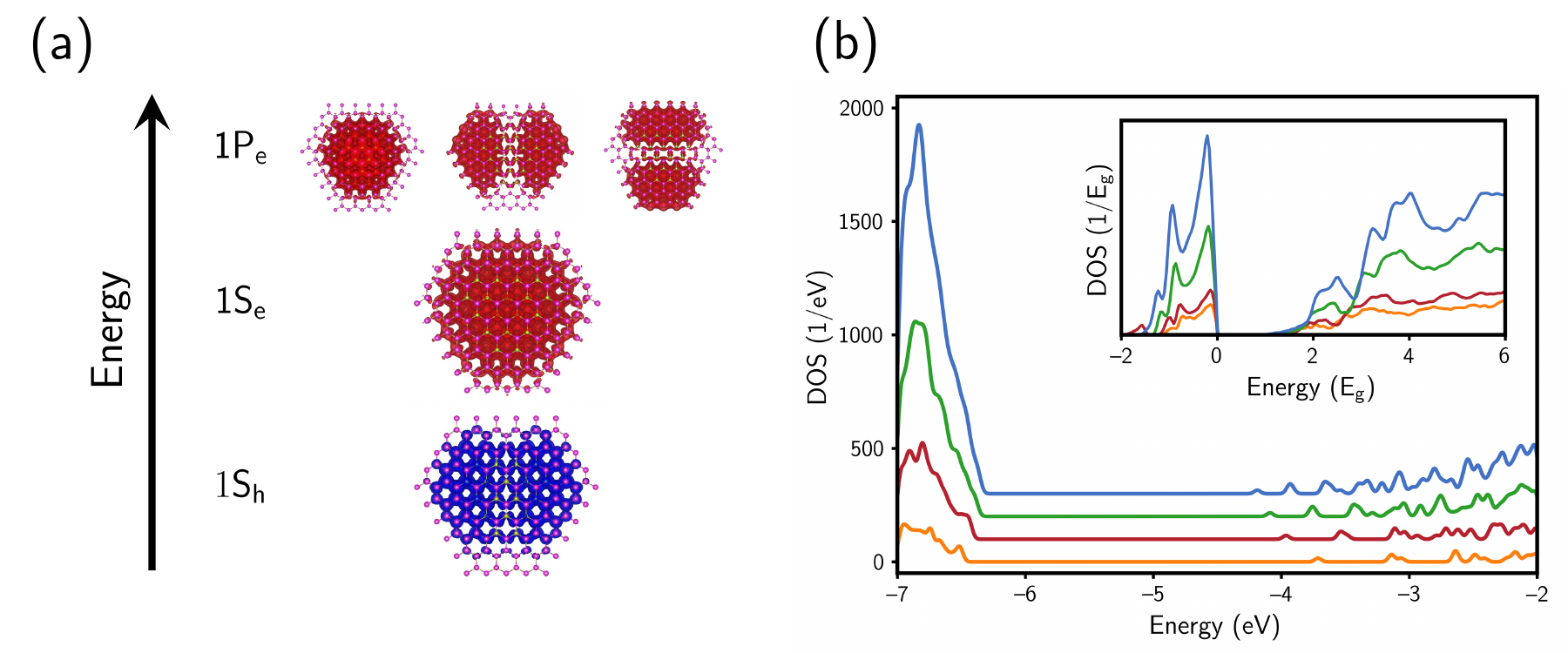}
\caption{\label{fig:states} (a) Densities of the quasi-electron (red) and quasi-hole (blue) wave functions reveal that they are periodic across several unit cells in the interior of the NC. The electron states are labeled based on the symmetry of the envelope function in analogy to effective mass descriptions. (b) The densities of single-particle states (DOS) for wurtzite CdSe NCs of different sizes shows the effects of quantum confinement and the larger density of hole states in these II-VI systems. The inset illustrates the density of states across a larger energy range (that is normalized to the fundamental gap, $E_g$, of each NC), where the continuum of high energy states can be seen.}
\end{figure*}

These considerations have informed our development of the semiempirical pseudopotential model as a sufficiently detailed description of NCs that can also tackle calculations of experimentally relevant systems. For example, a CdSe quantum dot only 4~nm in diameter has over $\sim$1000 atoms and $\sim$4000 valence electrons, so the conventional workhorses of quantum chemistry, such as DFT and related methods for excited states, despite making significant progress,\cite{Shulenberger2021,Song2022} are still far from being able to tackle this problem. On the other hand, continuum models based on the effective mass approximation have produced successful predictions for simple, linear spectroscopic observables~\cite{Efros2000} but are unable to capture many of the more complicated dynamic processes that determine the timescales of process like nonradiative exciton relaxation and AR. Furthermore, these continuum models are, by nature, blind to atomistic detail, such as defects, strain at heterostructure interfaces, and facet-dependent properties.\cite{Ondry2019, Cui2019}

Our approach is based on the semiempirical pseudopotential method,\cite{Wang1994,Rabani1999b,Williamson2000} which was first developed to characterize the band structures of simple bulk materials\cite{Cohen1966} and was later extended to describe the role of surfaces~\cite{Zunger1980} and confinement.\cite{Friesner1991,Wang1994}  The basic assumption made is that the bulk band structure can be described by a simple, non-interacting model Hamiltonian
\begin{equation}
    \hat{h}_{\text{qp}}=\hat{t}+\hat{v}(\bm{r})=\hat{t}+\sum_\mu \hat{v}_\mu(\bm{r})\,,
    \label{eqn:Hamil}
\end{equation} 
where $\hat{t}$ is the single-particle kinetic energy operator, and $\hat{v}(\bm{r})$ is the local (or non-local) pseudopotential, which is given by a sum over all atoms $\mu$ of a pseudopotential $\hat{v}_\mu(\bm{r})$ centered at the location of each atom $\bm{R}_\mu$. The parameters used to describe the pseudopotential of each atom are obtained by fitting the form factors to the bulk band structure obtained either from experimental measurements or high-accuracy electronic structure calculations, such as DFT+GW.\cite{Cohen1966,Hybertsen1986} Within the fitting procedure, we describe the real-space atomistic pseudopotential $\hat{v}_\mu(\bm{r})$ by its reciprocal-space counterpart, $\hat{\tilde{v}}_\mu(\bm{q})$. For example, one popular form of a local reciprocal-space pseudopotential is given by:\cite{Wang1999}
\begin{equation}
    \hat{\tilde{v}}_\mu(\bm{q})=\left[1+a_4\operatorname{Tr}\epsilon_\mu\right]\frac{a_{0}\left(q^{2}-a_{1}\right)}{a_{2}e^{a_{3}q^{2}}-1}\,,
    \label{eqn:LocalPotq}
\end{equation}
where $\epsilon_\mu$ is the local strain tensor around atom $\mu$, the parameters $\{a_0, a_1, a_2, a_3\}$ are used to fit the band structure at the equilibrium configuration, and $a_4$ is fit to match the absolute hydrostatic deformation potentials of the valence and conduction bands.\cite{Wei2006PRB_DefPot} The trace of the local strain tensor at each atom is approximated by the ratio of the volume of the tetrahedron formed by the nearest neighbors in the strained structure to that volume in the equilibrium bulk structure. For NCs with significant strain, such as core-shell QDs or other heterostructures,\cite{Grunwald2013} additional fitting parameters multiplying higher orders of the strain tensor can be included. Furthermore, note that this formalism accounts for hydrostratic strain that occurs due to isotropic compression or expansion of a material, such as the core in spherical core-shell QDs. In anisotropic core-shell nanoplatelets, however, the core experiences biaxial strain,\cite{Talapin2019} which may need to be incorporated into the model using additional terms.

\begin{figure*}[ht]
\includegraphics[width=\textwidth]{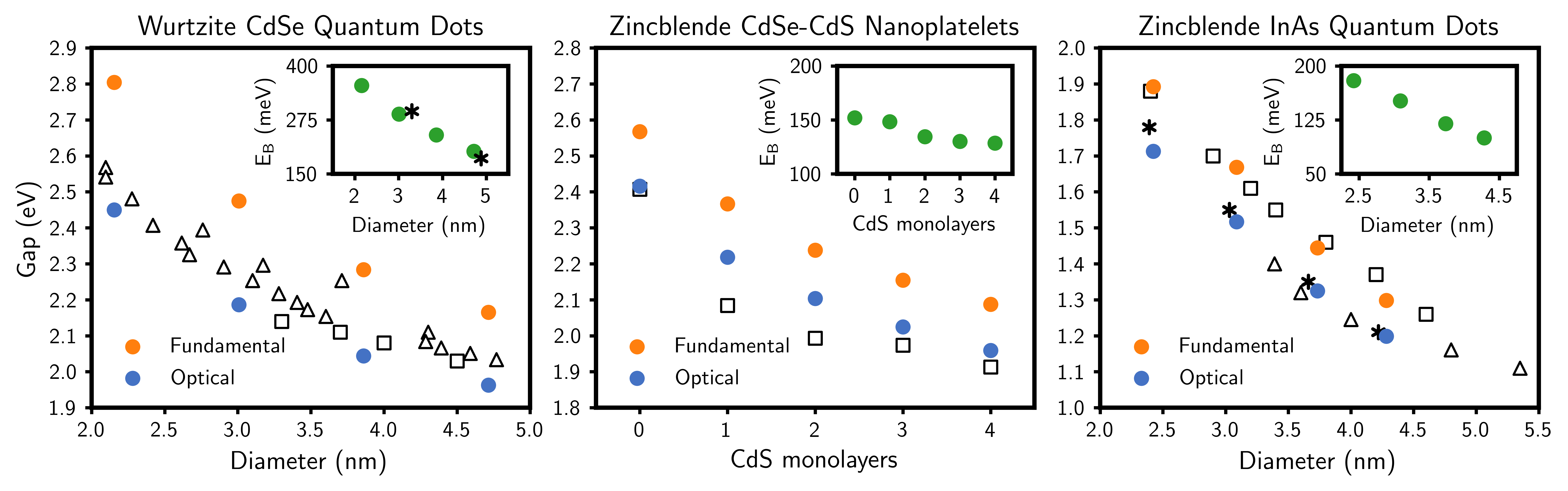}
\caption{\label{fig:validation} Gaps for wurtzite CdSe quantum dots of different sizes (left). The optical gaps computed by our semiempirical pseudopotential method agree with experimental measurements of the optical gap by Fan \textit{et al.}\cite{Zhang2015} (black squares) and Yu \textit{et al.}\cite{Peng2003} (black triangles). The inset shows the exciton binding energy, $E_{\text{B}}$, computed by our method and compared to values computed by Franceschetti and Zunger\cite{Zunger1997} (black asterisks). Gaps for zincblende CdSe-CdS core-shell nanoplatelets with different thicknesses of CdS shell (middle). The optical gaps calculated by our method compare favorably with those measured experimentally by Hazarika \textit{et al.}\cite{Talapin2019} (black squares). Gaps for zincblende InAs quantum dots of different sizes (right). The fundamental gaps calculated are in excellent agreement with those measured by Banin \textit{et al.}\cite{Millo1999} using scanning tunneling microscopy (black squares), and the optical gaps compare well with those measured by Guzelian \textit{et al.}\cite{Alivisatos1997} (black triangles) and computed by Franceschetti and Zunger\cite{Zunger2000} (black asterisks).}
\end{figure*} 

The fitting of parameters $\{a_0, a_1, a_2, a_3\}$ proceeds by comparing the generated band structure to the expected band structure with special care taken to correctly capture the band gaps and effective masses. As shown in Fig.~ \ref{fig:fitting}a, the model captures all band features and describes the band structure across the entire Brillouin zone. The real-space forms of the corresponding pseudopotentials are illustrated in Fig.~\ref{fig:fitting}b, where the psuedopotentials have been simultaneously fit to generate the correct band structures for both wurtzite and zincblende CdSe and CdS. The effects of strain are then incorporated through the $a_4$ parameter (and any necessary higher-order terms) to fit the absolute deformation potentials of both the conduction band minimum and valence band maximum. This fitting procedure ensures that hydrostatic deformation of the crystal alters the energies of the electron and hole levels in the correct manner. 

Once the pseudopotentials have been fit to describe bulk systems (the fits are not unique and often other physical measures are used to choose the best set of parameters~\cite{Wang1996}), they are used to construct the NC Hamiltonian. The central assumption made here is that the pseudopotentials that describe single particle properties in the bulk are adequate also when applied to quantum confined nanostructures. While this might seem a large leap, the error introduced by this assumption is relatively small compared to the fundamental  band gap.\cite{Ogut1997} A cross-section of the resulting pseudopotential for a wurtzite 3.9~nm CdSe NC is shown in the inset of Fig.~\ref{fig:fitting}b, illustrating both the near-periodic potential in the interior of the NC and the manner by which it is modified at the surface. The NC atomic configurations are obtained by first pruning the correctly sized NC from bulk. The atomic positions are then relaxed using molecular dynamics-based geometry optimization with previously-parameterized force fields,\cite{Rabani2002,Zhou2013} which includes two- and three-body terms to enforce tetrahedral bonding geometries, to produce NC configurations that are relatively crystalline in agreement with experiment.\cite{KelleyJCP2016} In the case of core-shell structures, the core is cut from bulk, and the shell material is grown on the surface using the lattice constant of the core material. The subsequent geometry optimization allows the shell to relax and results in compressive strain on the core to minimize the stress along the core-shell interface.\cite{Grunwald2013, Talapin2019}

The description of the surface of the NC presents a challenge, as simply terminating the NC will result in dangling bonds. These dangling bonds can give rise to localized electronic states within the band gap, which act as traps. For the II-VI and III-V families of semiconductors, we have found that dangling bonds from the non-metal atoms result in hole traps slightly above the valence band maximum, but metal dangling bonds do not result in electron traps due to the light electron effective mass relative to the hole effective mass.\cite{Jasrasaria2020,Enright2022} To passivate the surface of the NC, the outermost layer of atoms is replaced with passivation potentials that mimic the effect of organic ligands that terminate the surfaces of experimentally synthesized NCs, pushing the mid-gap electronic states out of the band gap.\cite{Wang1994} This procedure for building NC structures can be easily adapted to produce more complicated NCs, such as the core-shell NCs, nanorods, and nanoplatelets. Further modification, such as alloying, multi-layered NCs, dimer NC assemblies, and structural defects can also be modeled with atomistic detail. 

While a NC of experimentally relevant size will have many single-particle states (see Fig.~\ref{fig:states}b), only a few highest-energy, occupied and lowest-energy, unoccupied states are relevant to describing the optical properties near the band edge. These single-particle states are obtained using the filter diagonalization method,\cite{Wall1995,Toledo2002} which provides a framework to extract all the eigen-solutions within a specific energy window. This process can be done with nearly linear scaling with the system size due to the locality of the single-particle Hamiltonian, making feasible the calculation for NCs with volumes spanning several orders of magnitude. As the pseudopotentials are fit to reproduce quasiparticle band structures, the eigenstates of the pseudopotential Hamiltonian are assumed to describe the quasi-electron and quasi-hole wave functions of the NC. Examples of the quasi-electron and quasi-hole densities are shown in Fig.~\ref{fig:states}a. We see that both the electron and hole states show Bloch-like oscillations, which are significantly more pronounced for the hole, and the electron states show a progression of envelope functions with $s$- then $p$-type characteristics, in line with effective mass descriptions of NC electronic states.\cite{Wang1998,Efros2000}  

As previously stated, connection to experiments also requires an accurate description of the neutral excited states probed by optical spectroscopy. To account for electron-hole correlations, we use the single-particle eigenstates as the basis to solve the Bethe-Salpeter equation (BSE)~\cite{Rohlfing2000} for the correlated excitonic states using the static screening approximation.\cite{Eshet2013} This approach describes electron-hole correlations beyond the standard perturbation technique and is essential to describe excitons even in the strongly confined limit. We take the excitonic states to be a linear combination of noninteracting, electron-hole pair states:
\begin{equation}
    \vert \psi_n \rangle = \sum_{ai} c_{a,i}^n a_a^\dagger a_i \vert 0 \rangle\,,
    \label{eqn:PsiExciton}
\end{equation}
where $a_a^\dagger$ and $a_i$ are electron creation and annihilation operators in quasiparticle states $a$ and $i$, respectively. The indexes $a,b,c,\dots$ refer to quasi-electron (unoccupied) states while the indexes $i,j,k,\dots$ refer to quasi-hole (occupied) states. The expansion coefficients $c_{a,i}$ are determined by solving the eigenvalue equation~\cite{Rohlfing2000}
\begin{equation}
    (E_n-\Delta \varepsilon_{ai})c^n_{a,i}=\sum_{bj}\left(K^d_{ai;bj}+K^x_{ai;bj}\right)c^n_{b,j}\,,
    \label{eqn:BSEEig}
\end{equation}
which also determines the energy of exciton $n$, $E_n$, in terms of the direct and exchange parts of the electron-hole interaction kernel,\cite{Rohlfing2000} $K^d_{ai;bj}$ and $K^x_{ai;bj}$, respectively, and the quasiparticle energy difference $\Delta \varepsilon_{ai}=\varepsilon_a-\varepsilon_i$. The direct part of the kernel describes the main attractive interaction between quasi-electrons and quasi-holes while the exchange part controls details of the excitation spectrum, such as the singlet-triplet splittings. Importantly, the direct term is mediated by a screened Coulomb interaction,\cite{Rohlfing2000} which we approximate using the static screening limit with a dielectric constant that is obtained directly from the quasiparticle Hamiltonian\cite{Williamson2000} and that depends on the size and shape of the NC. The binding energy of the $n$th excitonic state, $E^n_\text{B}$, is calculated as
\begin{equation}
    E^n_\text{B}=\sum_{abij}\left(c^{n}_{a,i}\right)^{*}\left( K^d_{ai;bj}+K^x_{ai;bj}\right)c^{n}_{b,j} .
    \label{eqn:ExBind}
\end{equation}

As this model was built on semiempirical foundations, it is necessary to validate the resulting calculations on well-known NC properties before using the model to explore more complex phenomena. Furthermore, the fitting was carried out on pure bulk materials, so it is important to assess the performance of the model on different NCs across a range of sizes and compositions. One of the most fundamental properties we need to capture is the optical gap. As shown in Fig.~\ref{fig:validation}, we obtain results that compare favorably with experiments with respect to the magnitude of the gap and the scaling with NC size for several different NC compositions and geometries. We additionally validate properties, such as exciton binding energies,\cite{Philbin2018,Brumberg2019}, exciton fine structure effects on polarized emission,\cite{Hadar2017,Brumberg2019} radiative and AR lifetimes,\cite{Philbin2018,Philbin2020a,Philbin2020b} and optical signals of trapped carriers.\cite{Jasrasaria2020} The strong agreement we obtain between theoretical predictions and experimental observations across a variety of system sizes, compositions, and dimensionalities demonstrates that our approach is suitable for understanding and rationalizing trends across a wide range of nanomaterial systems. Additionally, as we will discuss in the following sections, this model is extremely versatile and lends itself to new development and expansion.

\section{Phonons and exciton-phonon couplings}
\label{sec:exph}
Electronic degrees of freedom couple to phonons in semiconductors, resulting in a diverse set of processes that affect electronic properties and dynamics. These electron-phonon interactions in bulk semiconductors tend to be weaker than electron-vibrational interactions in molecular systems because the relatively large dielectric screening in bulk semiconductors leads to delocalized, Wannier-Mott excitons, which do not depend as sensitively on the nuclear configuration as do the localized, Frenkel excitons in molecular systems.\cite{Scholes2006NatMater} Additionally, phonons in bulk semiconductors are also delocalized over the material, unlike localized vibrations in molecules.\cite{Mahan2000}

Semiconductor NCs lie somewhere in between the bulk and molecular limits. Excitons in NCs are delocalized over multiple atoms, but they are confined to the extent of the NC. Similarly, lattice vibrations resemble phonons in bulk semiconductors, but they are finite in number and spatially confined to the NC. The effects of confinement on the magnitude of exciton-phonon coupling (EXPC) are still poorly understood, and the challenges associated with studying EXPC, both experimentally and theoretically, have led to a set of outstanding questions regarding EXPC in semiconductor NCs. A detailed description of EXPC in NCs is essential for understanding the temperature dependence of excitonic properties\cite{Balan2017} and phenomena, such as exciton dephasing (\textit{i.e.}, homogeneous emission linewidths),\cite{Bawendi2016, Mack2017} phonon-mediated carrier relaxation,\cite{Nozik2001, Kilina2009,Peterson2014} and charge transfer.\cite{Kamat2011,Harris2016} 

Before delving into the role of confinement on EXPC, we will examine the phonon states in NCs. Phonon confinement to the spatial extent of the NC results in quantization of the phonon frequencies. This confinement introduces additional complicating factors, such as the role of the NC surface,\cite{Wood2016Nature, Mack2019} that motivate the need for an atomistic description of phonon modes.\cite{KelleyJCP2016} While DFT-based frozen phonon approaches have been used to compute phonon modes and frequencies,\cite{Chou1992, Bester2012, Bester2019} their computational expense restricts these methods to small NCs or to the computation of specific modes which are known \textit{a priori} to be relevant for the properties or processes of interest. Therefore, we model phonons using classical, atomic force fields, which allows for the computation of all phonon modes and frequencies in NC systems of experimentally relevant sizes. The dynamical matrix, or mass-weighted Hessian, can be computed at the equilibrium configuration of a NC:\cite{Kong2011}
\begin{equation}
    D_{\mu k, \mu^\prime k^\prime} = \frac{1}{\sqrt{m_\mu m_{\mu^\prime}}} \bigg( \frac{\partial^2 U(\boldsymbol{R})}{\partial u_{\mu k} \partial u_{\mu^\prime k^\prime}} \bigg)_{\boldsymbol{R}_0}\,,
    \label{eqn:dynamical}
\end{equation}
where $U(\boldsymbol{R})$ is the potential energy given by the force field, $u_{\mu k}=R_{\mu k} - R_{0,\mu k}$ is the displacement of nucleus $\mu$ away from its equilibrium position in the $k\in\{x,y,z\}$ direction, and $m_\mu$ is the mass of nucleus $\mu$. This $3N \times 3N$ dynamical matrix, where $N$ is the number of atoms in the NC, can be diagonalized to obtain the phonon mode frequencies and coordinates.

\begin{figure*}[ht]
\includegraphics[width=0.85\textwidth]{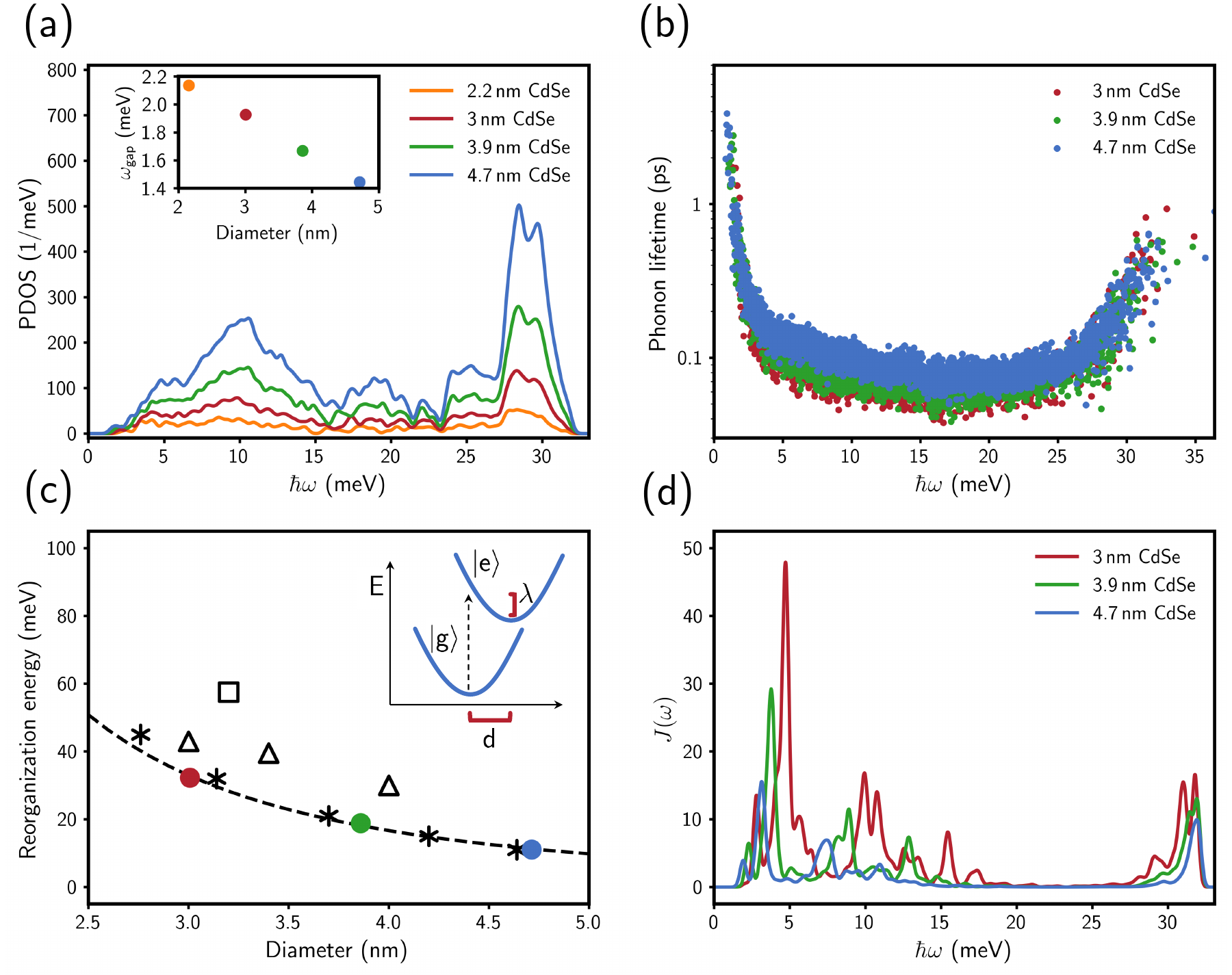}
\caption{\label{fig:ExPh} (a) The phonon densities of states (PDOS) calculated for wurtzite CdSe NCs of different sizes. The inset shows the phonon gap, $\omega_\text{gap}$, which decreases with increasing NC size. (b) Phonon lifetimes calculated for wurtzite CdSe NCs. For all systems, almost all modes have sub-picosecond lifetimes, indicating significant phonon-phonon coupling. (c) Reorganization energies for wurtzite CdSe NCs. Values calculated by our approach are compared to experimental measurements by Liptay \textit{et~al.}\cite{Bawendi2007} (black triangles) and Salvador \textit{et~al.}\cite{Scholes2006JCP} (black squares) as well as to calculations using an effective mass model by Kelley\cite{Kelley2011} (black asterisks). The inset schematically depicts the reorganization energy, $\lambda$, which corresponds to the energy of lattice rearrangement after vertical excitation from the electronic ground state $\vert g \rangle$ to the excited state $\vert e \rangle$. The ground and excited state minima are displaced along the phonon mode coordinate $q$ by a distance $d$. (d) The spectral densities, $J(\omega)$, calculated for wurtzite CdSe NCs show significant coupling to lower-frequency acoustic modes and to optical modes around $\hbar \omega \sim 30$~meV.}
\end{figure*}

The phonon densities of states (PDOS) for wurtzite CdSe NCs of different sizes computed using a previously-parameterized Stillinger-Weber interaction potential\cite{Zhou2013} are illustrated in Fig.~\ref{fig:ExPh}a. Acoustic modes, which involve in-phase motion of atoms, have lower frequencies (1\,THz\,$\sim$\,5\,meV or lower in CdSe NCs) while optical modes, which are made up of out-of-phase movements of atoms, have higher frequencies (4\,THz\,$\sim$\,16\,meV or higher).\cite{Grunwald2012} Modes at intermediate frequencies are difficult to characterize due to the overlap of acoustic and optical branches in the bulk phonon dispersion relation as well as the confounding effects of phonon confinement. Phonon confinement also leads to a gap in the PDOS near zero-frequency since the longest-wavelength (\textit{i.e.}, lowest-frequency) phonon mode in a NC is dictated by the NC size. As shown in the inset of Fig.~\ref{fig:ExPh}a, the lowest-frequency phonon mode in the system is inversely proportional to the NC diameter, as observed by Raman spectroscopy measurements.\cite{Weller2008, Tisdale2016}

This zero-frequency gap has led to a hypothesized hot phonon bottleneck in NCs, in which phonon-phonon scattering rates are slow because of the lack of low-frequency modes, and phonon thermalization becomes the rate-limiting step in processes, such as hot carrier relaxation.\cite{Ploog1988,Klimov1995,Li2017} However, we have found that phonons have significant coupling with one another (\textit{i.e.}, are anharmonic) at room temperature (300\,K).\cite{Guzelturk2021} We performed molecular dynamics simulations using a force field that consists of Lennard-Jones and Coulomb terms\cite{Rabani2002} to compute phonon relaxation lifetimes of CdSe NCs, shown in Fig.~\ref{fig:ExPh}b. Unlike the Stillinger-Weber potential, this force field includes long-range interactions that are necessary to accurately describe the splitting between acoustic and optical branches at the Brillouin zone edge in bulk polar semiconductors\cite{Bester2017} and, thus, the phonon lifetimes. Our calculations are within linear response theory, so they assume phonon modes are only weakly excited, but no assumptions are made about the strength of coupling between different phonon modes. The lifetimes, which are dictated by the phonon-phonon interactions, are sub-picosecond for all modes except for the lowest-energy acoustic modes, for which the lifetimes reach $\sim$4\,ps. Smaller NCs have shorter lifetimes because surface atoms, which are proportionally larger in number in smaller NCs, have increased anharmonic motion that leads to greater phonon-phonon coupling. For all systems, the phonon dynamics are overdamped, in agreement with experimental observations,\cite{Lounis2014,Lindenberg2015} and the relaxation timescales are less than the periods of the phonon modes. We expect that phonon modes that are strongly out of equilibrium would thermalize even more quickly, indicating that a hot phonon bottleneck is unlikely in these systems. (Note that the ``hot phonon bottleneck" is distinct from the ``phonon bottleneck", which describes slow carrier relaxation due to the mismatch between electronic gaps and phonon frequencies, that is discussed in Secs.~\ref{sec:intro} and \ref{sec:conclusions}.) 

With this discussion of NC phonons in mind, we now turn to discuss the EXPC terms, $V_{n,m}^{\alpha}$, that appear in Eq.~\eqref{eq:hamiltonian}. Historically, studies of EXPC in NCs have described electronic states within parameterized models and/or have described phonons as vibrations of an elastically isotropic sphere,\cite{Flytzanis1990,Kobayashi1992,Kartheuser1994, Takagahara1996, Zorkani2007, Kelley2011} leading to widely varying results, as the magnitude of EXPC is extremely sensitive to the descriptions of both excitons and phonons. Fully atomistic \textit{ab initio} methods have been used for small NCs of $\sim$100 atoms or fewer, for which the electron-phonon coupling can be inferred from fluctuations of the adiabatic electronic states that are generated ``on the fly" within DFT and time-dependent DFT frameworks.\cite{Prezhdo2005PRL, Prezhdo2013JCTC, Wood2018NL, Wood2021} While these methods have been moderately effective in modeling electron-phonon coupling, they are often limited to small systems due to the computational expense of DFT, and they ignore excitonic effects. One recent study that did include these excitonic effects using \textit{ab initio} methods was limited to clusters of tens of atoms.\cite{Zeng2021}

Instead, we rely on the semiempirical pseudopotential model to describe excitonic states and atomic force fields to describe phonons.\cite{Jasrasaria2021} Within this framework, the standard electron-nuclear matrix element to first order in the atomic displacements is given by:\cite{Giustino2017}
\begin{equation}
    V_{n,m}^{\mu k} \equiv \bigg\langle \psi_n \bigg\vert \bigg( \frac{\partial \hat{v}(\bm{r})}{\partial R_{\mu k}} \bigg)_{\boldsymbol{R}_0} \bigg\vert \psi_m \bigg\rangle\,,
\end{equation}
where $\vert \psi_{n} \rangle$ is the state of exciton $n$ (cf., Eq.~\eqref{eqn:PsiExciton}), $\hat{v}(\bm{r}) = \sum_\mu \hat{v}_\mu(\boldsymbol{r})$ is the sum over atomic pseudopotentials given in Eq.~(\ref{eqn:Hamil}), $R_{\mu k}$ is the position of atom $\mu$ in the $k\in \{x,y,z\}$ direction, and $\boldsymbol{R}_0$ is the equilibrium configuration of the NC. Using the static dielectric BSE approximation for the excitonic wave function, we can reduce the calculation of these matrix elements to a simpler form given by:\cite{Jasrasaria2021}
\begin{equation}
    V_{n,m}^{\mu k} = \sum_{abi} c_{a,i}^n c_{b,i}^m v_{ab,\mu}^\prime (R_{\mu k}) - \sum_{aij} c_{a,i}^n c_{a,j}^m v_{ij,\mu}^\prime (R_{\mu k})\,,
\label{eqn:ExPh_ElHoleChannels}
\end{equation}
where
\begin{equation}
    v_{rs,\mu}^\prime = \int d\boldsymbol{r} \phi^{*}_r (\boldsymbol{r}) \frac{\partial \hat{v}(\bm{r})}{\partial R_{\mu k}} \phi_s (\boldsymbol{r})\,.
\label{eqn:single-ElHoleChannels}
\end{equation}
Here, $c_{a,i}^n$ represent the Bethe-Salpeter coefficients introduced in Eq.~\eqref{eqn:PsiExciton}, and $\phi_a(\boldsymbol{r})$ are the real-space quasiparticle wave functions. The first term in Eq.~\eqref{eqn:ExPh_ElHoleChannels} represents the electron channel of EXPC, in which excitons comprised of different single-particle electron states are coupled. The second term describes the hole channel, in which excitons comprised of different single-particle hole states are coupled. These matrix elements can be transformed to phonon mode coordinates using the eigenvectors of the dynamical matrix:
\begin{equation}
    V_{n,m}^\alpha = \sum_{\mu k} \frac{1}{\sqrt{m_\mu}} e_{\alpha,\mu k}^{-1} V_{n,m}^{\mu k}\,,    
\end{equation}
where $e_{\alpha, \mu k}$ is the $\mu k$ element of the $\alpha$ eigenvector of the dynamical matrix given in Eq.~\eqref{eqn:dynamical}, and $m_{\mu}$ is the mass of atom $\mu$. The diagonal matrix elements $V_{n,n}^\alpha$ describe the renormalization of the energy of exciton $n$ through its interaction with phonon mode $\alpha$, and the off-diagonal matrix elements $V_{n,m}^\alpha$ describe the interaction of excitons $n$ and $m$ through the absorption or emission of a phonon of mode $\alpha$.

One measure of the EXPC is the reorganization energy,\cite{Ekimov1993, Scholes2006JCP} depicted schematically in the inset of Fig.~\ref{fig:ExPh}c, which is the energy associated with rearrangement of the NC lattice after exciton formation and is relevant for optical Stokes shifts, charge transfer processes, and NC-based device efficiencies. In the harmonic approximation, the total reorganization energy for a NC is the sum of reorganization energies for each mode, $\lambda = \sum_\alpha \lambda_\alpha$, where
\begin{equation}
    \lambda_\alpha = \frac{2}{Z} \sum_n e^{-\beta E_n} \bigg( \frac{1}{2\omega_\alpha} V_{n,n}^\alpha \bigg)^2\,.
\label{eq:renormalization}
\end{equation}
The above equation includes a Boltzmann-weighted average over excitonic states, where $Z=\sum_n e^{-\beta  E_n}$ is the partition function, $\beta = \frac{1}{k_{\rm B}T}$, and $T$ is the temperature. For wurtzite CdSe NCs ranging from 3 to 5~nm in diameter, we calculated the reorganization energy to be between 60 and 20~meV, which is in good agreement with experimentally measured values\cite{Scholes2006JCP, Bawendi2007} and previous effective mass model calculations\cite{Kelley2011}, as shown in Fig.~\ref{fig:ExPh}c. The remarkable agreement is an important validation of the semiempirical technique and is essential for describing the nonadiabatic transitions involved in exciton cooling. To investigate the contribution of each mode to the overall reorganization energy, we examine the spectral density, or weighted density of states, ($J(\omega) = \sum_\alpha \lambda_\alpha \delta (\omega - \omega_\alpha)$), which is illustrated in Fig.~\ref{fig:ExPh}d. Lower-frequency acoustic modes, which tend to involve collective motions of many atoms in the NC, are more significantly coupled to the exciton. Higher-frequency optical modes have weaker EXPC, but they have a large density of phonon states around 25~meV and 30~meV, which are at an energy scale that is more relevant for excitonic transitions. This feature suggests that optical modes may, in fact, be more important than acoustic modes for phonon-mediated exciton dynamics, but further assessment of the model and the EXCP is required for developing a better understanding of the exciton cooling process. The spectral density, however,  may explain discrepancies in experimental results, some of which find stronger exciton coupling to acoustic modes\cite{Scholes2006JCP, Anni2007, Kambhampati2008PRB, Kambhampati2008JPCC} while others observe stronger exciton coupling to optical modes.\cite{Kobayashi1992, Righini1996, Bimberg1999, Kelley2015ACSNano, Lin2015} Exciton formation causes NC lattice distortion primarily along the phonon coordinates of acoustic modes while optical modes may play a larger role in exciton dynamics.

Furthermore, our calculations show that excitons in all core and core-shell NCs are more strongly coupled to phonons \textit{via} the hole channel (\textit{i.e.}, the second term in Eq.~(\ref{eqn:ExPh_ElHoleChannels})) than through the electron channel.\cite{Jasrasaria2021} This effect is a result of the heavier effective hole mass, which makes hole states more sensitive to nuclear configuration and decreases the energy spacing between hole states, allowing them to couple more readily \textit{via} phonon absorption or emission. Moreover, we have found that phonon modes localized to the surface of the NC have significant contributions to the overall reorganization energy in small NCs because of increased surface strain and strong exciton confinement, which causes the exciton wave function to extend to the NC surface. This surface effect decreases drastically as the NC size increases.

Our framework includes electron-hole correlations, exciton-phonon coupling, and phonon-phonon interactions, enabling our ongoing work to address open questions regarding timescales and mechanisms of phonon-mediated exciton dynamics, such as hot exciton cooling. This atomistic theory can simultaneously study both hypothesized mechanisms -- the Auger decay mechanism, which would occur on the order of picoseconds, and the slower phonon-mediated transitions -- allowing a unified mechanism to emerge from the theory to explain experimental results in a range of NC systems.

\section{Auger recombination}
\label{sec:auger}

Auger recombination (AR) is the primary nonradiative, Coulomb-mediated, exciton-exciton decay channel of multiexcitons, in which one exciton recombines and transfers its energy to an additional electron-hole pair, as schematically illustrated in Fig.~\ref{fig:AllProcesses}, on timescales of a few hundreds of picoseconds.\cite{Klimov2014}  The energy given to the second exciton primarily excites one of the carriers, which then quickly dissipates the excess energy \textit{via} phonon emission.\cite{Achermann2006,Harvey2018} While fast AR in NCs is often responsible for decreased photoluminescence quantum yields and device efficiencies, it also makes NCs potentially useful as single photon sources.\cite{Correa2012,Utzat2019} AR lifetimes are commonly measured using time-resolved photoluminescence and transient absorption experiments.\cite{Klimov2000,Klimov2008,Robel2009,Baghani2015,Ben-Shahar2016,Li2016,Pelton2017} Recent experiments have demonstrated that for quasi-0D quantum dots (QDs),~\cite{Garcia-Santamaria2011} quasi-1D nanorods,~\cite{Stolle2017} and quasi-2D nanoplatelets,~\cite{She2015, Li2017, Philbin2020b} AR lifetimes increase linearly with the volume of the NC. Understanding the key factor that leads to this so-called ``universal volume scaling" of AR lifetimes is key to further controlling AR processes. Understanding the dependence of AR lifetimes on NC size, shape, and composition as well as on the number of excitons present in the NC is central to our understanding of this many-body relaxation process and will provide tools to control AR lifetimes, as further discussed below.

Prior to focusing on the microscopic origins of AR, we will discuss the nature of Coulomb-mediated interactions within and between excitons and their particular importance in confined semiconductors. The strength of Coulomb-mediated interactions within an exciton is normally characterized by the exciton binding energy (see Eq. \eqref{eqn:ExBind}), which is typically $\sim$10~meV 
in bulk semiconductors. Due the small sizes and reduced dimensionalities of NCs, Coulomb interactions are enhanced, leading to exciton binding energies that are greater than $100$~meV in QDs,~\cite{Franceschetti1997} nanorods,~\cite{Baskoutas2005,Rajadell2009,Royo2010} and nanoplatelets.\cite{Scholes2006NatMater,Scott2016,Rajadell2017,Brumberg2019} As this exciton binding energy is much greater than the thermal energy scale at room temperature ($k_{\text{B}}T\sim25$~meV), electrons and holes readily form bound excitons in NCs. The physics of electrons and holes forming bound, correlated electron-hole pairs impacts almost all physical processes, both radiative and nonradiative, in NCs. The enhancement of Coulomb interactions in NCs also affects interactions between excitons. Bulk materials require optical excitation from intense lasers to reach the exciton densities at which Coulomb-mediated exciton-exciton interactions are important. However, the small volumes of NCs lead to these large exciton densities even with just two excitons on a NC. Moreover, two excitons on a single NC will have significant wave function overlap, further increasing Coulomb interactions between excitons in confined systems. These enhanced Coulomb interactions are the primary reason for significant AR in NCs. 

We will now dive into the details of the approach we have developed for computing AR lifetimes, $\tau_{\text{AR}}$. AR is a process by which an initial biexcitonic state, $\left|B\right\rangle $, of energy $E_{B}$ decays into a final excitonic state, $\left|S\right\rangle$, of energy $E_{S}$ \textit{via} Coulomb scattering, $V$. AR lifetimes of a NC can be calculated using Fermi's golden rule, where we average over thermally distributed initial biexcitonic states and sum over all final decay channels into single excitonic states:
\begin{eqnarray}
\tau_{\text{AR}}^{-1} & = & \sum_{B}\frac{e^{-\beta E_{B}}}{Z_{B}}\frac{2\pi}{\hbar}\sum_{S}\left|\left\langle B\left|V\right|S\right\rangle \right|^{2}\delta\left(E_{B}-E_{S}\right)\,.\label{eq:fermisGoldenRule}
\end{eqnarray}
In the above, the Dirac delta function ($\delta\left(E_{B}-E_{S}\right)$) enforces energy conservation between the initial and final states and the partition function, $Z_{B}=\sum_{B}e^{-\beta E_{B}}$, describes a thermal average of initial biexcitonic states (assuming Boltzmann statistics for biexcitons). Despite the known fact that electron-hole interactions in NCs are significant, AR lifetimes had previously been calculated by approximating the initial biexcitonic state as two quasi-electrons and two quasi-holes, without any correlations between them.\cite{Chepic1990, Wang2003, Cragg2010, Vaxenburg2015, Vaxenburg2016} Mathematically, this approximation yields the initial biexcitonic state as 
\begin{eqnarray}
\left|B\right\rangle ^{(0)} & = & a_{b}^{\dagger}a_{j}a_{c}^{\dagger}a_{k}\left|0\right\rangle \otimes\left|\chi_{B}\right\rangle ,
\label{eq:nonintBiexciton}
\end{eqnarray}
and $E_B^{(0)} = \varepsilon_b-\varepsilon_j+\varepsilon_c-\varepsilon_k$ where the superscript ``$(0)$'' signifies that a noninteracting formalism is used.
In the above, $a^\dagger_b$ and $a_j$ are electron creation and annihilation operators in quasiparticle states $b$ and $j$, respectively, as defined in Sec.~\ref{sec:methods}, and $\left|\chi_{B}\right\rangle $ is the spin part of the biexciton wavefunction. 

\begin{figure*}[htb]
\includegraphics[width=\textwidth]{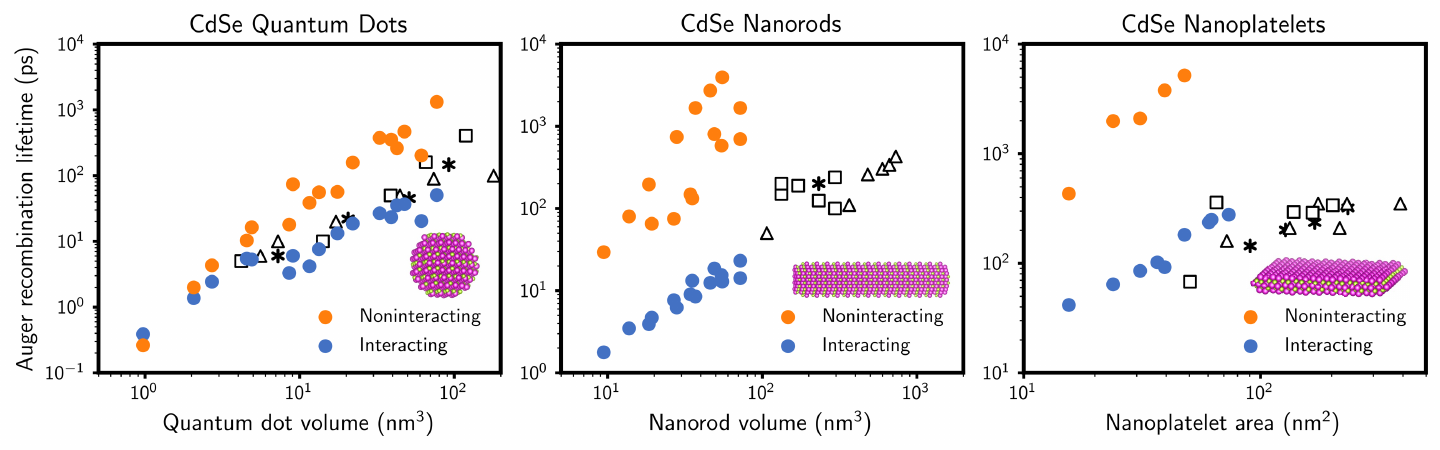}
\caption{\label{fig:AR}Biexciton Auger recombination (AR) lifetimes, $\tau_{\text{AR}}$, for CdSe quantum dots as a function of volume (left). Calculations\cite{Philbin2018} are compared to experimentally measured lifetimes by Taguchi \textit{et al.}\cite{Taguchi2011} (black squares), Htoon \textit{et al.}\cite{Htoon2003} (black triangles), and Klimov \textit{et al.}\cite{Klimov2000} (black asterisks), demonstrating excellent agreement between the interacting formalism and experiment. AR lifetimes for CdSe nanorods as a function of volume (middle). Calculated lifetimes\cite{Philbin2018} are shown with those measured by Taguchi \textit{et al.}\cite{Taguchi2011} (black squares), Htoon \textit{et al.}\cite{Htoon2003} (black triangles), and Zhu \textit{et al.}\cite{Zhu2012} (black asterisks). AR lifetimes for $4$~monolayer CdSe nanoplatelets as a function of nanoplatelet area. Calculated lifetimes\cite{Philbin2020a} compare well with measurements by Philbin \textit{et al.}\cite{Philbin2020b} (black squares), She \textit{et al.}\cite{She2015} (black triangles), and Li and Lian\cite{Li2017} (black asterisks). For all systems, the interacting formalism predicts the same volume scaling as experiment while the noninteracting formalism predicts a steeper scaling with NC volume.}
\end{figure*}

Intuitively, this approximation is only valid in the limit where the kinetic energy is much larger than the exciton binding energy (\textit{i.e.}, for very small quasi-0D QDs in the very strong confinement limit, as shown in Fig.~\ref{fig:AR}), and it quickly breaks down with increasing QD size.\cite{Philbin2018} Furthermore, this approximation results in computed AR lifetimes that are orders of magnitude too long for quasi-1D nanorods and quasi-2D nanoplatelets (Fig.~\ref{fig:AR}).\cite{Philbin2018, Philbin2020b} While this approximation to the initial biexcitonic state is conceptually and computationally simple, it leads to discrepancies between theoretical predictions and experimental measurements on the volume dependence of AR lifetimes in colloidal QDs that persisted for over $20$~years.

Beyond this approximation, the initial biexcitonic state can be written as a combination of two excitonic states that includes electron-hole correlations within each exciton\cite{Refaely-Abramson2017, Philbin2018} but that ignores correlations between excitons, which are typically two (or more) orders of magnitude weaker. The initial biexcitonic state within this formalism, which we previously termed the interacting formalism, is given by
\begin{eqnarray}
\left|B\right\rangle  & = & \sum_{b,j}\sum_{c,k}c_{b,j}^{n}c_{c,k}^{m}a_{b}^{\dagger}a_{j}a_{c}^{\dagger}a_{k}\left|0\right\rangle
\otimes\left|\chi_{B}\right\rangle ,
\label{eq:B I}
\end{eqnarray}
where the excitonic coefficients $c_{b,j}^{n}$ and $c_{b,j}^{m}$
are determined by solving the Bethe-Salpeter equation,\cite{Rohlfing2000} as detailed in Sec.~\ref{sec:methods}. In this formalism the energy of this biexcitonic state is $E_B = E_n+E_m$, which is simply the sum of the two exciton energies. Thus, a deterministic calculation of the AR lifetime can be performed using~\cite{Philbin2018}
\begin{widetext}
\begin{eqnarray}
\tau_{\text{AR}}^{-1} & = & \frac{2\pi}{\hbar Z_{B}}\sum_{B}e^{-\beta E_{B}}\sum_{a,i}\left|\sum_{b,c,k}c_{b,i}^{n}c_{c,k}^{m}V_{abck}\right|^{2}\delta\left(E_{B}-\varepsilon_{a}+\varepsilon_{i}\right)\label{eq:dIntAR}\\
 &  & +\frac{2\pi}{\hbar Z_{B}}\sum_{B}e^{-\beta E_{B}}\sum_{a,i}\left|\sum_{j,c,k}c_{a,j}^{n}c_{c,k}^{m}V_{ijck}\right|^{2}\delta\left(E_{B}-\varepsilon_{a}+\varepsilon_{i}\right).\nonumber 
\end{eqnarray}
\end{widetext}
In Eq.~\eqref{eq:dIntAR}, the first term on the right hand side indicates the electron channel, in which the electron of the final state is excited, and the second term corresponds to the hole channel, in which the hole of the final state is excited. The final states are still approximated by noninteracting electron-hole pairs, $\vert S \rangle = a_a^\dagger a_i \vert 0 \rangle$, for which $E_S = \varepsilon_{a}-\varepsilon_{i} $. This representation of the final state is a reasonable approximation given that the final states are high in energy (Fig.~\ref{fig:AllProcesses}), above the dissociation energy of the exciton. Eq.~\eqref{eq:dIntAR} was first shown to predict quantitatively accurate AR lifetimes for QDs and nanorods\cite{Philbin2018} and was then extended and applied to large core-shell QDs~\cite{Philbin2020a} and nanoplatelets~\cite{Philbin2020b} using stochastic orbital techniques to reduce the computational cost of the interacting formalism given by Eq.~\eqref{eq:dIntAR}. Specifically, stochastic orbitals were used to sample the final states \textit{via} the stochastic resolution of the identity\cite{Takeshita2017,Dou2019} and also to represent the Coulomb operator.\cite{Neuhauser2016} The overall computational scaling of the stochastic implementation of the interacting formalism was multiple factors of the system size lower than the deterministic implementation.\cite{Philbin2020a}   

We have yet to find a system in which Eq.~\eqref{eq:dIntAR} and the underlying approximation of treating the initial biexcitonic state as a product of two correlated excitonic states (Eq.~\eqref{eq:B I}) fail to agree with experimental AR lifetimes. However, future work may need to treat the initial state as a fully-correlated biexcitonic state given by
\begin{eqnarray}
\left|B\right\rangle  & = & \sum_{b,c,j,k}c_{b,c,j,k}a_{b}^{\dagger}a_{j}a_{c}^{\dagger}a_{k}\left|0\right\rangle \otimes\left|\chi_{B}\right\rangle .\label{eq:fullBiexciton}
\end{eqnarray}
In particular, there should be a size at which the AR lifetimes become independent of the volume of the QD (perhaps above the biexcitonic radius), where such exciton-exciton correlation becomes important. For example, recent advances in synthesizing single NCs that have multiple, spatially separated excitonic sites\cite{Cui2019,Koley2021,Philbin2021arxiv} may require the inclusion of all possible quasiparticle-quasiparticle correlations to accurately model the decay of biexcitonic states.

\begin{figure*}[htb]
\includegraphics[width=0.85\textwidth]{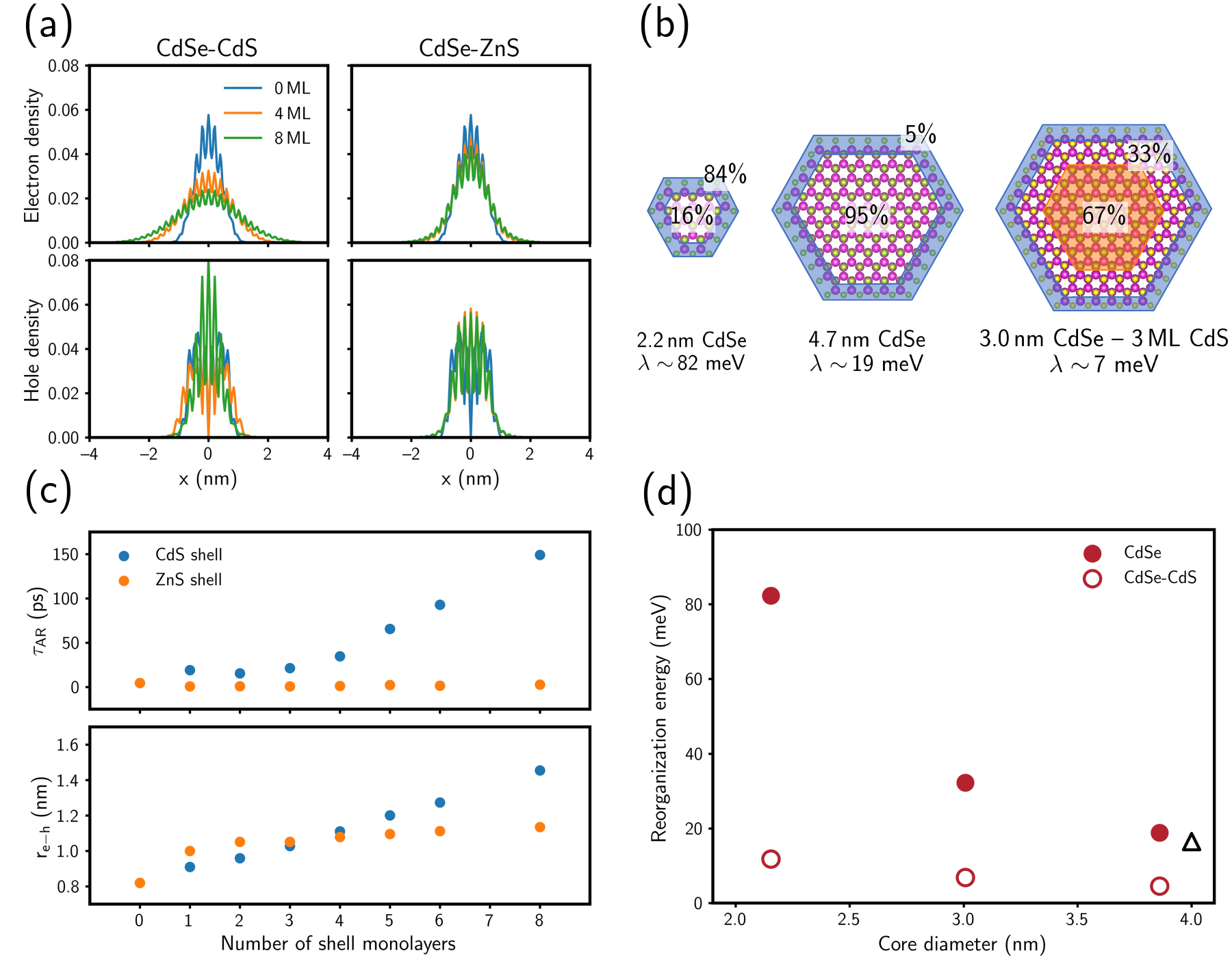}
\caption{\label{fig:Core-Shell} (a) Projected electron (top) and hole (bottom) carrier densities of the ground excitonic state for CdSe/CdS (left) and CdSe/ZnS (right) core-shell quantum dots with a core diameter of 2.2~nm and different shell thicknesses.\cite{Philbin2020a} (b) The large contribution of surface modes to the reorganization energy in small CdSe NCs can be mitigated by the addition of a passivating shell, lowering the overall reorganization energy. (c) Auger recombination lifetimes, $\tau_\text{AR}$, (top) and root-mean-square exciton radii, $r_\text{e-h} = \sqrt{\langle r_\text{e-h}^2 \rangle}$, (bottom) of CdSe/CdS and CdSe/ZnS core-shell quantum dots as a function of shell monolayers for a CdSe core diameter of 2.2~nm.\cite{Philbin2020a} (d) The reorganization energies of bare CdSe quantum dots are significantly larger than those of CdSe cores with 3 monolayers of CdS shell. The black triangle corresponds to the experimentally measured reorganization energy of a CdSe/CdS core-shell particle with a core diameter of 4~nm and 3~monolayers of shell by Talapin \textit{et al.}\cite{Weller2008}}
\end{figure*}

Returning to the scaling of the AR lifetime with NC size, the universal volume scaling of the AR lifetime with the volume of the QD ($\tau_{\text{AR}} \propto V_{\text{QD}}$) is shown in Fig.~\ref{fig:AR}. This volume scaling can be understood from analyzing the volume dependence of the Coulomb coupling and density of final states used to calculate the AR lifetime in Eq.~\eqref{eq:fermisGoldenRule}. The density of final states increases linearly with the volume of the QD, as it also does for nanorods and nanoplatelets.\cite{Rabani2010,Baer2012,Philbin2018,Philbin2020b} However, the decreasing Coulomb coupling between the initial biexcitonic state and final high energy excitonic states with increasing system size counteracts the increasing number of final states. We previously reported that these Coulomb couplings decrease with the square of the QD volume in the interacting formalism, such that the overall AR lifetime increases linearly with the volume of the QD, as found experimentally. In AR lifetime calculations that utilize noninteracting biexcitonic states (Eq.~\eqref{eq:nonintBiexciton}), the Coulomb couplings decrease too fast with the QD volume, which was responsible for the disagreement between theoretical predictions and experimental measurements of the scaling of the AR lifetime with QD volume. The inclusion of electron-hole correlations in the initial biexcitonic state leads to a less steep volume dependence of these Coulomb couplings, as they are related to the electron-hole overlap. Increasing this overlap by properly accounting for the attractive interactions between electrons and holes, as is done in the interacting formalism, leads to increased Coulomb couplings. 

Thus far, we have been concerned with understanding the decay of initial biexcitonic states \textit{via} AR. The decay of multiexcitonic states can be modeled by classical master equations (\textit{i.e.}, classical rate equations) that use decay rates of single excitons and biexcitons. These methods have proven to be surprisingly accurate for modeling the decay of a general number of excitons, $N_{\text{exc}}$, in a NC\cite{Ben-Shahar2018,Yan2021} as well as Auger heating,\cite{Guzelturk2021} or the long-lived heating of the NC lattice that occurs due to the sequence of AR events and subsequent hot carrier cooling.\cite{Achermann2006,Harvey2018}  To this end, the rate that $N_{\text{exc}}$ excitons decays to $(N_{\text{exc}}-1)$ excitons can be well-approximated by modeling AR as a bimolecular collision between excitons, such that the overall AR rate ($K_{\text{AR}}$) is given by 
\begin{eqnarray}
K_{\text{AR}} & = & {N_{\text{exc}} \choose 2} k_{\text{AR}}\,, \label{eq:totalARRate}
\end{eqnarray}
where $k_{\text{AR}}$ is the inverse of the biexciton AR lifetime ($k_{\text{AR}}=\tau_{\text{AR}}^{-1}$) and $N_{\text{exc}} \choose 2$ is the binomial coefficient equal to $N_{\text{exc}}(N_{\text{exc}}-1)/2$. 

The modeling of AR in terms of a bimolecular collision between excitons seems to be consistent with our findings that the interacting formalism, which includes the physics of electrons and holes forming correlated electron-hole pairs (excitons), predicts accurate AR lifetimes. The noninteracting formalism lends itself to modeling the total AR rate as a trimolecular collision between either two quasi-holes and a quasi-electron or two quasi-electrons and a quasi-hole. Given that the noninteracting formalism predicts biexciton AR lifetimes that are far too long, we believe that modeling the total AR decay rate as a trimolecular collision is inappropriate in semiconductor NCs, especially in QDs with radii that are comparable to the exciton Bohr radius of the material and in all nanorods and nanoplatelets.\cite{Zhu2012}

\section{Role of interfaces in nanocrystals}
\label{sec:interfaces}
The decay channels described thus far, such as hot exciton cooling and AR, are dictated by electron-hole correlations, exciton-phonon couplings, and exciton-exciton interactions. In addition to size, dimensionality, and material composition, these interactions can be tuned in NCs by the synthesis of heterostructures, such as core-shell NCs.\cite{Frederick2010,Jain2016,Sagar2020} The core-shell interface introduces another point of control that enables wave function engineering.

For example, CdSe/CdS NCs have a quasi-type II band alignment due to the valence band offset between these bulk materials. The interplay of quantum confinement, band alignment, and electron-hole correlation confines the hole density to the core while the electron density delocalizes into the CdS shell.\cite{Kong2018} On the other hand, CdSe/ZnS core-shell systems have a type I band alignment that confines both the electron and hole to the core.\cite{Zhu2010} These behaviors are well-captured by our atomistic electronic structure framework, as illustrated in Fig.~\ref{fig:Core-Shell}a. The qualitative differences in wave functions in single-material NCs versus heterostructures have large effects on the magnitudes of both EXPC and AR lifetimes.

As described in Sec.~\ref{sec:exph}, phonon modes localized to the surface of NCs have significant contributions to the overall EXPC, especially in NCs that are in the strongly confined regime (Fig.~\ref{fig:Core-Shell}b). This result suggests that EXPC can be mitigated by treatment of the NC surface, such as through the introduction of a passivating shell. For CdSe/CdS core-shell QDs, the overall reorganization energy can be almost an order of magnitude smaller than that of bare CdSe cores, as shown in Fig.~\ref{fig:Core-Shell}d, depending on the core size. This effect is a direct consequence of the quasi-type II band alignment that confines the exciton hole to the CdSe core. As the hole channel is the dominant channel for EXPC,\cite{Jasrasaria2021} hole localization suppresses coupling of the exciton to surface modes and low-frequency acoustic modes that are delocalized over the NC.

Growth of a passivating shell on CdSe core NCs also has a profound effect on the AR lifetimes. Fig.~\ref{fig:Core-Shell}c highlights the dramatic differences between AR lifetimes in CdSe/CdS versus CdSe/ZnS systems.\cite{Philbin2020a} The type I band alignment in CdSe/ZnS means that the root-mean-square exciton radius, or average electron-hole radial coordinate, is relatively independent of ZnS shell thickness after the growth of one shell monolayer. In the quasi-type II systems, however, the root-mean-square exciton radius grows linearly with the number of CdS shell monolayers as the electron delocalizes over the CdS shell. These results directly affect the AR lifetime, which depends on electron-hole wavefunction overlap \textit{via} the Coulomb coupling. The CdSe/ZnS QDs show AR lifetimes that do not change with growth of ZnS shell while those of CdSe/CdS QDs increase dramatically with growth of CdS shell.

In addition to core-shell QDs, dimers and superlattices of NCs and NC heterostructures are being developed and studied theoretically.\cite{Williams2009,Evers2015,Lee2018,Ondry2019,Ondry2021,Notot2022} These materials offer the potential for significant engineering and control of both electronic and phononic properties, enabling the realization of new phenomena.


\section{Outlook}
\label{sec:conclusions}

\begin{figure}[htb]
\includegraphics[width=0.5\textwidth]{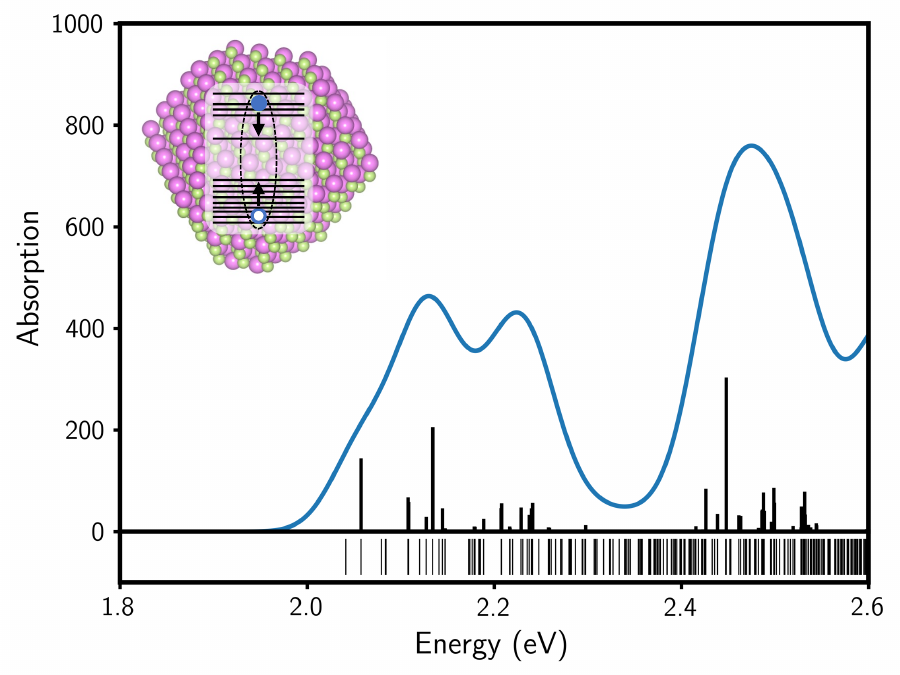}
\caption{\label{fig:Cooling} The calculated absorption spectrum (top) and density of excitonic states (bottom) for a wurtzite 3.9~nm CdSe quantum dot. The vertical lines in the top panel indicate the oscillator strength of the transition from the ground state to that excitonic state. The inset depicts the cooling process schematically, indicating that exciton cooling occurs through a cascade of states.}
\end{figure}

Thus far, we have described our framework for computing the electronic/vibronic properties of confined semiconductor NCs of experimentally relevant sizes. Our approach includes electron-hole correlations, which are key to accurately describing excited-state phenomena, and exciton-phonon coupling, which are essential for understanding room-temperature optical properties and phonon-mediated exciton dynamics. Calculations using our approach yield very good agreement with experimental measurements of observables, such as fundamental and optical gaps, phonon lifetimes, reorganization energies, and AR lifetimes, for II-VI and III-V materials of a variety of sizes, dimensionalities, and compositions.

The main short-term goal is to use these tools to address the longstanding controversy surrounding the phonon bottleneck. The original hypothesis~\cite{Nozik2001} of the phonon bottleneck in NCs is based on a single-particle picture of the \textit{electronic} states, for which the energy spacing between states near the band edge becomes several hundreds of meV. In this case, because the phonon frequencies in these systems are $\sim$30~meV and lower, a multiphonon process would be required for phonon-mediated relaxation, which would be extremely slow. In this electron-hole picture, one requires Auger-like, Coulomb-mediated coupling to break the phonon bottleneck. This picture is translated to a relatively high density of \textit{excitonic } states due to the dense spectrum of holes. Fig.~\ref{fig:Cooling} illustrates the calculated absorption spectrum of a wurtzite 3.9~nm CdSe quantum dot, which is made up of a few very bright excitonic states that have large oscillator strengths and several dim excitonic states that have small oscillator strengths. While the energy spectrum is relatively sparse near the ground excitonic state, the largest excitonic energy gap in this system is 25~meV, and the energy spacing quickly decreases for states higher in energy.\cite{Zunger1997} This would result in a cascade of relaxation events to dark/bright excitons that would be relatively fast due to the small energy spacing, breaking the phonon bottleneck.

Future work will focus on dynamical processes to describe spectral lineshapes, providing means to further assess and improve the approach, as well as integrating this model with a framework for nonadiabatic dynamics to simulate the exciton cooling process and delineate the timescales and mechanism of cooling as a function of excitation energy, NC dimensionality, and NC size. 

\begin{acknowledgments}
E.R. acknowledges support from the Department of Energy, Photonics at the Thermodynamic Limits Energy Frontier Research Center, under grant no. DESC0019140. Methods used in this work were provided by the Center for Computational Study of Excited State Phenomena in Energy Materials (C2SEPEM), which is funded by the U.S. Department of Energy, Office of Science, Basic Energy Sciences, Materials Sciences and Engineering Division, via contract no. DE-AC02-05CH11231, as part of the Computational Materials Sciences Program. Computational resources were provided by the National Energy Research Scientific Computing Center (NERSC), a U.S. Department of Energy Office of Science User Facility operated under contract no. DE-AC02-05CH11231. D.J. acknowledges the support of the Computational Science Graduate Fellowship from the U.S. Department of Energy under grant no. DE-SC0019323. J.P.P. acknowledges support from the Harvard University Center for the Environment.
\end{acknowledgments}


\begin{thebibliography}{192}%
\makeatletter
\providecommand \@ifxundefined [1]{%
 \@ifx{#1\undefined}
}%
\providecommand \@ifnum [1]{%
 \ifnum #1\expandafter \@firstoftwo
 \else \expandafter \@secondoftwo
 \fi
}%
\providecommand \@ifx [1]{%
 \ifx #1\expandafter \@firstoftwo
 \else \expandafter \@secondoftwo
 \fi
}%
\providecommand \natexlab [1]{#1}%
\providecommand \enquote  [1]{``#1''}%
\providecommand \bibnamefont  [1]{#1}%
\providecommand \bibfnamefont [1]{#1}%
\providecommand \citenamefont [1]{#1}%
\providecommand \href@noop [0]{\@secondoftwo}%
\providecommand \href [0]{\begingroup \@sanitize@url \@href}%
\providecommand \@href[1]{\@@startlink{#1}\@@href}%
\providecommand \@@href[1]{\endgroup#1\@@endlink}%
\providecommand \@sanitize@url [0]{\catcode `\\12\catcode `\$12\catcode
  `\&12\catcode `\#12\catcode `\^12\catcode `\_12\catcode `\%12\relax}%
\providecommand \@@startlink[1]{}%
\providecommand \@@endlink[0]{}%
\providecommand \url  [0]{\begingroup\@sanitize@url \@url }%
\providecommand \@url [1]{\endgroup\@href {#1}{\urlprefix }}%
\providecommand \urlprefix  [0]{URL }%
\providecommand \Eprint [0]{\href }%
\providecommand \doibase [0]{http://dx.doi.org/}%
\providecommand \selectlanguage [0]{\@gobble}%
\providecommand \bibinfo  [0]{\@secondoftwo}%
\providecommand \bibfield  [0]{\@secondoftwo}%
\providecommand \translation [1]{[#1]}%
\providecommand \BibitemOpen [0]{}%
\providecommand \bibitemStop [0]{}%
\providecommand \bibitemNoStop [0]{.\EOS\space}%
\providecommand \EOS [0]{\spacefactor3000\relax}%
\providecommand \BibitemShut  [1]{\csname bibitem#1\endcsname}%
\let\auto@bib@innerbib\@empty
\bibitem [{\citenamefont {Alivisatos}(1996)}]{Alivisatos1996}%
  \BibitemOpen
  \bibfield  {author} {\bibinfo {author} {\bibfnamefont {A.~P.}\ \bibnamefont
  {Alivisatos}},\ }\href {\doibase 10.1126/science.271.5251.933} {\bibfield
  {journal} {\bibinfo  {journal} {Science}\ }\textbf {\bibinfo {volume}
  {271}},\ \bibinfo {pages} {933} (\bibinfo {year} {1996})}\BibitemShut
  {NoStop}%
\bibitem [{\citenamefont {Scholes}\ and\ \citenamefont
  {Rumbles}(2006)}]{Scholes2006NatMater}%
  \BibitemOpen
  \bibfield  {author} {\bibinfo {author} {\bibfnamefont {G.~D.}\ \bibnamefont
  {Scholes}}\ and\ \bibinfo {author} {\bibfnamefont {G.}~\bibnamefont
  {Rumbles}},\ }\href {\doibase 10.1038/nmat1710} {\bibfield  {journal}
  {\bibinfo  {journal} {Nat. Mater.}\ }\textbf {\bibinfo {volume} {5}},\
  \bibinfo {pages} {683} (\bibinfo {year} {2006})}\BibitemShut {NoStop}%
\bibitem [{\citenamefont {Klimov}(2014)}]{Klimov2014}%
  \BibitemOpen
  \bibfield  {author} {\bibinfo {author} {\bibfnamefont {V.~I.}\ \bibnamefont
  {Klimov}},\ }\href {\doibase 10.1146/annurev-conmatphys-031113-133900}
  {\bibfield  {journal} {\bibinfo  {journal} {Annu. Rev. Condens. Matter
  Phys.}\ }\textbf {\bibinfo {volume} {5}},\ \bibinfo {pages} {285} (\bibinfo
  {year} {2014})}\BibitemShut {NoStop}%
\bibitem [{\citenamefont {Efros}\ and\ \citenamefont
  {Brus}(2021)}]{EfrosBrus2021}%
  \BibitemOpen
  \bibfield  {author} {\bibinfo {author} {\bibfnamefont {A.~L.}\ \bibnamefont
  {Efros}}\ and\ \bibinfo {author} {\bibfnamefont {L.~E.}\ \bibnamefont
  {Brus}},\ }\href {\doibase 10.1021/acsnano.1c01399} {\bibfield  {journal}
  {\bibinfo  {journal} {ACS Nano}\ }\textbf {\bibinfo {volume} {15}},\ \bibinfo
  {pages} {6192} (\bibinfo {year} {2021})}\BibitemShut {NoStop}%
\bibitem [{\citenamefont {Melnychuk}\ and\ \citenamefont
  {Guyot-Sionnest}(2021)}]{PGS2021}%
  \BibitemOpen
  \bibfield  {author} {\bibinfo {author} {\bibfnamefont {C.}~\bibnamefont
  {Melnychuk}}\ and\ \bibinfo {author} {\bibfnamefont {P.}~\bibnamefont
  {Guyot-Sionnest}},\ }\href {\doibase 10.1021/acs.chemrev.0c00931} {\bibfield
  {journal} {\bibinfo  {journal} {Chem. Rev.}\ }\textbf {\bibinfo {volume}
  {121}},\ \bibinfo {pages} {2325} (\bibinfo {year} {2021})}\BibitemShut
  {NoStop}%
\bibitem [{\citenamefont {Kagan}\ \emph {et~al.}(2021)\citenamefont {Kagan},
  \citenamefont {Bassett}, \citenamefont {Murray},\ and\ \citenamefont
  {Thompson}}]{Kagan2021}%
  \BibitemOpen
  \bibfield  {author} {\bibinfo {author} {\bibfnamefont {C.~R.}\ \bibnamefont
  {Kagan}}, \bibinfo {author} {\bibfnamefont {L.~C.}\ \bibnamefont {Bassett}},
  \bibinfo {author} {\bibfnamefont {C.~B.}\ \bibnamefont {Murray}}, \ and\
  \bibinfo {author} {\bibfnamefont {S.~M.}\ \bibnamefont {Thompson}},\ }\href
  {\doibase 10.1021/acs.chemrev.0c00831} {\bibfield  {journal} {\bibinfo
  {journal} {Chem. Rev.}\ }\textbf {\bibinfo {volume} {121}},\ \bibinfo {pages}
  {3186} (\bibinfo {year} {2021})}\BibitemShut {NoStop}%
\bibitem [{\citenamefont {Efros}\ and\ \citenamefont
  {Efros}(1982)}]{Efros1982}%
  \BibitemOpen
  \bibfield  {author} {\bibinfo {author} {\bibfnamefont {A.}~\bibnamefont
  {Efros}}\ and\ \bibinfo {author} {\bibfnamefont {A.}~\bibnamefont {Efros}},\
  }\href@noop {} {\bibfield  {journal} {\bibinfo  {journal} {Soviet physics.
  Semiconductors}\ }\textbf {\bibinfo {volume} {16}},\ \bibinfo {pages} {772}
  (\bibinfo {year} {1982})}\BibitemShut {NoStop}%
\bibitem [{\citenamefont {Murray}, \citenamefont {Norris},\ and\ \citenamefont
  {Bawendi}(1993)}]{Murray1993}%
  \BibitemOpen
  \bibfield  {author} {\bibinfo {author} {\bibfnamefont {C.~B.}\ \bibnamefont
  {Murray}}, \bibinfo {author} {\bibfnamefont {D.~J.}\ \bibnamefont {Norris}},
  \ and\ \bibinfo {author} {\bibfnamefont {M.~G.}\ \bibnamefont {Bawendi}},\
  }\href {\doibase 10.1021/ja00072a025} {\bibfield  {journal} {\bibinfo
  {journal} {J. Am. Chem. Soc.}\ }\textbf {\bibinfo {volume} {115}},\ \bibinfo
  {pages} {8706} (\bibinfo {year} {1993})}\BibitemShut {NoStop}%
\bibitem [{\citenamefont {Wang}\ and\ \citenamefont {Zunger}(1994)}]{Wang1994}%
  \BibitemOpen
  \bibfield  {author} {\bibinfo {author} {\bibfnamefont {L.~W.}\ \bibnamefont
  {Wang}}\ and\ \bibinfo {author} {\bibfnamefont {A.}~\bibnamefont {Zunger}},\
  }\href {\doibase 10.1021/j100059a032} {\bibfield  {journal} {\bibinfo
  {journal} {J. Phys. Chem.}\ }\textbf {\bibinfo {volume} {98}},\ \bibinfo
  {pages} {2158} (\bibinfo {year} {1994})}\BibitemShut {NoStop}%
\bibitem [{\citenamefont {Dabbousi}\ \emph {et~al.}(1997)\citenamefont
  {Dabbousi}, \citenamefont {Rodriguez-Viejo}, \citenamefont {Mikulec},
  \citenamefont {Heine}, \citenamefont {Mattoussi}, \citenamefont {Ober},
  \citenamefont {Jensen},\ and\ \citenamefont {Bawendi}}]{Dabbousi1997}%
  \BibitemOpen
  \bibfield  {author} {\bibinfo {author} {\bibfnamefont {B.~O.}\ \bibnamefont
  {Dabbousi}}, \bibinfo {author} {\bibfnamefont {J.}~\bibnamefont
  {Rodriguez-Viejo}}, \bibinfo {author} {\bibfnamefont {F.~V.}\ \bibnamefont
  {Mikulec}}, \bibinfo {author} {\bibfnamefont {J.~R.}\ \bibnamefont {Heine}},
  \bibinfo {author} {\bibfnamefont {H.}~\bibnamefont {Mattoussi}}, \bibinfo
  {author} {\bibfnamefont {R.}~\bibnamefont {Ober}}, \bibinfo {author}
  {\bibfnamefont {K.~F.}\ \bibnamefont {Jensen}}, \ and\ \bibinfo {author}
  {\bibfnamefont {M.~G.}\ \bibnamefont {Bawendi}},\ }\href {\doibase
  10.1021/jp971091y} {\bibfield  {journal} {\bibinfo  {journal} {J. Phys. Chem.
  B}\ }\textbf {\bibinfo {volume} {101}},\ \bibinfo {pages} {9463} (\bibinfo
  {year} {1997})}\BibitemShut {NoStop}%
\bibitem [{\citenamefont {Efros}\ and\ \citenamefont
  {Rosen}(2000)}]{Efros2000}%
  \BibitemOpen
  \bibfield  {author} {\bibinfo {author} {\bibfnamefont {A.~L.}\ \bibnamefont
  {Efros}}\ and\ \bibinfo {author} {\bibfnamefont {M.}~\bibnamefont {Rosen}},\
  }\href {\doibase 10.1146/annurev.matsci.30.1.475} {\bibfield  {journal}
  {\bibinfo  {journal} {Annu. Rev. Mater. Sci.}\ }\textbf {\bibinfo {volume}
  {30}},\ \bibinfo {pages} {475} (\bibinfo {year} {2000})}\BibitemShut
  {NoStop}%
\bibitem [{\citenamefont {Rabani}\ and\ \citenamefont
  {Baer}(2010)}]{Rabani2010}%
  \BibitemOpen
  \bibfield  {author} {\bibinfo {author} {\bibfnamefont {E.}~\bibnamefont
  {Rabani}}\ and\ \bibinfo {author} {\bibfnamefont {R.}~\bibnamefont {Baer}},\
  }\href {\doibase 10.1016/j.cplett.2010.07.059} {\bibfield  {journal}
  {\bibinfo  {journal} {Chem. Phys. Lett.}\ }\textbf {\bibinfo {volume}
  {496}},\ \bibinfo {pages} {227} (\bibinfo {year} {2010})}\BibitemShut
  {NoStop}%
\bibitem [{\citenamefont {Boles}, \citenamefont {Engel},\ and\ \citenamefont
  {Talapin}(2016)}]{Boles2016}%
  \BibitemOpen
  \bibfield  {author} {\bibinfo {author} {\bibfnamefont {M.~A.}\ \bibnamefont
  {Boles}}, \bibinfo {author} {\bibfnamefont {M.}~\bibnamefont {Engel}}, \ and\
  \bibinfo {author} {\bibfnamefont {D.~V.}\ \bibnamefont {Talapin}},\ }\href
  {\doibase 10.1021/acs.chemrev.6b00196} {\bibfield  {journal} {\bibinfo
  {journal} {Chem. Rev.}\ }\textbf {\bibinfo {volume} {116}},\ \bibinfo {pages}
  {11220} (\bibinfo {year} {2016})}\BibitemShut {NoStop}%
\bibitem [{\citenamefont {Weiss}(2021)}]{Weiss2021}%
  \BibitemOpen
  \bibfield  {author} {\bibinfo {author} {\bibfnamefont {E.~A.}\ \bibnamefont
  {Weiss}},\ }\href {\doibase 10.1021/acsnano.1c01337} {\bibfield  {journal}
  {\bibinfo  {journal} {ACS Nano}\ }\textbf {\bibinfo {volume} {15}},\ \bibinfo
  {pages} {3568} (\bibinfo {year} {2021})}\BibitemShut {NoStop}%
\bibitem [{\citenamefont {Rossetti}, \citenamefont {Nakahara},\ and\
  \citenamefont {Brus}(1983)}]{Rossetti1983}%
  \BibitemOpen
  \bibfield  {author} {\bibinfo {author} {\bibfnamefont {R.}~\bibnamefont
  {Rossetti}}, \bibinfo {author} {\bibfnamefont {S.}~\bibnamefont {Nakahara}},
  \ and\ \bibinfo {author} {\bibfnamefont {L.~E.}\ \bibnamefont {Brus}},\
  }\href {\doibase 10.1063/1.445834} {\bibfield  {journal} {\bibinfo  {journal}
  {J. Chem. Phys.}\ }\textbf {\bibinfo {volume} {79}},\ \bibinfo {pages} {1086}
  (\bibinfo {year} {1983})}\BibitemShut {NoStop}%
\bibitem [{\citenamefont {Ekimov}\ \emph {et~al.}(1993)\citenamefont {Ekimov},
  \citenamefont {Hache}, \citenamefont {Schanne-Klein}, \citenamefont {Ricard},
  \citenamefont {Flytzanis}, \citenamefont {Kudryavtsev}, \citenamefont
  {Yazeva}, \citenamefont {Rodina},\ and\ \citenamefont {Efros}}]{Ekimov1993}%
  \BibitemOpen
  \bibfield  {author} {\bibinfo {author} {\bibfnamefont {A.~I.}\ \bibnamefont
  {Ekimov}}, \bibinfo {author} {\bibfnamefont {F.}~\bibnamefont {Hache}},
  \bibinfo {author} {\bibfnamefont {M.~C.}\ \bibnamefont {Schanne-Klein}},
  \bibinfo {author} {\bibfnamefont {D.}~\bibnamefont {Ricard}}, \bibinfo
  {author} {\bibfnamefont {C.}~\bibnamefont {Flytzanis}}, \bibinfo {author}
  {\bibfnamefont {I.~A.}\ \bibnamefont {Kudryavtsev}}, \bibinfo {author}
  {\bibfnamefont {T.~V.}\ \bibnamefont {Yazeva}}, \bibinfo {author}
  {\bibfnamefont {A.~V.}\ \bibnamefont {Rodina}}, \ and\ \bibinfo {author}
  {\bibfnamefont {A.~L.}\ \bibnamefont {Efros}},\ }\href@noop {} {\bibfield
  {journal} {\bibinfo  {journal} {J. Opt. Soc. Am. B}\ }\textbf {\bibinfo
  {volume} {10}},\ \bibinfo {pages} {100} (\bibinfo {year} {1993})}\BibitemShut
  {NoStop}%
\bibitem [{\citenamefont {Norris}\ \emph {et~al.}(1994)\citenamefont {Norris},
  \citenamefont {Sacra}, \citenamefont {Murray},\ and\ \citenamefont
  {Bawendi}}]{Norris1994}%
  \BibitemOpen
  \bibfield  {author} {\bibinfo {author} {\bibfnamefont {D.~J.}\ \bibnamefont
  {Norris}}, \bibinfo {author} {\bibfnamefont {A.}~\bibnamefont {Sacra}},
  \bibinfo {author} {\bibfnamefont {C.~B.}\ \bibnamefont {Murray}}, \ and\
  \bibinfo {author} {\bibfnamefont {M.~G.}\ \bibnamefont {Bawendi}},\ }\href
  {\doibase 10.1103/PhysRevLett.72.2612} {\bibfield  {journal} {\bibinfo
  {journal} {Phys. Rev. Lett.}\ }\textbf {\bibinfo {volume} {72}},\ \bibinfo
  {pages} {2612} (\bibinfo {year} {1994})}\BibitemShut {NoStop}%
\bibitem [{\citenamefont {Norris}\ and\ \citenamefont
  {Bawendi}(1996)}]{Norris1996a}%
  \BibitemOpen
  \bibfield  {author} {\bibinfo {author} {\bibfnamefont {D.~J.}\ \bibnamefont
  {Norris}}\ and\ \bibinfo {author} {\bibfnamefont {M.~G.}\ \bibnamefont
  {Bawendi}},\ }\href {\doibase 10.1103/PhysRevB.53.16338} {\bibfield
  {journal} {\bibinfo  {journal} {Phys. Rev. B}\ }\textbf {\bibinfo {volume}
  {53}},\ \bibinfo {pages} {16338} (\bibinfo {year} {1996})}\BibitemShut
  {NoStop}%
\bibitem [{\citenamefont {Nirmal}, \citenamefont {Murray},\ and\ \citenamefont
  {Bawendi}(1994)}]{Nirmal1994}%
  \BibitemOpen
  \bibfield  {author} {\bibinfo {author} {\bibfnamefont {M.}~\bibnamefont
  {Nirmal}}, \bibinfo {author} {\bibfnamefont {C.~B.}\ \bibnamefont {Murray}},
  \ and\ \bibinfo {author} {\bibfnamefont {M.~G.}\ \bibnamefont {Bawendi}},\
  }\href {\doibase 10.1103/PhysRevB.50.2293} {\bibfield  {journal} {\bibinfo
  {journal} {Phys. Rev. B}\ }\textbf {\bibinfo {volume} {50}},\ \bibinfo
  {pages} {2293} (\bibinfo {year} {1994})}\BibitemShut {NoStop}%
\bibitem [{\citenamefont {Nirmal}\ \emph {et~al.}(1995)\citenamefont {Nirmal},
  \citenamefont {Norris}, \citenamefont {Kuno}, \citenamefont {Bawendi},
  \citenamefont {Efros},\ and\ \citenamefont {Rosen}}]{Nirmal1995}%
  \BibitemOpen
  \bibfield  {author} {\bibinfo {author} {\bibfnamefont {M.}~\bibnamefont
  {Nirmal}}, \bibinfo {author} {\bibfnamefont {D.~J.}\ \bibnamefont {Norris}},
  \bibinfo {author} {\bibfnamefont {M.}~\bibnamefont {Kuno}}, \bibinfo {author}
  {\bibfnamefont {M.~G.}\ \bibnamefont {Bawendi}}, \bibinfo {author}
  {\bibfnamefont {A.~L.}\ \bibnamefont {Efros}}, \ and\ \bibinfo {author}
  {\bibfnamefont {M.}~\bibnamefont {Rosen}},\ }\href {\doibase
  10.1103/PhysRevLett.75.3728} {\bibfield  {journal} {\bibinfo  {journal}
  {Phys. Rev. Lett.}\ }\textbf {\bibinfo {volume} {75}},\ \bibinfo {pages}
  {3728} (\bibinfo {year} {1995})}\BibitemShut {NoStop}%
\bibitem [{\citenamefont {Norris}\ \emph {et~al.}(1996)\citenamefont {Norris},
  \citenamefont {Efros}, \citenamefont {Rosen},\ and\ \citenamefont
  {Bawendi}}]{Norris1996b}%
  \BibitemOpen
  \bibfield  {author} {\bibinfo {author} {\bibfnamefont {D.~J.}\ \bibnamefont
  {Norris}}, \bibinfo {author} {\bibfnamefont {A.~L.}\ \bibnamefont {Efros}},
  \bibinfo {author} {\bibfnamefont {M.}~\bibnamefont {Rosen}}, \ and\ \bibinfo
  {author} {\bibfnamefont {M.~G.}\ \bibnamefont {Bawendi}},\ }\href {\doibase
  10.1103/PhysRevB.53.16347} {\bibfield  {journal} {\bibinfo  {journal} {Phys.
  Rev. B}\ }\textbf {\bibinfo {volume} {53}},\ \bibinfo {pages} {16347}
  (\bibinfo {year} {1996})}\BibitemShut {NoStop}%
\bibitem [{\citenamefont {Efros}\ \emph {et~al.}(1996)\citenamefont {Efros},
  \citenamefont {Rosen}, \citenamefont {Kuno}, \citenamefont {Nirmal},
  \citenamefont {Norris},\ and\ \citenamefont {Bawendi}}]{Efros1996}%
  \BibitemOpen
  \bibfield  {author} {\bibinfo {author} {\bibfnamefont {A.~L.}\ \bibnamefont
  {Efros}}, \bibinfo {author} {\bibfnamefont {M.}~\bibnamefont {Rosen}},
  \bibinfo {author} {\bibfnamefont {M.}~\bibnamefont {Kuno}}, \bibinfo {author}
  {\bibfnamefont {M.}~\bibnamefont {Nirmal}}, \bibinfo {author} {\bibfnamefont
  {D.~J.}\ \bibnamefont {Norris}}, \ and\ \bibinfo {author} {\bibfnamefont
  {M.}~\bibnamefont {Bawendi}},\ }\href {\doibase 10.1103/PhysRevB.54.4843}
  {\bibfield  {journal} {\bibinfo  {journal} {Phys. Rev. B}\ }\textbf {\bibinfo
  {volume} {54}},\ \bibinfo {pages} {4843} (\bibinfo {year}
  {1996})}\BibitemShut {NoStop}%
\bibitem [{\citenamefont {Cohen}\ and\ \citenamefont
  {Bergstresser}(1966)}]{Cohen1966}%
  \BibitemOpen
  \bibfield  {author} {\bibinfo {author} {\bibfnamefont {M.~L.}\ \bibnamefont
  {Cohen}}\ and\ \bibinfo {author} {\bibfnamefont {T.~K.}\ \bibnamefont
  {Bergstresser}},\ }\href {\doibase 10.1103/PhysRev.141.789} {\bibfield
  {journal} {\bibinfo  {journal} {Phys. Rev.}\ }\textbf {\bibinfo {volume}
  {141}},\ \bibinfo {pages} {789} (\bibinfo {year} {1966})}\BibitemShut
  {NoStop}%
\bibitem [{\citenamefont {Rama~Krishna}\ and\ \citenamefont
  {Friesner}(1991)}]{Friesner1991}%
  \BibitemOpen
  \bibfield  {author} {\bibinfo {author} {\bibfnamefont {M.~V.}\ \bibnamefont
  {Rama~Krishna}}\ and\ \bibinfo {author} {\bibfnamefont {R.~A.}\ \bibnamefont
  {Friesner}},\ }\href {\doibase 10.1103/PhysRevLett.67.629} {\bibfield
  {journal} {\bibinfo  {journal} {Phys. Rev. Lett.}\ }\textbf {\bibinfo
  {volume} {67}},\ \bibinfo {pages} {629} (\bibinfo {year} {1991})}\BibitemShut
  {NoStop}%
\bibitem [{\citenamefont {Wang}\ and\ \citenamefont {Zunger}(1996)}]{Wang1996}%
  \BibitemOpen
  \bibfield  {author} {\bibinfo {author} {\bibfnamefont {L.-W.}\ \bibnamefont
  {Wang}}\ and\ \bibinfo {author} {\bibfnamefont {A.}~\bibnamefont {Zunger}},\
  }\href {\doibase 10.1103/PhysRevB.53.9579} {\bibfield  {journal} {\bibinfo
  {journal} {Phys. Rev. B}\ }\textbf {\bibinfo {volume} {53}},\ \bibinfo
  {pages} {9579} (\bibinfo {year} {1996})}\BibitemShut {NoStop}%
\bibitem [{\citenamefont {Rabani}\ \emph {et~al.}(1999)\citenamefont {Rabani},
  \citenamefont {Hetenyi}, \citenamefont {Berne},\ and\ \citenamefont
  {Brus}}]{Rabani1999b}%
  \BibitemOpen
  \bibfield  {author} {\bibinfo {author} {\bibfnamefont {E.}~\bibnamefont
  {Rabani}}, \bibinfo {author} {\bibfnamefont {B.}~\bibnamefont {Hetenyi}},
  \bibinfo {author} {\bibfnamefont {B.~J.}\ \bibnamefont {Berne}}, \ and\
  \bibinfo {author} {\bibfnamefont {L.~E.}\ \bibnamefont {Brus}},\ }\href
  {\doibase 10.1063/1.478431} {\bibfield  {journal} {\bibinfo  {journal} {J.
  Chem. Phys.}\ }\textbf {\bibinfo {volume} {110}},\ \bibinfo {pages} {5355}
  (\bibinfo {year} {1999})}\BibitemShut {NoStop}%
\bibitem [{\citenamefont {Califano}, \citenamefont {Franceschetti},\ and\
  \citenamefont {Zunger}(2005)}]{Califano2005}%
  \BibitemOpen
  \bibfield  {author} {\bibinfo {author} {\bibfnamefont {M.}~\bibnamefont
  {Califano}}, \bibinfo {author} {\bibfnamefont {A.}~\bibnamefont
  {Franceschetti}}, \ and\ \bibinfo {author} {\bibfnamefont {A.}~\bibnamefont
  {Zunger}},\ }\href {\doibase 10.1021/nl051027p} {\bibfield  {journal}
  {\bibinfo  {journal} {Nano Lett.}\ }\textbf {\bibinfo {volume} {5}},\
  \bibinfo {pages} {2360} (\bibinfo {year} {2005})}\BibitemShut {NoStop}%
\bibitem [{\citenamefont {Jasrasaria}\ \emph {et~al.}(2020)\citenamefont
  {Jasrasaria}, \citenamefont {Philbin}, \citenamefont {Yan}, \citenamefont
  {Weinberg}, \citenamefont {Alivisatos},\ and\ \citenamefont
  {Rabani}}]{Jasrasaria2020}%
  \BibitemOpen
  \bibfield  {author} {\bibinfo {author} {\bibfnamefont {D.}~\bibnamefont
  {Jasrasaria}}, \bibinfo {author} {\bibfnamefont {J.~P.}\ \bibnamefont
  {Philbin}}, \bibinfo {author} {\bibfnamefont {C.}~\bibnamefont {Yan}},
  \bibinfo {author} {\bibfnamefont {D.}~\bibnamefont {Weinberg}}, \bibinfo
  {author} {\bibfnamefont {A.~P.}\ \bibnamefont {Alivisatos}}, \ and\ \bibinfo
  {author} {\bibfnamefont {E.}~\bibnamefont {Rabani}},\ }\href {\doibase
  10.1021/acs.jpcc.0c04746} {\bibfield  {journal} {\bibinfo  {journal} {J.
  Phys. Chem. C}\ }\textbf {\bibinfo {volume} {124}},\ \bibinfo {pages} {17372}
  (\bibinfo {year} {2020})}\BibitemShut {NoStop}%
\bibitem [{\citenamefont {Wang}, \citenamefont {Kim},\ and\ \citenamefont
  {Zunger}(1999)}]{Wang1999}%
  \BibitemOpen
  \bibfield  {author} {\bibinfo {author} {\bibfnamefont {L.-W.}\ \bibnamefont
  {Wang}}, \bibinfo {author} {\bibfnamefont {J.}~\bibnamefont {Kim}}, \ and\
  \bibinfo {author} {\bibfnamefont {A.}~\bibnamefont {Zunger}},\ }\href
  {\doibase 10.1103/PhysRevB.59.5678} {\bibfield  {journal} {\bibinfo
  {journal} {Phys. Rev. B}\ }\textbf {\bibinfo {volume} {59}},\ \bibinfo
  {pages} {5678} (\bibinfo {year} {1999})}\BibitemShut {NoStop}%
\bibitem [{\citenamefont {Mattila}, \citenamefont {Wang},\ and\ \citenamefont
  {Zunger}(1999)}]{Mattila1999}%
  \BibitemOpen
  \bibfield  {author} {\bibinfo {author} {\bibfnamefont {T.}~\bibnamefont
  {Mattila}}, \bibinfo {author} {\bibfnamefont {L.-W.}\ \bibnamefont {Wang}}, \
  and\ \bibinfo {author} {\bibfnamefont {A.}~\bibnamefont {Zunger}},\ }\href
  {\doibase 10.1103/PhysRevB.59.15270} {\bibfield  {journal} {\bibinfo
  {journal} {Phys. Rev. B}\ }\textbf {\bibinfo {volume} {59}},\ \bibinfo
  {pages} {15270} (\bibinfo {year} {1999})}\BibitemShut {NoStop}%
\bibitem [{\citenamefont {Guyot-Sionnest}\ \emph {et~al.}(1999)\citenamefont
  {Guyot-Sionnest}, \citenamefont {Shim}, \citenamefont {Matranga},\ and\
  \citenamefont {Hines}}]{GS1999}%
  \BibitemOpen
  \bibfield  {author} {\bibinfo {author} {\bibfnamefont {P.}~\bibnamefont
  {Guyot-Sionnest}}, \bibinfo {author} {\bibfnamefont {M.}~\bibnamefont
  {Shim}}, \bibinfo {author} {\bibfnamefont {C.}~\bibnamefont {Matranga}}, \
  and\ \bibinfo {author} {\bibfnamefont {M.}~\bibnamefont {Hines}},\ }\href
  {\doibase 10.1103/PhysRevB.60.R2181} {\bibfield  {journal} {\bibinfo
  {journal} {Phys. Rev. B}\ }\textbf {\bibinfo {volume} {60}},\ \bibinfo
  {pages} {R2181} (\bibinfo {year} {1999})}\BibitemShut {NoStop}%
\bibitem [{\citenamefont {Klimov}\ \emph
  {et~al.}(2000{\natexlab{a}})\citenamefont {Klimov}, \citenamefont
  {Mikhailovsky}, \citenamefont {McBranch}, \citenamefont {Leatherdale},\ and\
  \citenamefont {Bawendi}}]{Klimov2000}%
  \BibitemOpen
  \bibfield  {author} {\bibinfo {author} {\bibfnamefont {V.~I.}\ \bibnamefont
  {Klimov}}, \bibinfo {author} {\bibfnamefont {A.~A.}\ \bibnamefont
  {Mikhailovsky}}, \bibinfo {author} {\bibfnamefont {D.~W.}\ \bibnamefont
  {McBranch}}, \bibinfo {author} {\bibfnamefont {C.~A.}\ \bibnamefont
  {Leatherdale}}, \ and\ \bibinfo {author} {\bibfnamefont {M.~G.}\ \bibnamefont
  {Bawendi}},\ }\href {\doibase 10.1126/science.287.5455.1011} {\bibfield
  {journal} {\bibinfo  {journal} {Science}\ }\textbf {\bibinfo {volume}
  {287}},\ \bibinfo {pages} {1011} (\bibinfo {year}
  {2000}{\natexlab{a}})}\BibitemShut {NoStop}%
\bibitem [{\citenamefont {Guyot-Sionnest}, \citenamefont {Wehrenberg},\ and\
  \citenamefont {Yu}(2005)}]{GS2005}%
  \BibitemOpen
  \bibfield  {author} {\bibinfo {author} {\bibfnamefont {P.}~\bibnamefont
  {Guyot-Sionnest}}, \bibinfo {author} {\bibfnamefont {B.}~\bibnamefont
  {Wehrenberg}}, \ and\ \bibinfo {author} {\bibfnamefont {D.}~\bibnamefont
  {Yu}},\ }\href {\doibase 10.1063/1.2004818} {\bibfield  {journal} {\bibinfo
  {journal} {J. Chem. Phys.}\ }\textbf {\bibinfo {volume} {123}},\ \bibinfo
  {pages} {074709} (\bibinfo {year} {2005})}\BibitemShut {NoStop}%
\bibitem [{\citenamefont {Oron}, \citenamefont {Kazes},\ and\ \citenamefont
  {Banin}(2007)}]{Oron2007}%
  \BibitemOpen
  \bibfield  {author} {\bibinfo {author} {\bibfnamefont {D.}~\bibnamefont
  {Oron}}, \bibinfo {author} {\bibfnamefont {M.}~\bibnamefont {Kazes}}, \ and\
  \bibinfo {author} {\bibfnamefont {U.}~\bibnamefont {Banin}},\ }\href
  {\doibase 10.1103/PhysRevB.75.035330} {\bibfield  {journal} {\bibinfo
  {journal} {Phys. Rev. B}\ }\textbf {\bibinfo {volume} {75}},\ \bibinfo
  {pages} {1} (\bibinfo {year} {2007})}\BibitemShut {NoStop}%
\bibitem [{\citenamefont {Pandey}\ and\ \citenamefont
  {Guyot-Sionnest}(2008)}]{Pandey2008}%
  \BibitemOpen
  \bibfield  {author} {\bibinfo {author} {\bibfnamefont {A.}~\bibnamefont
  {Pandey}}\ and\ \bibinfo {author} {\bibfnamefont {P.}~\bibnamefont
  {Guyot-Sionnest}},\ }\href {\doibase 10.1126/science.1159832} {\bibfield
  {journal} {\bibinfo  {journal} {Science}\ }\textbf {\bibinfo {volume}
  {322}},\ \bibinfo {pages} {929} (\bibinfo {year} {2008})}\BibitemShut
  {NoStop}%
\bibitem [{\citenamefont {Jones}, \citenamefont {Lo},\ and\ \citenamefont
  {Scholes}(2009)}]{Jones2009}%
  \BibitemOpen
  \bibfield  {author} {\bibinfo {author} {\bibfnamefont {M.}~\bibnamefont
  {Jones}}, \bibinfo {author} {\bibfnamefont {S.~S.}\ \bibnamefont {Lo}}, \
  and\ \bibinfo {author} {\bibfnamefont {G.~D.}\ \bibnamefont {Scholes}},\
  }\href {\doibase 10.1073/pnas.0809316106} {\bibfield  {journal} {\bibinfo
  {journal} {Proc. Natl. Acad. Sci. U.S.A.}\ }\textbf {\bibinfo {volume}
  {106}},\ \bibinfo {pages} {3011} (\bibinfo {year} {2009})}\BibitemShut
  {NoStop}%
\bibitem [{\citenamefont {Sukhovatkin}\ \emph {et~al.}(2009)\citenamefont
  {Sukhovatkin}, \citenamefont {Hinds}, \citenamefont {Brzozowski},\ and\
  \citenamefont {Sargent}}]{Sukhovatkin2009}%
  \BibitemOpen
  \bibfield  {author} {\bibinfo {author} {\bibfnamefont {V.}~\bibnamefont
  {Sukhovatkin}}, \bibinfo {author} {\bibfnamefont {S.}~\bibnamefont {Hinds}},
  \bibinfo {author} {\bibfnamefont {L.}~\bibnamefont {Brzozowski}}, \ and\
  \bibinfo {author} {\bibfnamefont {E.~H.}\ \bibnamefont {Sargent}},\ }\href
  {\doibase 10.1126/science.1173812} {\bibfield  {journal} {\bibinfo  {journal}
  {Science}\ }\textbf {\bibinfo {volume} {324}},\ \bibinfo {pages} {1542}
  (\bibinfo {year} {2009})}\BibitemShut {NoStop}%
\bibitem [{\citenamefont {McArthur}\ \emph {et~al.}(2010)\citenamefont
  {McArthur}, \citenamefont {Morris-Cohen}, \citenamefont {Knowles},\ and\
  \citenamefont {Weiss}}]{McArthur2010}%
  \BibitemOpen
  \bibfield  {author} {\bibinfo {author} {\bibfnamefont {E.~A.}\ \bibnamefont
  {McArthur}}, \bibinfo {author} {\bibfnamefont {A.~J.}\ \bibnamefont
  {Morris-Cohen}}, \bibinfo {author} {\bibfnamefont {K.~E.}\ \bibnamefont
  {Knowles}}, \ and\ \bibinfo {author} {\bibfnamefont {E.~A.}\ \bibnamefont
  {Weiss}},\ }\href {\doibase 10.1021/jp102101f} {\bibfield  {journal}
  {\bibinfo  {journal} {J. Phys. Chem. B}\ }\textbf {\bibinfo {volume} {114}},\
  \bibinfo {pages} {14514} (\bibinfo {year} {2010})}\BibitemShut {NoStop}%
\bibitem [{\citenamefont {Ulbricht}\ \emph {et~al.}(2011)\citenamefont
  {Ulbricht}, \citenamefont {Hendry}, \citenamefont {Shan}, \citenamefont
  {Heinz},\ and\ \citenamefont {Bonn}}]{Ulbricht2011}%
  \BibitemOpen
  \bibfield  {author} {\bibinfo {author} {\bibfnamefont {R.}~\bibnamefont
  {Ulbricht}}, \bibinfo {author} {\bibfnamefont {E.}~\bibnamefont {Hendry}},
  \bibinfo {author} {\bibfnamefont {J.}~\bibnamefont {Shan}}, \bibinfo {author}
  {\bibfnamefont {T.~F.}\ \bibnamefont {Heinz}}, \ and\ \bibinfo {author}
  {\bibfnamefont {M.}~\bibnamefont {Bonn}},\ }\href {\doibase
  10.1103/RevModPhys.83.543} {\bibfield  {journal} {\bibinfo  {journal} {Rev.
  Mod. Phys.}\ }\textbf {\bibinfo {volume} {83}},\ \bibinfo {pages} {543}
  (\bibinfo {year} {2011})}\BibitemShut {NoStop}%
\bibitem [{\citenamefont {Knowles}, \citenamefont {McArthur},\ and\
  \citenamefont {Weiss}(2011)}]{Knowles2011}%
  \BibitemOpen
  \bibfield  {author} {\bibinfo {author} {\bibfnamefont {K.~E.}\ \bibnamefont
  {Knowles}}, \bibinfo {author} {\bibfnamefont {E.~A.}\ \bibnamefont
  {McArthur}}, \ and\ \bibinfo {author} {\bibfnamefont {E.~A.}\ \bibnamefont
  {Weiss}},\ }\href {\doibase 10.1021/nn2002689} {\bibfield  {journal}
  {\bibinfo  {journal} {ACS Nano}\ }\textbf {\bibinfo {volume} {5}},\ \bibinfo
  {pages} {2026} (\bibinfo {year} {2011})}\BibitemShut {NoStop}%
\bibitem [{\citenamefont {Bae}\ \emph {et~al.}(2013)\citenamefont {Bae},
  \citenamefont {Padilha}, \citenamefont {Park}, \citenamefont {McDaniel},
  \citenamefont {Robel}, \citenamefont {Pietryga},\ and\ \citenamefont
  {Klimov}}]{Bae2013}%
  \BibitemOpen
  \bibfield  {author} {\bibinfo {author} {\bibfnamefont {W.~K.}\ \bibnamefont
  {Bae}}, \bibinfo {author} {\bibfnamefont {L.~A.}\ \bibnamefont {Padilha}},
  \bibinfo {author} {\bibfnamefont {Y.~S.}\ \bibnamefont {Park}}, \bibinfo
  {author} {\bibfnamefont {H.}~\bibnamefont {McDaniel}}, \bibinfo {author}
  {\bibfnamefont {I.}~\bibnamefont {Robel}}, \bibinfo {author} {\bibfnamefont
  {J.~M.}\ \bibnamefont {Pietryga}}, \ and\ \bibinfo {author} {\bibfnamefont
  {V.~I.}\ \bibnamefont {Klimov}},\ }\href {\doibase 10.1021/nn4002825}
  {\bibfield  {journal} {\bibinfo  {journal} {ACS Nano}\ }\textbf {\bibinfo
  {volume} {7}},\ \bibinfo {pages} {3411} (\bibinfo {year} {2013})}\BibitemShut
  {NoStop}%
\bibitem [{\citenamefont {Qin}, \citenamefont {Liu},\ and\ \citenamefont
  {Guyot-Sionnest}(2014)}]{Qin2014}%
  \BibitemOpen
  \bibfield  {author} {\bibinfo {author} {\bibfnamefont {W.}~\bibnamefont
  {Qin}}, \bibinfo {author} {\bibfnamefont {H.}~\bibnamefont {Liu}}, \ and\
  \bibinfo {author} {\bibfnamefont {P.}~\bibnamefont {Guyot-Sionnest}},\ }\href
  {\doibase 10.1021/nn403893b} {\bibfield  {journal} {\bibinfo  {journal} {ACS
  Nano}\ }\textbf {\bibinfo {volume} {8}},\ \bibinfo {pages} {283} (\bibinfo
  {year} {2014})}\BibitemShut {NoStop}%
\bibitem [{\citenamefont {Kambhampati}(2015)}]{Kambhampati2015}%
  \BibitemOpen
  \bibfield  {author} {\bibinfo {author} {\bibfnamefont {P.}~\bibnamefont
  {Kambhampati}},\ }\href {\doibase 10.1016/j.chemphys.2014.11.008} {\bibfield
  {journal} {\bibinfo  {journal} {Chem. Phys.}\ }\textbf {\bibinfo {volume}
  {446}},\ \bibinfo {pages} {92} (\bibinfo {year} {2015})}\BibitemShut
  {NoStop}%
\bibitem [{\citenamefont {Wu}\ \emph {et~al.}(2016)\citenamefont {Wu},
  \citenamefont {Congreve}, \citenamefont {Wilson}, \citenamefont {Jean},
  \citenamefont {Geva}, \citenamefont {Welborn}, \citenamefont {Van~Voorhis},
  \citenamefont {Bulovic}, \citenamefont {Bawendi},\ and\ \citenamefont
  {Baldo}}]{Wu2016}%
  \BibitemOpen
  \bibfield  {author} {\bibinfo {author} {\bibfnamefont {M.}~\bibnamefont
  {Wu}}, \bibinfo {author} {\bibfnamefont {D.~N.}\ \bibnamefont {Congreve}},
  \bibinfo {author} {\bibfnamefont {M.~W.~B.}\ \bibnamefont {Wilson}}, \bibinfo
  {author} {\bibfnamefont {J.}~\bibnamefont {Jean}}, \bibinfo {author}
  {\bibfnamefont {N.}~\bibnamefont {Geva}}, \bibinfo {author} {\bibfnamefont
  {M.}~\bibnamefont {Welborn}}, \bibinfo {author} {\bibfnamefont
  {T.}~\bibnamefont {Van~Voorhis}}, \bibinfo {author} {\bibfnamefont
  {V.}~\bibnamefont {Bulovic}}, \bibinfo {author} {\bibfnamefont {M.~G.}\
  \bibnamefont {Bawendi}}, \ and\ \bibinfo {author} {\bibfnamefont {M.~A.}\
  \bibnamefont {Baldo}},\ }\href {\doibase 10.1038/NPHOTON.2015.226} {\bibfield
   {journal} {\bibinfo  {journal} {Nat. Photonics}\ }\textbf {\bibinfo {volume}
  {10}},\ \bibinfo {pages} {31} (\bibinfo {year} {2016})}\BibitemShut {NoStop}%
\bibitem [{\citenamefont {Li}\ \emph {et~al.}(2017)\citenamefont {Li},
  \citenamefont {Zhou}, \citenamefont {McBride},\ and\ \citenamefont
  {Lian}}]{Li2017c}%
  \BibitemOpen
  \bibfield  {author} {\bibinfo {author} {\bibfnamefont {Q.}~\bibnamefont
  {Li}}, \bibinfo {author} {\bibfnamefont {B.}~\bibnamefont {Zhou}}, \bibinfo
  {author} {\bibfnamefont {J.~R.}\ \bibnamefont {McBride}}, \ and\ \bibinfo
  {author} {\bibfnamefont {T.}~\bibnamefont {Lian}},\ }\href {\doibase
  10.1021/acsenergylett.6b00634} {\bibfield  {journal} {\bibinfo  {journal}
  {ACS Energy Lett.}\ }\textbf {\bibinfo {volume} {2}},\ \bibinfo {pages} {174}
  (\bibinfo {year} {2017})}\BibitemShut {NoStop}%
\bibitem [{\citenamefont {Kaledin}\ \emph {et~al.}(2018)\citenamefont
  {Kaledin}, \citenamefont {Kong}, \citenamefont {Wu}, \citenamefont {Lian},\
  and\ \citenamefont {Musaev}}]{Kaledin2018}%
  \BibitemOpen
  \bibfield  {author} {\bibinfo {author} {\bibfnamefont {A.~L.}\ \bibnamefont
  {Kaledin}}, \bibinfo {author} {\bibfnamefont {D.}~\bibnamefont {Kong}},
  \bibinfo {author} {\bibfnamefont {K.}~\bibnamefont {Wu}}, \bibinfo {author}
  {\bibfnamefont {T.}~\bibnamefont {Lian}}, \ and\ \bibinfo {author}
  {\bibfnamefont {D.~G.}\ \bibnamefont {Musaev}},\ }\href {\doibase
  10.1021/acs.jpcc.8b04874} {\bibfield  {journal} {\bibinfo  {journal} {J.
  Phys. Chem. C}\ }\textbf {\bibinfo {volume} {122}},\ \bibinfo {pages} {18742}
  (\bibinfo {year} {2018})}\BibitemShut {NoStop}%
\bibitem [{\citenamefont {Li}\ \emph {et~al.}(2019)\citenamefont {Li},
  \citenamefont {Yang}, \citenamefont {Que},\ and\ \citenamefont
  {Lian}}]{Li2019}%
  \BibitemOpen
  \bibfield  {author} {\bibinfo {author} {\bibfnamefont {Q.}~\bibnamefont
  {Li}}, \bibinfo {author} {\bibfnamefont {Y.}~\bibnamefont {Yang}}, \bibinfo
  {author} {\bibfnamefont {W.}~\bibnamefont {Que}}, \ and\ \bibinfo {author}
  {\bibfnamefont {T.}~\bibnamefont {Lian}},\ }\href@noop {} {\bibfield
  {journal} {\bibinfo  {journal} {Nano Lett.}\ }\textbf {\bibinfo {volume}
  {19}},\ \bibinfo {pages} {5620} (\bibinfo {year} {2019})}\BibitemShut
  {NoStop}%
\bibitem [{\citenamefont {Prezhdo}(2009)}]{Oleg2009}%
  \BibitemOpen
  \bibfield  {author} {\bibinfo {author} {\bibfnamefont {O.~V.}\ \bibnamefont
  {Prezhdo}},\ }\href {\doibase 10.1021/ar900157s} {\bibfield  {journal}
  {\bibinfo  {journal} {Acc. Chem. Res.}\ }\textbf {\bibinfo {volume} {42}},\
  \bibinfo {pages} {2005} (\bibinfo {year} {2009})}\BibitemShut {NoStop}%
\bibitem [{\citenamefont {Gfroerer}\ \emph {et~al.}(1996)\citenamefont
  {Gfroerer}, \citenamefont {Sturge}, \citenamefont {Kash}, \citenamefont
  {Yater}, \citenamefont {Plaut}, \citenamefont {Lin}, \citenamefont {Florez},
  \citenamefont {Harbison}, \citenamefont {Das},\ and\ \citenamefont
  {Lebrun}}]{Gfroerer1996}%
  \BibitemOpen
  \bibfield  {author} {\bibinfo {author} {\bibfnamefont {T.~H.}\ \bibnamefont
  {Gfroerer}}, \bibinfo {author} {\bibfnamefont {M.~D.}\ \bibnamefont
  {Sturge}}, \bibinfo {author} {\bibfnamefont {K.}~\bibnamefont {Kash}},
  \bibinfo {author} {\bibfnamefont {J.~A.}\ \bibnamefont {Yater}}, \bibinfo
  {author} {\bibfnamefont {A.~S.}\ \bibnamefont {Plaut}}, \bibinfo {author}
  {\bibfnamefont {P.~S.~D.}\ \bibnamefont {Lin}}, \bibinfo {author}
  {\bibfnamefont {L.~T.}\ \bibnamefont {Florez}}, \bibinfo {author}
  {\bibfnamefont {J.~P.}\ \bibnamefont {Harbison}}, \bibinfo {author}
  {\bibfnamefont {S.~R.}\ \bibnamefont {Das}}, \ and\ \bibinfo {author}
  {\bibfnamefont {L.}~\bibnamefont {Lebrun}},\ }\href {\doibase
  10.1103/PhysRevB.53.16474} {\bibfield  {journal} {\bibinfo  {journal} {Phys.
  Rev. B}\ }\textbf {\bibinfo {volume} {53}},\ \bibinfo {pages} {16474}
  (\bibinfo {year} {1996})}\BibitemShut {NoStop}%
\bibitem [{\citenamefont {Yu}\ \emph {et~al.}(1996)\citenamefont {Yu},
  \citenamefont {Lycett}, \citenamefont {Roberts},\ and\ \citenamefont
  {Murray}}]{Haiping1996}%
  \BibitemOpen
  \bibfield  {author} {\bibinfo {author} {\bibfnamefont {H.}~\bibnamefont
  {Yu}}, \bibinfo {author} {\bibfnamefont {S.}~\bibnamefont {Lycett}}, \bibinfo
  {author} {\bibfnamefont {C.}~\bibnamefont {Roberts}}, \ and\ \bibinfo
  {author} {\bibfnamefont {R.}~\bibnamefont {Murray}},\ }\href {\doibase
  10.1063/1.117827} {\bibfield  {journal} {\bibinfo  {journal} {Appl. Phys.
  Lett.}\ }\textbf {\bibinfo {volume} {69}},\ \bibinfo {pages} {4087} (\bibinfo
  {year} {1996})}\BibitemShut {NoStop}%
\bibitem [{\citenamefont {Heitz}\ \emph {et~al.}(1997)\citenamefont {Heitz},
  \citenamefont {Veit}, \citenamefont {Ledentsov}, \citenamefont {Hoffmann},
  \citenamefont {Bimberg}, \citenamefont {Ustinov}, \citenamefont {Kop'ev},\
  and\ \citenamefont {Alferov}}]{Heitz1997}%
  \BibitemOpen
  \bibfield  {author} {\bibinfo {author} {\bibfnamefont {R.}~\bibnamefont
  {Heitz}}, \bibinfo {author} {\bibfnamefont {M.}~\bibnamefont {Veit}},
  \bibinfo {author} {\bibfnamefont {N.~N.}\ \bibnamefont {Ledentsov}}, \bibinfo
  {author} {\bibfnamefont {A.}~\bibnamefont {Hoffmann}}, \bibinfo {author}
  {\bibfnamefont {D.}~\bibnamefont {Bimberg}}, \bibinfo {author} {\bibfnamefont
  {V.~M.}\ \bibnamefont {Ustinov}}, \bibinfo {author} {\bibfnamefont {P.~S.}\
  \bibnamefont {Kop'ev}}, \ and\ \bibinfo {author} {\bibfnamefont {Z.~I.}\
  \bibnamefont {Alferov}},\ }\href {\doibase 10.1103/PhysRevB.56.10435}
  {\bibfield  {journal} {\bibinfo  {journal} {Phys. Rev. B}\ }\textbf {\bibinfo
  {volume} {56}},\ \bibinfo {pages} {10435} (\bibinfo {year}
  {1997})}\BibitemShut {NoStop}%
\bibitem [{\citenamefont {Sosnowski}\ \emph {et~al.}(1998)\citenamefont
  {Sosnowski}, \citenamefont {Norris}, \citenamefont {Jiang}, \citenamefont
  {Singh}, \citenamefont {Kamath},\ and\ \citenamefont
  {Bhattacharya}}]{Sosnowski1998}%
  \BibitemOpen
  \bibfield  {author} {\bibinfo {author} {\bibfnamefont {T.~S.}\ \bibnamefont
  {Sosnowski}}, \bibinfo {author} {\bibfnamefont {T.~B.}\ \bibnamefont
  {Norris}}, \bibinfo {author} {\bibfnamefont {H.}~\bibnamefont {Jiang}},
  \bibinfo {author} {\bibfnamefont {J.}~\bibnamefont {Singh}}, \bibinfo
  {author} {\bibfnamefont {K.}~\bibnamefont {Kamath}}, \ and\ \bibinfo {author}
  {\bibfnamefont {P.}~\bibnamefont {Bhattacharya}},\ }\href {\doibase
  10.1103/PhysRevB.57.R9423} {\bibfield  {journal} {\bibinfo  {journal} {Phys.
  Rev. B}\ }\textbf {\bibinfo {volume} {57}},\ \bibinfo {pages} {R9423}
  (\bibinfo {year} {1998})}\BibitemShut {NoStop}%
\bibitem [{\citenamefont {Mukai}\ and\ \citenamefont
  {Sugawara}(1998)}]{Mukai1998}%
  \BibitemOpen
  \bibfield  {author} {\bibinfo {author} {\bibfnamefont {K.}~\bibnamefont
  {Mukai}}\ and\ \bibinfo {author} {\bibfnamefont {M.}~\bibnamefont
  {Sugawara}},\ }\href {\doibase 10.1143/jjap.37.5451} {\bibfield  {journal}
  {\bibinfo  {journal} {Jpn J. Appl. Phys.}\ }\textbf {\bibinfo {volume}
  {37}},\ \bibinfo {pages} {5451} (\bibinfo {year} {1998})}\BibitemShut
  {NoStop}%
\bibitem [{\citenamefont {Klimov}\ \emph {et~al.}(1999)\citenamefont {Klimov},
  \citenamefont {McBranch}, \citenamefont {Leatherdale},\ and\ \citenamefont
  {Bawendi}}]{Klimov1999}%
  \BibitemOpen
  \bibfield  {author} {\bibinfo {author} {\bibfnamefont {V.~I.}\ \bibnamefont
  {Klimov}}, \bibinfo {author} {\bibfnamefont {D.~W.}\ \bibnamefont
  {McBranch}}, \bibinfo {author} {\bibfnamefont {C.~A.}\ \bibnamefont
  {Leatherdale}}, \ and\ \bibinfo {author} {\bibfnamefont {M.~G.}\ \bibnamefont
  {Bawendi}},\ }\href {\doibase 10.1103/PhysRevB.60.13740} {\bibfield
  {journal} {\bibinfo  {journal} {Phys. Rev. B}\ }\textbf {\bibinfo {volume}
  {60}},\ \bibinfo {pages} {13740} (\bibinfo {year} {1999})}\BibitemShut
  {NoStop}%
\bibitem [{\citenamefont {Klimov}\ \emph
  {et~al.}(2000{\natexlab{b}})\citenamefont {Klimov}, \citenamefont
  {Mikhailovsky}, \citenamefont {McBranch}, \citenamefont {Leatherdale},\ and\
  \citenamefont {Bawendi}}]{Klimov2000a}%
  \BibitemOpen
  \bibfield  {author} {\bibinfo {author} {\bibfnamefont {V.~I.}\ \bibnamefont
  {Klimov}}, \bibinfo {author} {\bibfnamefont {A.~A.}\ \bibnamefont
  {Mikhailovsky}}, \bibinfo {author} {\bibfnamefont {D.~W.}\ \bibnamefont
  {McBranch}}, \bibinfo {author} {\bibfnamefont {C.~A.}\ \bibnamefont
  {Leatherdale}}, \ and\ \bibinfo {author} {\bibfnamefont {M.~G.}\ \bibnamefont
  {Bawendi}},\ }\href {\doibase 10.1103/PhysRevB.61.R13349} {\bibfield
  {journal} {\bibinfo  {journal} {Phys. Rev. B}\ }\textbf {\bibinfo {volume}
  {61}},\ \bibinfo {pages} {R13349} (\bibinfo {year}
  {2000}{\natexlab{b}})}\BibitemShut {NoStop}%
\bibitem [{\citenamefont {Harbold}\ \emph {et~al.}(2005)\citenamefont
  {Harbold}, \citenamefont {Du}, \citenamefont {Krauss}, \citenamefont {Cho},
  \citenamefont {Murray},\ and\ \citenamefont {Wise}}]{Harbold2005}%
  \BibitemOpen
  \bibfield  {author} {\bibinfo {author} {\bibfnamefont {J.~M.}\ \bibnamefont
  {Harbold}}, \bibinfo {author} {\bibfnamefont {H.}~\bibnamefont {Du}},
  \bibinfo {author} {\bibfnamefont {T.~D.}\ \bibnamefont {Krauss}}, \bibinfo
  {author} {\bibfnamefont {K.-S.}\ \bibnamefont {Cho}}, \bibinfo {author}
  {\bibfnamefont {C.~B.}\ \bibnamefont {Murray}}, \ and\ \bibinfo {author}
  {\bibfnamefont {F.~W.}\ \bibnamefont {Wise}},\ }\href {\doibase
  10.1103/PhysRevB.72.195312} {\bibfield  {journal} {\bibinfo  {journal} {Phys.
  Rev. B}\ }\textbf {\bibinfo {volume} {72}},\ \bibinfo {pages} {195312}
  (\bibinfo {year} {2005})}\BibitemShut {NoStop}%
\bibitem [{\citenamefont {Nozik}(2001)}]{Nozik2001}%
  \BibitemOpen
  \bibfield  {author} {\bibinfo {author} {\bibfnamefont {A.~J.}\ \bibnamefont
  {Nozik}},\ }\href@noop {} {\bibfield  {journal} {\bibinfo  {journal} {Annu.
  Rev. Phys. Chem.}\ ,\ \bibinfo {pages} {193}} (\bibinfo {year}
  {2001})}\BibitemShut {NoStop}%
\bibitem [{\citenamefont {Efros}(2003)}]{Efros2003}%
  \BibitemOpen
  \bibfield  {author} {\bibinfo {author} {\bibfnamefont {A.}~\bibnamefont
  {Efros}},\ }\enquote {\bibinfo {title} {Auger processes in nanosize
  semiconductor crystals},}\ in\ \href {\doibase 10.1007/978-1-4757-3677-9_2}
  {\emph {\bibinfo {booktitle} {Semiconductor Nanocrystals: From Basic
  Principles to Applications}}},\ \bibinfo {editor} {edited by\ \bibinfo
  {editor} {\bibfnamefont {A.~L.}\ \bibnamefont {Efros}}, \bibinfo {editor}
  {\bibfnamefont {D.~J.}\ \bibnamefont {Lockwood}}, \ and\ \bibinfo {editor}
  {\bibfnamefont {L.}~\bibnamefont {Tsybeskov}}}\ (\bibinfo  {publisher}
  {Springer US},\ \bibinfo {address} {Boston, MA},\ \bibinfo {year} {2003})\
  pp.\ \bibinfo {pages} {52--72}\BibitemShut {NoStop}%
\bibitem [{\citenamefont {Wang}\ \emph {et~al.}(2003)\citenamefont {Wang},
  \citenamefont {Califano}, \citenamefont {Zunger},\ and\ \citenamefont
  {Franceschetti}}]{Wang2003}%
  \BibitemOpen
  \bibfield  {author} {\bibinfo {author} {\bibfnamefont {L.-W.}\ \bibnamefont
  {Wang}}, \bibinfo {author} {\bibfnamefont {M.}~\bibnamefont {Califano}},
  \bibinfo {author} {\bibfnamefont {A.}~\bibnamefont {Zunger}}, \ and\ \bibinfo
  {author} {\bibfnamefont {A.}~\bibnamefont {Franceschetti}},\ }\href {\doibase
  10.1103/PhysRevLett.91.056404} {\bibfield  {journal} {\bibinfo  {journal}
  {Phys. Rev. Lett.}\ }\textbf {\bibinfo {volume} {91}},\ \bibinfo {pages}
  {056404} (\bibinfo {year} {2003})}\BibitemShut {NoStop}%
\bibitem [{\citenamefont {Hendry}\ \emph {et~al.}(2006)\citenamefont {Hendry},
  \citenamefont {Koeberg}, \citenamefont {Wang}, \citenamefont {Zhang},
  \citenamefont {de~Mello~Doneg\'a}, \citenamefont {Vanmaekelbergh},\ and\
  \citenamefont {Bonn}}]{Hendry2006}%
  \BibitemOpen
  \bibfield  {author} {\bibinfo {author} {\bibfnamefont {E.}~\bibnamefont
  {Hendry}}, \bibinfo {author} {\bibfnamefont {M.}~\bibnamefont {Koeberg}},
  \bibinfo {author} {\bibfnamefont {F.}~\bibnamefont {Wang}}, \bibinfo {author}
  {\bibfnamefont {H.}~\bibnamefont {Zhang}}, \bibinfo {author} {\bibfnamefont
  {C.}~\bibnamefont {de~Mello~Doneg\'a}}, \bibinfo {author} {\bibfnamefont
  {D.}~\bibnamefont {Vanmaekelbergh}}, \ and\ \bibinfo {author} {\bibfnamefont
  {M.}~\bibnamefont {Bonn}},\ }\href {\doibase 10.1103/PhysRevLett.96.057408}
  {\bibfield  {journal} {\bibinfo  {journal} {Phys. Rev. Lett.}\ }\textbf
  {\bibinfo {volume} {96}},\ \bibinfo {pages} {057408} (\bibinfo {year}
  {2006})}\BibitemShut {NoStop}%
\bibitem [{\citenamefont {Kilina}, \citenamefont {Kilin},\ and\ \citenamefont
  {Prezhdo}(2009)}]{Kilina2009}%
  \BibitemOpen
  \bibfield  {author} {\bibinfo {author} {\bibfnamefont {S.~V.}\ \bibnamefont
  {Kilina}}, \bibinfo {author} {\bibfnamefont {D.~S.}\ \bibnamefont {Kilin}}, \
  and\ \bibinfo {author} {\bibfnamefont {O.~V.}\ \bibnamefont {Prezhdo}},\
  }\href {\doibase 10.1021/nn800674n} {\bibfield  {journal} {\bibinfo
  {journal} {ACS Nano}\ }\textbf {\bibinfo {volume} {3}},\ \bibinfo {pages}
  {93} (\bibinfo {year} {2009})}\BibitemShut {NoStop}%
\bibitem [{\citenamefont {Robel}\ \emph {et~al.}(2009)\citenamefont {Robel},
  \citenamefont {Gresback}, \citenamefont {Kortshagen}, \citenamefont
  {Schaller},\ and\ \citenamefont {Klimov}}]{Robel2009}%
  \BibitemOpen
  \bibfield  {author} {\bibinfo {author} {\bibfnamefont {I.}~\bibnamefont
  {Robel}}, \bibinfo {author} {\bibfnamefont {R.}~\bibnamefont {Gresback}},
  \bibinfo {author} {\bibfnamefont {U.}~\bibnamefont {Kortshagen}}, \bibinfo
  {author} {\bibfnamefont {R.~D.}\ \bibnamefont {Schaller}}, \ and\ \bibinfo
  {author} {\bibfnamefont {V.~I.}\ \bibnamefont {Klimov}},\ }\href {\doibase
  10.1103/PhysRevLett.102.177404} {\bibfield  {journal} {\bibinfo  {journal}
  {Phys. Rev. Lett.}\ }\textbf {\bibinfo {volume} {102}},\ \bibinfo {pages}
  {177404} (\bibinfo {year} {2009})}\BibitemShut {NoStop}%
\bibitem [{\citenamefont {Li}\ and\ \citenamefont {Lian}(2019)}]{Li2019a}%
  \BibitemOpen
  \bibfield  {author} {\bibinfo {author} {\bibfnamefont {Q.}~\bibnamefont
  {Li}}\ and\ \bibinfo {author} {\bibfnamefont {T.}~\bibnamefont {Lian}},\
  }\href {\doibase 10.1021/acs.accounts.9b00252} {\bibfield  {journal}
  {\bibinfo  {journal} {Acc. Chem. Res.}\ }\textbf {\bibinfo {volume} {52}},\
  \bibinfo {pages} {2684} (\bibinfo {year} {2019})}\BibitemShut {NoStop}%
\bibitem [{\citenamefont {Chepic}\ \emph {et~al.}(1990)\citenamefont {Chepic},
  \citenamefont {Efros}, \citenamefont {Ekimov}, \citenamefont {Ivanov},
  \citenamefont {Kharchenko}, \citenamefont {Kudriavtsev},\ and\ \citenamefont
  {Yazeva}}]{Chepic1990}%
  \BibitemOpen
  \bibfield  {author} {\bibinfo {author} {\bibfnamefont {D.~I.}\ \bibnamefont
  {Chepic}}, \bibinfo {author} {\bibfnamefont {A.~L.}\ \bibnamefont {Efros}},
  \bibinfo {author} {\bibfnamefont {A.~I.}\ \bibnamefont {Ekimov}}, \bibinfo
  {author} {\bibfnamefont {M.~G.}\ \bibnamefont {Ivanov}}, \bibinfo {author}
  {\bibfnamefont {V.~A.}\ \bibnamefont {Kharchenko}}, \bibinfo {author}
  {\bibfnamefont {I.~A.}\ \bibnamefont {Kudriavtsev}}, \ and\ \bibinfo {author}
  {\bibfnamefont {T.~V.}\ \bibnamefont {Yazeva}},\ }\href {\doibase
  10.1016/0022-2313(90)90007-X} {\bibfield  {journal} {\bibinfo  {journal} {J.
  Lumin.}\ }\textbf {\bibinfo {volume} {47}},\ \bibinfo {pages} {113} (\bibinfo
  {year} {1990})}\BibitemShut {NoStop}%
\bibitem [{\citenamefont {Vaxenburg}\ \emph {et~al.}(2015)\citenamefont
  {Vaxenburg}, \citenamefont {Rodina}, \citenamefont {Shabaev}, \citenamefont
  {Lifshitz},\ and\ \citenamefont {Efros}}]{Vaxenburg2015}%
  \BibitemOpen
  \bibfield  {author} {\bibinfo {author} {\bibfnamefont {R.}~\bibnamefont
  {Vaxenburg}}, \bibinfo {author} {\bibfnamefont {A.}~\bibnamefont {Rodina}},
  \bibinfo {author} {\bibfnamefont {A.}~\bibnamefont {Shabaev}}, \bibinfo
  {author} {\bibfnamefont {E.}~\bibnamefont {Lifshitz}}, \ and\ \bibinfo
  {author} {\bibfnamefont {A.~L.}\ \bibnamefont {Efros}},\ }\href {\doibase
  10.1021/nl504987h} {\bibfield  {journal} {\bibinfo  {journal} {Nano Lett.}\
  }\textbf {\bibinfo {volume} {15}},\ \bibinfo {pages} {2092} (\bibinfo {year}
  {2015})}\BibitemShut {NoStop}%
\bibitem [{\citenamefont {Vaxenburg}\ \emph {et~al.}(2016)\citenamefont
  {Vaxenburg}, \citenamefont {Rodina}, \citenamefont {Lifshitz},\ and\
  \citenamefont {Efros}}]{Vaxenburg2016}%
  \BibitemOpen
  \bibfield  {author} {\bibinfo {author} {\bibfnamefont {R.}~\bibnamefont
  {Vaxenburg}}, \bibinfo {author} {\bibfnamefont {A.}~\bibnamefont {Rodina}},
  \bibinfo {author} {\bibfnamefont {E.}~\bibnamefont {Lifshitz}}, \ and\
  \bibinfo {author} {\bibfnamefont {A.~L.}\ \bibnamefont {Efros}},\ }\href
  {\doibase 10.1021/acs.nanolett.6b00066} {\bibfield  {journal} {\bibinfo
  {journal} {Nano Lett.}\ }\textbf {\bibinfo {volume} {16}},\ \bibinfo {pages}
  {2503} (\bibinfo {year} {2016})}\BibitemShut {NoStop}%
\bibitem [{\citenamefont {Chelikowsky}, \citenamefont {Kronik},\ and\
  \citenamefont {Vasiliev}(2003)}]{Chelikowsky2003}%
  \BibitemOpen
  \bibfield  {author} {\bibinfo {author} {\bibfnamefont {J.~R.}\ \bibnamefont
  {Chelikowsky}}, \bibinfo {author} {\bibfnamefont {L.}~\bibnamefont {Kronik}},
  \ and\ \bibinfo {author} {\bibfnamefont {I.}~\bibnamefont {Vasiliev}},\
  }\href {\doibase 10.1088/0953-8984/15/35/201} {\bibfield  {journal} {\bibinfo
   {journal} {J. Phys. Condens. Matter}\ }\textbf {\bibinfo {volume} {15}},\
  \bibinfo {pages} {R1517} (\bibinfo {year} {2003})}\BibitemShut {NoStop}%
\bibitem [{\citenamefont {Shulenberger}\ \emph {et~al.}(2021)\citenamefont
  {Shulenberger}, \citenamefont {Coppieters~‘t Wallant}, \citenamefont
  {Klein}, \citenamefont {McIsaac}, \citenamefont {Goldzak}, \citenamefont
  {Berkinsky}, \citenamefont {Utzat}, \citenamefont {Barotov}, \citenamefont
  {Van~Voorhis},\ and\ \citenamefont {Bawendi}}]{Shulenberger2021}%
  \BibitemOpen
  \bibfield  {author} {\bibinfo {author} {\bibfnamefont {K.~E.}\ \bibnamefont
  {Shulenberger}}, \bibinfo {author} {\bibfnamefont {S.~C.}\ \bibnamefont
  {Coppieters~‘t Wallant}}, \bibinfo {author} {\bibfnamefont {M.~D.}\
  \bibnamefont {Klein}}, \bibinfo {author} {\bibfnamefont {A.~R.}\ \bibnamefont
  {McIsaac}}, \bibinfo {author} {\bibfnamefont {T.}~\bibnamefont {Goldzak}},
  \bibinfo {author} {\bibfnamefont {D.~B.}\ \bibnamefont {Berkinsky}}, \bibinfo
  {author} {\bibfnamefont {H.}~\bibnamefont {Utzat}}, \bibinfo {author}
  {\bibfnamefont {U.}~\bibnamefont {Barotov}}, \bibinfo {author} {\bibfnamefont
  {T.}~\bibnamefont {Van~Voorhis}}, \ and\ \bibinfo {author} {\bibfnamefont
  {M.~G.}\ \bibnamefont {Bawendi}},\ }\href {\doibase
  10.1021/acs.nanolett.0c05109} {\bibfield  {journal} {\bibinfo  {journal}
  {Nano Lett.}\ }\textbf {\bibinfo {volume} {21}},\ \bibinfo {pages} {7457}
  (\bibinfo {year} {2021})}\BibitemShut {NoStop}%
\bibitem [{\citenamefont {Song}\ \emph {et~al.}(2022)\citenamefont {Song},
  \citenamefont {Liu}, \citenamefont {Wang}, \citenamefont {Xu}, \citenamefont
  {Ma}, \citenamefont {Fan}, \citenamefont {Voznyy},\ and\ \citenamefont
  {Du}}]{Song2022}%
  \BibitemOpen
  \bibfield  {author} {\bibinfo {author} {\bibfnamefont {Y.}~\bibnamefont
  {Song}}, \bibinfo {author} {\bibfnamefont {R.}~\bibnamefont {Liu}}, \bibinfo
  {author} {\bibfnamefont {Z.}~\bibnamefont {Wang}}, \bibinfo {author}
  {\bibfnamefont {H.}~\bibnamefont {Xu}}, \bibinfo {author} {\bibfnamefont
  {Y.}~\bibnamefont {Ma}}, \bibinfo {author} {\bibfnamefont {F.}~\bibnamefont
  {Fan}}, \bibinfo {author} {\bibfnamefont {O.}~\bibnamefont {Voznyy}}, \ and\
  \bibinfo {author} {\bibfnamefont {J.}~\bibnamefont {Du}},\ }\href {\doibase
  10.1126/sciadv.abl8219} {\bibfield  {journal} {\bibinfo  {journal} {Sci.
  Adv.}\ }\textbf {\bibinfo {volume} {8}},\ \bibinfo {pages} {eabl8219}
  (\bibinfo {year} {2022})}\BibitemShut {NoStop}%
\bibitem [{\citenamefont {Degoli}\ \emph {et~al.}(2009)\citenamefont {Degoli},
  \citenamefont {Guerra}, \citenamefont {Iori}, \citenamefont {Magri},
  \citenamefont {Marri}, \citenamefont {Pulci}, \citenamefont {Bisi},\ and\
  \citenamefont {Ossicini}}]{Degoli2009}%
  \BibitemOpen
  \bibfield  {author} {\bibinfo {author} {\bibfnamefont {E.}~\bibnamefont
  {Degoli}}, \bibinfo {author} {\bibfnamefont {R.}~\bibnamefont {Guerra}},
  \bibinfo {author} {\bibfnamefont {F.}~\bibnamefont {Iori}}, \bibinfo {author}
  {\bibfnamefont {R.}~\bibnamefont {Magri}}, \bibinfo {author} {\bibfnamefont
  {I.}~\bibnamefont {Marri}}, \bibinfo {author} {\bibfnamefont
  {O.}~\bibnamefont {Pulci}}, \bibinfo {author} {\bibfnamefont
  {O.}~\bibnamefont {Bisi}}, \ and\ \bibinfo {author} {\bibfnamefont
  {S.}~\bibnamefont {Ossicini}},\ }\href {\doibase 10.1016/j.crhy.2008.09.003}
  {\bibfield  {journal} {\bibinfo  {journal} {Comptes Rendus Phisique}\
  }\textbf {\bibinfo {volume} {10}},\ \bibinfo {pages} {575} (\bibinfo {year}
  {2009})}\BibitemShut {NoStop}%
\bibitem [{\citenamefont {Friesner}(2005)}]{Friesner2005}%
  \BibitemOpen
  \bibfield  {author} {\bibinfo {author} {\bibfnamefont {R.~A.}\ \bibnamefont
  {Friesner}},\ }\href {\doibase 10.1073/pnas.0408036102} {\bibfield  {journal}
  {\bibinfo  {journal} {Proc. Natl. Acad. Sci. U.S.A.}\ }\textbf {\bibinfo
  {volume} {102}},\ \bibinfo {pages} {6648} (\bibinfo {year}
  {2005})}\BibitemShut {NoStop}%
\bibitem [{\citenamefont {Voznyy}\ \emph {et~al.}(2016)\citenamefont {Voznyy},
  \citenamefont {Mokkath}, \citenamefont {Jain}, \citenamefont {Sargent},\ and\
  \citenamefont {Schwingenschlögl}}]{Voznyy2016}%
  \BibitemOpen
  \bibfield  {author} {\bibinfo {author} {\bibfnamefont {O.}~\bibnamefont
  {Voznyy}}, \bibinfo {author} {\bibfnamefont {J.~H.}\ \bibnamefont {Mokkath}},
  \bibinfo {author} {\bibfnamefont {A.}~\bibnamefont {Jain}}, \bibinfo {author}
  {\bibfnamefont {E.~H.}\ \bibnamefont {Sargent}}, \ and\ \bibinfo {author}
  {\bibfnamefont {U.}~\bibnamefont {Schwingenschlögl}},\ }\href {\doibase
  10.1021/acs.jpcc.5b10908} {\bibfield  {journal} {\bibinfo  {journal} {J.
  Phys. Chem. C}\ }\textbf {\bibinfo {volume} {120}},\ \bibinfo {pages} {10015}
  (\bibinfo {year} {2016})}\BibitemShut {NoStop}%
\bibitem [{\citenamefont {Williamson}\ and\ \citenamefont
  {Zunger}(2000)}]{Williamson2000}%
  \BibitemOpen
  \bibfield  {author} {\bibinfo {author} {\bibfnamefont {A.}~\bibnamefont
  {Williamson}}\ and\ \bibinfo {author} {\bibfnamefont {A.}~\bibnamefont
  {Zunger}},\ }\href {\doibase 10.1103/PhysRevB.61.1978} {\bibfield  {journal}
  {\bibinfo  {journal} {Phys. Rev. B}\ }\textbf {\bibinfo {volume} {61}},\
  \bibinfo {pages} {1978} (\bibinfo {year} {2000})}\BibitemShut {NoStop}%
\bibitem [{\citenamefont {Wall}\ and\ \citenamefont
  {Neuhauser}(1995)}]{Wall1995}%
  \BibitemOpen
  \bibfield  {author} {\bibinfo {author} {\bibfnamefont {M.~R.}\ \bibnamefont
  {Wall}}\ and\ \bibinfo {author} {\bibfnamefont {D.}~\bibnamefont
  {Neuhauser}},\ }\href {\doibase 10.1063/1.468999} {\bibfield  {journal}
  {\bibinfo  {journal} {J. Chem. Phys.}\ }\textbf {\bibinfo {volume} {102}},\
  \bibinfo {pages} {8011} (\bibinfo {year} {1995})}\BibitemShut {NoStop}%
\bibitem [{\citenamefont {Toledo}\ and\ \citenamefont
  {Rabani}(2002)}]{Toledo2002}%
  \BibitemOpen
  \bibfield  {author} {\bibinfo {author} {\bibfnamefont {S.}~\bibnamefont
  {Toledo}}\ and\ \bibinfo {author} {\bibfnamefont {E.}~\bibnamefont
  {Rabani}},\ }\href {\doibase 10.1006/jcph.2002.7090} {\bibfield  {journal}
  {\bibinfo  {journal} {J. Comput. Phys.}\ }\textbf {\bibinfo {volume} {180}},\
  \bibinfo {pages} {256} (\bibinfo {year} {2002})}\BibitemShut {NoStop}%
\bibitem [{\citenamefont {Rohlfing}\ and\ \citenamefont
  {Louie}(2000)}]{Rohlfing2000}%
  \BibitemOpen
  \bibfield  {author} {\bibinfo {author} {\bibfnamefont {M.}~\bibnamefont
  {Rohlfing}}\ and\ \bibinfo {author} {\bibfnamefont {S.~G.}\ \bibnamefont
  {Louie}},\ }\href {\doibase 10.1103/PhysRevB.62.4927} {\bibfield  {journal}
  {\bibinfo  {journal} {Phys. Rev. B}\ }\textbf {\bibinfo {volume} {62}},\
  \bibinfo {pages} {4927} (\bibinfo {year} {2000})}\BibitemShut {NoStop}%
\bibitem [{\citenamefont {Philbin}\ and\ \citenamefont
  {Rabani}(2018)}]{Philbin2018}%
  \BibitemOpen
  \bibfield  {author} {\bibinfo {author} {\bibfnamefont {J.~P.}\ \bibnamefont
  {Philbin}}\ and\ \bibinfo {author} {\bibfnamefont {E.}~\bibnamefont
  {Rabani}},\ }\href {\doibase 10.1021/acs.nanolett.8b03715} {\bibfield
  {journal} {\bibinfo  {journal} {Nano Lett.}\ }\textbf {\bibinfo {volume}
  {18}},\ \bibinfo {pages} {7889} (\bibinfo {year} {2018})}\BibitemShut
  {NoStop}%
\bibitem [{\citenamefont {Giustino}(2017)}]{Giustino2017}%
  \BibitemOpen
  \bibfield  {author} {\bibinfo {author} {\bibfnamefont {F.}~\bibnamefont
  {Giustino}},\ }\href {\doibase 10.1103/RevModPhys.89.015003} {\bibfield
  {journal} {\bibinfo  {journal} {Rev. Mod. Phys.}\ }\textbf {\bibinfo {volume}
  {89}},\ \bibinfo {pages} {1} (\bibinfo {year} {2017})}\BibitemShut {NoStop}%
\bibitem [{\citenamefont {Zhou}\ \emph {et~al.}(2013)\citenamefont {Zhou},
  \citenamefont {Ward}, \citenamefont {Martin}, \citenamefont {{Van Swol}},
  \citenamefont {Cruz-Campa},\ and\ \citenamefont {Zubia}}]{Zhou2013}%
  \BibitemOpen
  \bibfield  {author} {\bibinfo {author} {\bibfnamefont {X.~W.}\ \bibnamefont
  {Zhou}}, \bibinfo {author} {\bibfnamefont {D.~K.}\ \bibnamefont {Ward}},
  \bibinfo {author} {\bibfnamefont {J.~E.}\ \bibnamefont {Martin}}, \bibinfo
  {author} {\bibfnamefont {F.~B.}\ \bibnamefont {{Van Swol}}}, \bibinfo
  {author} {\bibfnamefont {J.~L.}\ \bibnamefont {Cruz-Campa}}, \ and\ \bibinfo
  {author} {\bibfnamefont {D.}~\bibnamefont {Zubia}},\ }\href {\doibase
  10.1103/PhysRevB.88.085309} {\bibfield  {journal} {\bibinfo  {journal} {Phys.
  Rev. B}\ }\textbf {\bibinfo {volume} {88}},\ \bibinfo {pages} {085309}
  (\bibinfo {year} {2013})}\BibitemShut {NoStop}%
\bibitem [{\citenamefont {Philbin}\ and\ \citenamefont
  {Rabani}(2020)}]{Philbin2020a}%
  \BibitemOpen
  \bibfield  {author} {\bibinfo {author} {\bibfnamefont {J.~P.}\ \bibnamefont
  {Philbin}}\ and\ \bibinfo {author} {\bibfnamefont {E.}~\bibnamefont
  {Rabani}},\ }\href {\doibase 0.1021/acs.jpclett.0c01460} {\bibfield
  {journal} {\bibinfo  {journal} {J. Phys. Chem. Lett.}\ }\textbf {\bibinfo
  {volume} {11}},\ \bibinfo {pages} {5132} (\bibinfo {year}
  {2020})}\BibitemShut {NoStop}%
\bibitem [{\citenamefont {Takeshita}\ \emph {et~al.}(2017)\citenamefont
  {Takeshita}, \citenamefont {Jong}, \citenamefont {Neuhauser}, \citenamefont
  {Baer},\ and\ \citenamefont {Rabani}}]{Takeshita2017}%
  \BibitemOpen
  \bibfield  {author} {\bibinfo {author} {\bibfnamefont {T.~Y.}\ \bibnamefont
  {Takeshita}}, \bibinfo {author} {\bibfnamefont {W.~A.~D.}\ \bibnamefont
  {Jong}}, \bibinfo {author} {\bibfnamefont {D.}~\bibnamefont {Neuhauser}},
  \bibinfo {author} {\bibfnamefont {R.}~\bibnamefont {Baer}}, \ and\ \bibinfo
  {author} {\bibfnamefont {E.}~\bibnamefont {Rabani}},\ }\href {\doibase
  10.1021/acs.jctc.7b00343} {\bibfield  {journal} {\bibinfo  {journal} {J.
  Chem. Theory Comput.}\ }\textbf {\bibinfo {volume} {13}},\ \bibinfo {pages}
  {4605} (\bibinfo {year} {2017})}\BibitemShut {NoStop}%
\bibitem [{\citenamefont {Dou}\ \emph {et~al.}(2019)\citenamefont {Dou},
  \citenamefont {Takeshita}, \citenamefont {Chen}, \citenamefont {Baer},
  \citenamefont {Neuhauser},\ and\ \citenamefont {Rabani}}]{Dou2019}%
  \BibitemOpen
  \bibfield  {author} {\bibinfo {author} {\bibfnamefont {W.}~\bibnamefont
  {Dou}}, \bibinfo {author} {\bibfnamefont {T.~Y.}\ \bibnamefont {Takeshita}},
  \bibinfo {author} {\bibfnamefont {M.}~\bibnamefont {Chen}}, \bibinfo {author}
  {\bibfnamefont {R.}~\bibnamefont {Baer}}, \bibinfo {author} {\bibfnamefont
  {D.}~\bibnamefont {Neuhauser}}, \ and\ \bibinfo {author} {\bibfnamefont
  {E.}~\bibnamefont {Rabani}},\ }\href {\doibase 10.1021/acs.jctc.9b00918}
  {\bibfield  {journal} {\bibinfo  {journal} {J. Chem. Theory Comput.}\
  }\textbf {\bibinfo {volume} {15}},\ \bibinfo {pages} {6703} (\bibinfo {year}
  {2019})}\BibitemShut {NoStop}%
\bibitem [{\citenamefont {Bergstresser}\ and\ \citenamefont
  {Cohen}(1967)}]{Bergstresser1967}%
  \BibitemOpen
  \bibfield  {author} {\bibinfo {author} {\bibfnamefont {T.~K.}\ \bibnamefont
  {Bergstresser}}\ and\ \bibinfo {author} {\bibfnamefont {M.~L.}\ \bibnamefont
  {Cohen}},\ }\href {\doibase 10.1103/PhysRev.164.1069} {\bibfield  {journal}
  {\bibinfo  {journal} {Phys. Rev.}\ }\textbf {\bibinfo {volume} {164}},\
  \bibinfo {pages} {1069} (\bibinfo {year} {1967})}\BibitemShut {NoStop}%
\bibitem [{\citenamefont {Cao}\ and\ \citenamefont
  {Banin}(2000)}]{CaoBanin200}%
  \BibitemOpen
  \bibfield  {author} {\bibinfo {author} {\bibnamefont {Cao}}\ and\ \bibinfo
  {author} {\bibfnamefont {U.}~\bibnamefont {Banin}},\ }\href {\doibase
  10.1021/ja001386g} {\bibfield  {journal} {\bibinfo  {journal} {J. Am. Chem.
  Sci.}\ }\textbf {\bibinfo {volume} {122}},\ \bibinfo {pages} {9692} (\bibinfo
  {year} {2000})}\BibitemShut {NoStop}%
\bibitem [{\citenamefont {Reiss}, \citenamefont {Protière},\ and\
  \citenamefont {Li}(2009)}]{ReissSmall2009}%
  \BibitemOpen
  \bibfield  {author} {\bibinfo {author} {\bibfnamefont {P.}~\bibnamefont
  {Reiss}}, \bibinfo {author} {\bibfnamefont {M.}~\bibnamefont {Protière}}, \
  and\ \bibinfo {author} {\bibfnamefont {L.}~\bibnamefont {Li}},\ }\href
  {\doibase https://doi.org/10.1002/smll.200800841} {\bibfield  {journal}
  {\bibinfo  {journal} {Small}\ }\textbf {\bibinfo {volume} {5}},\ \bibinfo
  {pages} {154} (\bibinfo {year} {2009})}\BibitemShut {NoStop}%
\bibitem [{\citenamefont {Wuister}, \citenamefont {de~Mello~Donegá},\ and\
  \citenamefont {Meijerink}(2004)}]{Wuister2004}%
  \BibitemOpen
  \bibfield  {author} {\bibinfo {author} {\bibfnamefont {S.~F.}\ \bibnamefont
  {Wuister}}, \bibinfo {author} {\bibfnamefont {C.}~\bibnamefont
  {de~Mello~Donegá}}, \ and\ \bibinfo {author} {\bibfnamefont
  {A.}~\bibnamefont {Meijerink}},\ }\href {\doibase 10.1021/jp047078c}
  {\bibfield  {journal} {\bibinfo  {journal} {J. Phys. Chem. B}\ }\textbf
  {\bibinfo {volume} {108}},\ \bibinfo {pages} {17393} (\bibinfo {year}
  {2004})}\BibitemShut {NoStop}%
\bibitem [{\citenamefont {Guzelturk}\ \emph {et~al.}(2021)\citenamefont
  {Guzelturk}, \citenamefont {Cotts}, \citenamefont {Jasrasaria}, \citenamefont
  {Philbin}, \citenamefont {Hanifi}, \citenamefont {Koscher}, \citenamefont
  {Balan}, \citenamefont {Curling}, \citenamefont {Zajac}, \citenamefont
  {Park}, \citenamefont {Yazdani}, \citenamefont {Nyby}, \citenamefont
  {Kamysbayev}, \citenamefont {Fischer}, \citenamefont {Nett}, \citenamefont
  {Shen}, \citenamefont {Kozina}, \citenamefont {Lin}, \citenamefont {Reid},
  \citenamefont {Weathersby}, \citenamefont {Schaller}, \citenamefont {Wood},
  \citenamefont {Wang}, \citenamefont {Dionne}, \citenamefont {Talapin},
  \citenamefont {Alivisatos}, \citenamefont {Salleo}, \citenamefont {Rabani},\
  and\ \citenamefont {Lindenberg}}]{Guzelturk2021}%
  \BibitemOpen
  \bibfield  {author} {\bibinfo {author} {\bibfnamefont {B.}~\bibnamefont
  {Guzelturk}}, \bibinfo {author} {\bibfnamefont {B.~L.}\ \bibnamefont
  {Cotts}}, \bibinfo {author} {\bibfnamefont {D.}~\bibnamefont {Jasrasaria}},
  \bibinfo {author} {\bibfnamefont {J.~P.}\ \bibnamefont {Philbin}}, \bibinfo
  {author} {\bibfnamefont {D.~A.}\ \bibnamefont {Hanifi}}, \bibinfo {author}
  {\bibfnamefont {B.~A.}\ \bibnamefont {Koscher}}, \bibinfo {author}
  {\bibfnamefont {A.~D.}\ \bibnamefont {Balan}}, \bibinfo {author}
  {\bibfnamefont {E.}~\bibnamefont {Curling}}, \bibinfo {author} {\bibfnamefont
  {M.}~\bibnamefont {Zajac}}, \bibinfo {author} {\bibfnamefont
  {S.}~\bibnamefont {Park}}, \bibinfo {author} {\bibfnamefont {N.}~\bibnamefont
  {Yazdani}}, \bibinfo {author} {\bibfnamefont {C.}~\bibnamefont {Nyby}},
  \bibinfo {author} {\bibfnamefont {V.}~\bibnamefont {Kamysbayev}}, \bibinfo
  {author} {\bibfnamefont {S.}~\bibnamefont {Fischer}}, \bibinfo {author}
  {\bibfnamefont {Z.}~\bibnamefont {Nett}}, \bibinfo {author} {\bibfnamefont
  {X.}~\bibnamefont {Shen}}, \bibinfo {author} {\bibfnamefont {M.~E.}\
  \bibnamefont {Kozina}}, \bibinfo {author} {\bibfnamefont {M.~F.}\
  \bibnamefont {Lin}}, \bibinfo {author} {\bibfnamefont {A.~H.}\ \bibnamefont
  {Reid}}, \bibinfo {author} {\bibfnamefont {S.~P.}\ \bibnamefont
  {Weathersby}}, \bibinfo {author} {\bibfnamefont {R.~D.}\ \bibnamefont
  {Schaller}}, \bibinfo {author} {\bibfnamefont {V.}~\bibnamefont {Wood}},
  \bibinfo {author} {\bibfnamefont {X.}~\bibnamefont {Wang}}, \bibinfo {author}
  {\bibfnamefont {J.~A.}\ \bibnamefont {Dionne}}, \bibinfo {author}
  {\bibfnamefont {D.~V.}\ \bibnamefont {Talapin}}, \bibinfo {author}
  {\bibfnamefont {A.~P.}\ \bibnamefont {Alivisatos}}, \bibinfo {author}
  {\bibfnamefont {A.}~\bibnamefont {Salleo}}, \bibinfo {author} {\bibfnamefont
  {E.}~\bibnamefont {Rabani}}, \ and\ \bibinfo {author} {\bibfnamefont {A.~M.}\
  \bibnamefont {Lindenberg}},\ }\href {\doibase 10.1038/s41467-021-22116-0}
  {\bibfield  {journal} {\bibinfo  {journal} {Nat. Commun.}\ }\textbf {\bibinfo
  {volume} {12}},\ \bibinfo {pages} {1} (\bibinfo {year} {2021})}\BibitemShut
  {NoStop}%
\bibitem [{\citenamefont {Enright}\ \emph {et~al.}(2022)\citenamefont
  {Enright}, \citenamefont {Jasrasaria}, \citenamefont {Hanchard},
  \citenamefont {Needell}, \citenamefont {Phelan}, \citenamefont {Weinberg},
  \citenamefont {McDowell}, \citenamefont {Hsiao}, \citenamefont {Akbari},
  \citenamefont {Kottwitz}, \citenamefont {Potter}, \citenamefont {Wong},
  \citenamefont {Zuo}, \citenamefont {Atwater}, \citenamefont {Rabani},\ and\
  \citenamefont {Nuzzo}}]{Enright2022}%
  \BibitemOpen
  \bibfield  {author} {\bibinfo {author} {\bibfnamefont {M.~J.}\ \bibnamefont
  {Enright}}, \bibinfo {author} {\bibfnamefont {D.}~\bibnamefont {Jasrasaria}},
  \bibinfo {author} {\bibfnamefont {M.~M.}\ \bibnamefont {Hanchard}}, \bibinfo
  {author} {\bibfnamefont {D.~R.}\ \bibnamefont {Needell}}, \bibinfo {author}
  {\bibfnamefont {M.~E.}\ \bibnamefont {Phelan}}, \bibinfo {author}
  {\bibfnamefont {D.}~\bibnamefont {Weinberg}}, \bibinfo {author}
  {\bibfnamefont {B.~E.}\ \bibnamefont {McDowell}}, \bibinfo {author}
  {\bibfnamefont {H.-W.}\ \bibnamefont {Hsiao}}, \bibinfo {author}
  {\bibfnamefont {H.}~\bibnamefont {Akbari}}, \bibinfo {author} {\bibfnamefont
  {M.}~\bibnamefont {Kottwitz}}, \bibinfo {author} {\bibfnamefont {M.~M.}\
  \bibnamefont {Potter}}, \bibinfo {author} {\bibfnamefont {J.}~\bibnamefont
  {Wong}}, \bibinfo {author} {\bibfnamefont {J.-M.}\ \bibnamefont {Zuo}},
  \bibinfo {author} {\bibfnamefont {H.~A.}\ \bibnamefont {Atwater}}, \bibinfo
  {author} {\bibfnamefont {E.}~\bibnamefont {Rabani}}, \ and\ \bibinfo {author}
  {\bibfnamefont {R.~G.}\ \bibnamefont {Nuzzo}},\ }\href@noop {} {\bibfield
  {journal} {\bibinfo  {journal} {J. Phys. Chem. C}\ }\textbf {\bibinfo
  {volume} {126}},\ \bibinfo {pages} {7576} (\bibinfo {year}
  {2022})}\BibitemShut {NoStop}%
\bibitem [{\citenamefont {Jasrasaria}\ and\ \citenamefont
  {Rabani}(2021)}]{Jasrasaria2021}%
  \BibitemOpen
  \bibfield  {author} {\bibinfo {author} {\bibfnamefont {D.}~\bibnamefont
  {Jasrasaria}}\ and\ \bibinfo {author} {\bibfnamefont {E.}~\bibnamefont
  {Rabani}},\ }\href {\doibase 10.1021/acs.nanolett.1c02953} {\bibfield
  {journal} {\bibinfo  {journal} {Nano Lett.}\ }\textbf {\bibinfo {volume}
  {21}},\ \bibinfo {pages} {8741} (\bibinfo {year} {2021})}\BibitemShut
  {NoStop}%
\bibitem [{\citenamefont {Ekimov}, \citenamefont {Efros},\ and\ \citenamefont
  {Onushchenko}(1985)}]{Ekimov1985}%
  \BibitemOpen
  \bibfield  {author} {\bibinfo {author} {\bibfnamefont {A.~I.}\ \bibnamefont
  {Ekimov}}, \bibinfo {author} {\bibfnamefont {A.~L.}\ \bibnamefont {Efros}}, \
  and\ \bibinfo {author} {\bibfnamefont {A.~A.}\ \bibnamefont {Onushchenko}},\
  }\href {\doibase 10.1016/S0038-1098(85)80025-9} {\bibfield  {journal}
  {\bibinfo  {journal} {Solid State Commun.}\ }\textbf {\bibinfo {volume}
  {56}},\ \bibinfo {pages} {921} (\bibinfo {year} {1985})}\BibitemShut
  {NoStop}%
\bibitem [{\citenamefont {Ondry}\ \emph {et~al.}(2019)\citenamefont {Ondry},
  \citenamefont {Philbin}, \citenamefont {Lostica}, \citenamefont {Rabani},\
  and\ \citenamefont {Alivisatos}}]{Ondry2019}%
  \BibitemOpen
  \bibfield  {author} {\bibinfo {author} {\bibfnamefont {J.~C.}\ \bibnamefont
  {Ondry}}, \bibinfo {author} {\bibfnamefont {J.~P.}\ \bibnamefont {Philbin}},
  \bibinfo {author} {\bibfnamefont {M.}~\bibnamefont {Lostica}}, \bibinfo
  {author} {\bibfnamefont {E.}~\bibnamefont {Rabani}}, \ and\ \bibinfo {author}
  {\bibfnamefont {A.~P.}\ \bibnamefont {Alivisatos}},\ }\href {\doibase
  10.1021/acsnano.9b03052} {\bibfield  {journal} {\bibinfo  {journal} {ACS
  Nano}\ }\textbf {\bibinfo {volume} {13}},\ \bibinfo {pages} {12322} (\bibinfo
  {year} {2019})}\BibitemShut {NoStop}%
\bibitem [{\citenamefont {Cui}\ \emph {et~al.}(2019)\citenamefont {Cui},
  \citenamefont {Panfil}, \citenamefont {Koley}, \citenamefont {Shamalia},
  \citenamefont {Waiskopf}, \citenamefont {Remennik}, \citenamefont {Popov},
  \citenamefont {Oded},\ and\ \citenamefont {Banin}}]{Cui2019}%
  \BibitemOpen
  \bibfield  {author} {\bibinfo {author} {\bibfnamefont {J.}~\bibnamefont
  {Cui}}, \bibinfo {author} {\bibfnamefont {Y.~E.}\ \bibnamefont {Panfil}},
  \bibinfo {author} {\bibfnamefont {S.}~\bibnamefont {Koley}}, \bibinfo
  {author} {\bibfnamefont {D.}~\bibnamefont {Shamalia}}, \bibinfo {author}
  {\bibfnamefont {N.}~\bibnamefont {Waiskopf}}, \bibinfo {author}
  {\bibfnamefont {S.}~\bibnamefont {Remennik}}, \bibinfo {author}
  {\bibfnamefont {I.}~\bibnamefont {Popov}}, \bibinfo {author} {\bibfnamefont
  {M.}~\bibnamefont {Oded}}, \ and\ \bibinfo {author} {\bibfnamefont
  {U.}~\bibnamefont {Banin}},\ }\href {\doibase 10.1038/s41467-019-13349-1}
  {\bibfield  {journal} {\bibinfo  {journal} {Nat. Commun.}\ }\textbf {\bibinfo
  {volume} {10}},\ \bibinfo {pages} {5401} (\bibinfo {year}
  {2019})}\BibitemShut {NoStop}%
\bibitem [{\citenamefont {Zunger}(1980)}]{Zunger1980}%
  \BibitemOpen
  \bibfield  {author} {\bibinfo {author} {\bibfnamefont {A.}~\bibnamefont
  {Zunger}},\ }\href {\doibase 10.1103/PhysRevB.22.959} {\bibfield  {journal}
  {\bibinfo  {journal} {Phys. Rev. B}\ }\textbf {\bibinfo {volume} {22}},\
  \bibinfo {pages} {959} (\bibinfo {year} {1980})}\BibitemShut {NoStop}%
\bibitem [{\citenamefont {Hybertsen}\ and\ \citenamefont
  {Louie}(1986)}]{Hybertsen1986}%
  \BibitemOpen
  \bibfield  {author} {\bibinfo {author} {\bibfnamefont {M.~S.}\ \bibnamefont
  {Hybertsen}}\ and\ \bibinfo {author} {\bibfnamefont {S.~G.}\ \bibnamefont
  {Louie}},\ }\href {\doibase 10.1103/PhysRevB.34.5390} {\bibfield  {journal}
  {\bibinfo  {journal} {Phys. Rev. B}\ }\textbf {\bibinfo {volume} {34}},\
  \bibinfo {pages} {5390} (\bibinfo {year} {1986})}\BibitemShut {NoStop}%
\bibitem [{\citenamefont {Li}, \citenamefont {Gong},\ and\ \citenamefont
  {Wei}(2006)}]{Wei2006PRB_DefPot}%
  \BibitemOpen
  \bibfield  {author} {\bibinfo {author} {\bibfnamefont {Y.-H.}\ \bibnamefont
  {Li}}, \bibinfo {author} {\bibfnamefont {X.~G.}\ \bibnamefont {Gong}}, \ and\
  \bibinfo {author} {\bibfnamefont {S.-H.}\ \bibnamefont {Wei}},\ }\href
  {\doibase 10.1103/PhysRevB.73.245206} {\bibfield  {journal} {\bibinfo
  {journal} {Phys. Rev. B}\ }\textbf {\bibinfo {volume} {73}},\ \bibinfo
  {pages} {245206} (\bibinfo {year} {2006})}\BibitemShut {NoStop}%
\bibitem [{\citenamefont {Grünwald}\ \emph {et~al.}(2013)\citenamefont
  {Grünwald}, \citenamefont {Lutker}, \citenamefont {Alivisatos},
  \citenamefont {Rabani},\ and\ \citenamefont {Geissler}}]{Grunwald2013}%
  \BibitemOpen
  \bibfield  {author} {\bibinfo {author} {\bibfnamefont {M.}~\bibnamefont
  {Grünwald}}, \bibinfo {author} {\bibfnamefont {K.}~\bibnamefont {Lutker}},
  \bibinfo {author} {\bibfnamefont {A.~P.}\ \bibnamefont {Alivisatos}},
  \bibinfo {author} {\bibfnamefont {E.}~\bibnamefont {Rabani}}, \ and\ \bibinfo
  {author} {\bibfnamefont {P.~L.}\ \bibnamefont {Geissler}},\ }\href {\doibase
  10.1021/nl3007165} {\bibfield  {journal} {\bibinfo  {journal} {Nano Lett.}\
  }\textbf {\bibinfo {volume} {13}},\ \bibinfo {pages} {1367} (\bibinfo {year}
  {2013})}\BibitemShut {NoStop}%
\bibitem [{\citenamefont {Hazarika}\ \emph {et~al.}(2019)\citenamefont
  {Hazarika}, \citenamefont {Fedin}, \citenamefont {Hong}, \citenamefont {Guo},
  \citenamefont {Srivastava}, \citenamefont {Cho}, \citenamefont {Coropceanu},
  \citenamefont {Portner}, \citenamefont {Diroll}, \citenamefont {Philbin},
  \citenamefont {Rabani}, \citenamefont {Klie},\ and\ \citenamefont
  {Talapin}}]{Talapin2019}%
  \BibitemOpen
  \bibfield  {author} {\bibinfo {author} {\bibfnamefont {A.}~\bibnamefont
  {Hazarika}}, \bibinfo {author} {\bibfnamefont {I.}~\bibnamefont {Fedin}},
  \bibinfo {author} {\bibfnamefont {L.}~\bibnamefont {Hong}}, \bibinfo {author}
  {\bibfnamefont {J.}~\bibnamefont {Guo}}, \bibinfo {author} {\bibfnamefont
  {V.}~\bibnamefont {Srivastava}}, \bibinfo {author} {\bibfnamefont
  {W.}~\bibnamefont {Cho}}, \bibinfo {author} {\bibfnamefont {I.}~\bibnamefont
  {Coropceanu}}, \bibinfo {author} {\bibfnamefont {J.}~\bibnamefont {Portner}},
  \bibinfo {author} {\bibfnamefont {B.~T.}\ \bibnamefont {Diroll}}, \bibinfo
  {author} {\bibfnamefont {J.~P.}\ \bibnamefont {Philbin}}, \bibinfo {author}
  {\bibfnamefont {E.}~\bibnamefont {Rabani}}, \bibinfo {author} {\bibfnamefont
  {R.}~\bibnamefont {Klie}}, \ and\ \bibinfo {author} {\bibfnamefont {D.~V.}\
  \bibnamefont {Talapin}},\ }\href {\doibase 10.1021/jacs.9b04866} {\bibfield
  {journal} {\bibinfo  {journal} {J. Am. Chem. Soc.}\ }\textbf {\bibinfo
  {volume} {141}},\ \bibinfo {pages} {13487} (\bibinfo {year}
  {2019})}\BibitemShut {NoStop}%
\bibitem [{\citenamefont {Fan}\ \emph {et~al.}(2015)\citenamefont {Fan},
  \citenamefont {Liao}, \citenamefont {Xu}, \citenamefont {Zhang},
  \citenamefont {Cui},\ and\ \citenamefont {Zhang}}]{Zhang2015}%
  \BibitemOpen
  \bibfield  {author} {\bibinfo {author} {\bibfnamefont {K.}~\bibnamefont
  {Fan}}, \bibinfo {author} {\bibfnamefont {C.}~\bibnamefont {Liao}}, \bibinfo
  {author} {\bibfnamefont {R.}~\bibnamefont {Xu}}, \bibinfo {author}
  {\bibfnamefont {H.}~\bibnamefont {Zhang}}, \bibinfo {author} {\bibfnamefont
  {Y.}~\bibnamefont {Cui}}, \ and\ \bibinfo {author} {\bibfnamefont
  {J.}~\bibnamefont {Zhang}},\ }\href {\doibase
  https://doi.org/10.1016/j.cplett.2015.05.006} {\bibfield  {journal} {\bibinfo
   {journal} {Chem. Phys. Lett.}\ }\textbf {\bibinfo {volume} {633}},\ \bibinfo
  {pages} {1} (\bibinfo {year} {2015})}\BibitemShut {NoStop}%
\bibitem [{\citenamefont {Yu}\ \emph {et~al.}(2003)\citenamefont {Yu},
  \citenamefont {Qu}, \citenamefont {Guo},\ and\ \citenamefont
  {Peng}}]{Peng2003}%
  \BibitemOpen
  \bibfield  {author} {\bibinfo {author} {\bibfnamefont {W.~W.}\ \bibnamefont
  {Yu}}, \bibinfo {author} {\bibfnamefont {L.}~\bibnamefont {Qu}}, \bibinfo
  {author} {\bibfnamefont {W.}~\bibnamefont {Guo}}, \ and\ \bibinfo {author}
  {\bibfnamefont {X.}~\bibnamefont {Peng}},\ }\href {\doibase
  10.1021/cm034081k} {\bibfield  {journal} {\bibinfo  {journal} {Chem. Mater.}\
  }\textbf {\bibinfo {volume} {15}},\ \bibinfo {pages} {2854} (\bibinfo {year}
  {2003})}\BibitemShut {NoStop}%
\bibitem [{\citenamefont {Franceschetti}\ and\ \citenamefont
  {Zunger}(1997{\natexlab{a}})}]{Zunger1997}%
  \BibitemOpen
  \bibfield  {author} {\bibinfo {author} {\bibfnamefont {A.}~\bibnamefont
  {Franceschetti}}\ and\ \bibinfo {author} {\bibfnamefont {A.}~\bibnamefont
  {Zunger}},\ }\href {\doibase 10.1103/PhysRevLett.78.915} {\bibfield
  {journal} {\bibinfo  {journal} {Phys. Rev. Lett.}\ }\textbf {\bibinfo
  {volume} {78}},\ \bibinfo {pages} {915} (\bibinfo {year}
  {1997}{\natexlab{a}})}\BibitemShut {NoStop}%
\bibitem [{\citenamefont {Banin}\ \emph {et~al.}(1999)\citenamefont {Banin},
  \citenamefont {Cao}, \citenamefont {Katz},\ and\ \citenamefont
  {Millo}}]{Millo1999}%
  \BibitemOpen
  \bibfield  {author} {\bibinfo {author} {\bibfnamefont {U.}~\bibnamefont
  {Banin}}, \bibinfo {author} {\bibfnamefont {Y.}~\bibnamefont {Cao}}, \bibinfo
  {author} {\bibfnamefont {D.}~\bibnamefont {Katz}}, \ and\ \bibinfo {author}
  {\bibfnamefont {O.}~\bibnamefont {Millo}},\ }\href@noop {} {\bibfield
  {journal} {\bibinfo  {journal} {Nature}\ }\textbf {\bibinfo {volume} {400}},\
  \bibinfo {pages} {542} (\bibinfo {year} {1999})}\BibitemShut {NoStop}%
\bibitem [{\citenamefont {Guzelian}\ \emph {et~al.}(1996)\citenamefont
  {Guzelian}, \citenamefont {Banin}, \citenamefont {Kadavanich}, \citenamefont
  {Peng},\ and\ \citenamefont {Alivisatos}}]{Alivisatos1997}%
  \BibitemOpen
  \bibfield  {author} {\bibinfo {author} {\bibfnamefont {A.~A.}\ \bibnamefont
  {Guzelian}}, \bibinfo {author} {\bibfnamefont {U.}~\bibnamefont {Banin}},
  \bibinfo {author} {\bibfnamefont {A.~V.}\ \bibnamefont {Kadavanich}},
  \bibinfo {author} {\bibfnamefont {X.}~\bibnamefont {Peng}}, \ and\ \bibinfo
  {author} {\bibfnamefont {A.~P.}\ \bibnamefont {Alivisatos}},\ }\href
  {\doibase 10.1063/1.117605} {\bibfield  {journal} {\bibinfo  {journal} {Appl.
  Phys. Lett.}\ }\textbf {\bibinfo {volume} {69}},\ \bibinfo {pages} {1432}
  (\bibinfo {year} {1996})}\BibitemShut {NoStop}%
\bibitem [{\citenamefont {Franceschetti}\ and\ \citenamefont
  {Zunger}(2000)}]{Zunger2000}%
  \BibitemOpen
  \bibfield  {author} {\bibinfo {author} {\bibfnamefont {A.}~\bibnamefont
  {Franceschetti}}\ and\ \bibinfo {author} {\bibfnamefont {A.}~\bibnamefont
  {Zunger}},\ }\href {\doibase 10.1103/PhysRevB.62.2614} {\bibfield  {journal}
  {\bibinfo  {journal} {Phys. Rev. B}\ }\textbf {\bibinfo {volume} {62}},\
  \bibinfo {pages} {2614} (\bibinfo {year} {2000})}\BibitemShut {NoStop}%
\bibitem [{\citenamefont {\"O\ifmmode~\breve{g}\else \u{g}\fi{}\"ut},
  \citenamefont {Chelikowsky},\ and\ \citenamefont {Louie}(1997)}]{Ogut1997}%
  \BibitemOpen
  \bibfield  {author} {\bibinfo {author} {\bibfnamefont {S.}~\bibnamefont
  {\"O\ifmmode~\breve{g}\else \u{g}\fi{}\"ut}}, \bibinfo {author}
  {\bibfnamefont {J.~R.}\ \bibnamefont {Chelikowsky}}, \ and\ \bibinfo {author}
  {\bibfnamefont {S.~G.}\ \bibnamefont {Louie}},\ }\href {\doibase
  10.1103/PhysRevLett.79.1770} {\bibfield  {journal} {\bibinfo  {journal}
  {Phys. Rev. Lett.}\ }\textbf {\bibinfo {volume} {79}},\ \bibinfo {pages}
  {1770} (\bibinfo {year} {1997})}\BibitemShut {NoStop}%
\bibitem [{\citenamefont {Rabani}(2002)}]{Rabani2002}%
  \BibitemOpen
  \bibfield  {author} {\bibinfo {author} {\bibfnamefont {E.}~\bibnamefont
  {Rabani}},\ }\href {\doibase 10.1063/1.1424321} {\bibfield  {journal}
  {\bibinfo  {journal} {J. Chem. Phys.}\ }\textbf {\bibinfo {volume} {116}},\
  \bibinfo {pages} {258} (\bibinfo {year} {2002})}\BibitemShut {NoStop}%
\bibitem [{\citenamefont {Kelley}(2016)}]{KelleyJCP2016}%
  \BibitemOpen
  \bibfield  {author} {\bibinfo {author} {\bibfnamefont {A.~M.}\ \bibnamefont
  {Kelley}},\ }\href {\doibase 10.1063/1.4952990} {\bibfield  {journal}
  {\bibinfo  {journal} {J. Chem. Phys.}\ }\textbf {\bibinfo {volume} {144}}
  (\bibinfo {year} {2016}),\ 10.1063/1.4952990}\BibitemShut {NoStop}%
\bibitem [{\citenamefont {Wang}\ and\ \citenamefont {Zunger}(1998)}]{Wang1998}%
  \BibitemOpen
  \bibfield  {author} {\bibinfo {author} {\bibfnamefont {L.-W.}\ \bibnamefont
  {Wang}}\ and\ \bibinfo {author} {\bibfnamefont {A.}~\bibnamefont {Zunger}},\
  }\href {\doibase 10.1021/jp981018n} {\bibfield  {journal} {\bibinfo
  {journal} {J. Phys. Chem. B}\ }\textbf {\bibinfo {volume} {102}},\ \bibinfo
  {pages} {6449} (\bibinfo {year} {1998})}\BibitemShut {NoStop}%
\bibitem [{\citenamefont {Eshet}, \citenamefont {Gr{\"{u}}nwald},\ and\
  \citenamefont {Rabani}(2013)}]{Eshet2013}%
  \BibitemOpen
  \bibfield  {author} {\bibinfo {author} {\bibfnamefont {H.}~\bibnamefont
  {Eshet}}, \bibinfo {author} {\bibfnamefont {M.}~\bibnamefont
  {Gr{\"{u}}nwald}}, \ and\ \bibinfo {author} {\bibfnamefont {E.}~\bibnamefont
  {Rabani}},\ }\href {\doibase 10.1021/nl402722n} {\bibfield  {journal}
  {\bibinfo  {journal} {Nano Lett.}\ }\textbf {\bibinfo {volume} {13}},\
  \bibinfo {pages} {5880} (\bibinfo {year} {2013})}\BibitemShut {NoStop}%
\bibitem [{\citenamefont {Brumberg}\ \emph {et~al.}(2019)\citenamefont
  {Brumberg}, \citenamefont {Harvey}, \citenamefont {Philbin}, \citenamefont
  {Diroll}, \citenamefont {Crooker}, \citenamefont {Wasielewski}, \citenamefont
  {Rabani},\ and\ \citenamefont {Schaller}}]{Brumberg2019}%
  \BibitemOpen
  \bibfield  {author} {\bibinfo {author} {\bibfnamefont {A.}~\bibnamefont
  {Brumberg}}, \bibinfo {author} {\bibfnamefont {S.~M.}\ \bibnamefont
  {Harvey}}, \bibinfo {author} {\bibfnamefont {J.~P.}\ \bibnamefont {Philbin}},
  \bibinfo {author} {\bibfnamefont {B.~T.}\ \bibnamefont {Diroll}}, \bibinfo
  {author} {\bibfnamefont {S.~A.}\ \bibnamefont {Crooker}}, \bibinfo {author}
  {\bibfnamefont {M.~R.}\ \bibnamefont {Wasielewski}}, \bibinfo {author}
  {\bibfnamefont {E.}~\bibnamefont {Rabani}}, \ and\ \bibinfo {author}
  {\bibfnamefont {R.~D.}\ \bibnamefont {Schaller}},\ }\href {\doibase
  10.1021/acsnano.9b02008} {\bibfield  {journal} {\bibinfo  {journal} {ACS
  Nano}\ }\textbf {\bibinfo {volume} {13}},\ \bibinfo {pages} {8589} (\bibinfo
  {year} {2019})}\BibitemShut {NoStop}%
\bibitem [{\citenamefont {Hadar}\ \emph {et~al.}(2017)\citenamefont {Hadar},
  \citenamefont {Philbin}, \citenamefont {Panfil}, \citenamefont {Neyshtadt},
  \citenamefont {Lieberman}, \citenamefont {Eshet}, \citenamefont {Lazar},
  \citenamefont {Rabani},\ and\ \citenamefont {Banin}}]{Hadar2017}%
  \BibitemOpen
  \bibfield  {author} {\bibinfo {author} {\bibfnamefont {I.}~\bibnamefont
  {Hadar}}, \bibinfo {author} {\bibfnamefont {J.~P.}\ \bibnamefont {Philbin}},
  \bibinfo {author} {\bibfnamefont {Y.~E.}\ \bibnamefont {Panfil}}, \bibinfo
  {author} {\bibfnamefont {S.}~\bibnamefont {Neyshtadt}}, \bibinfo {author}
  {\bibfnamefont {I.}~\bibnamefont {Lieberman}}, \bibinfo {author}
  {\bibfnamefont {H.}~\bibnamefont {Eshet}}, \bibinfo {author} {\bibfnamefont
  {S.}~\bibnamefont {Lazar}}, \bibinfo {author} {\bibfnamefont
  {E.}~\bibnamefont {Rabani}}, \ and\ \bibinfo {author} {\bibfnamefont
  {U.}~\bibnamefont {Banin}},\ }\href {\doibase 10.1021/acs.nanolett.7b00254}
  {\bibfield  {journal} {\bibinfo  {journal} {Nano Lett.}\ }\textbf {\bibinfo
  {volume} {17}},\ \bibinfo {pages} {2524} (\bibinfo {year}
  {2017})}\BibitemShut {NoStop}%
\bibitem [{\citenamefont {Philbin}\ \emph {et~al.}(2020)\citenamefont
  {Philbin}, \citenamefont {Brumberg}, \citenamefont {Diroll}, \citenamefont
  {Cho}, \citenamefont {Talapin}, \citenamefont {Schaller},\ and\ \citenamefont
  {Rabani}}]{Philbin2020b}%
  \BibitemOpen
  \bibfield  {author} {\bibinfo {author} {\bibfnamefont {J.~P.}\ \bibnamefont
  {Philbin}}, \bibinfo {author} {\bibfnamefont {A.}~\bibnamefont {Brumberg}},
  \bibinfo {author} {\bibfnamefont {B.~T.}\ \bibnamefont {Diroll}}, \bibinfo
  {author} {\bibfnamefont {W.}~\bibnamefont {Cho}}, \bibinfo {author}
  {\bibfnamefont {D.~V.}\ \bibnamefont {Talapin}}, \bibinfo {author}
  {\bibfnamefont {R.~D.}\ \bibnamefont {Schaller}}, \ and\ \bibinfo {author}
  {\bibfnamefont {E.}~\bibnamefont {Rabani}},\ }\href {\doibase
  10.1063/5.0012973} {\bibfield  {journal} {\bibinfo  {journal} {J. Chem.
  Phys.}\ }\textbf {\bibinfo {volume} {153}},\ \bibinfo {pages} {054104}
  (\bibinfo {year} {2020})}\BibitemShut {NoStop}%
\bibitem [{\citenamefont {Mahan}(2000)}]{Mahan2000}%
  \BibitemOpen
  \bibfield  {author} {\bibinfo {author} {\bibfnamefont {G.~D.}\ \bibnamefont
  {Mahan}},\ }\href {\doibase 10.1007/978-1-4757-5714-9} {\emph {\bibinfo
  {title} {Many-Particle Physics}}}\ (\bibinfo  {publisher} {Springer},\
  \bibinfo {address} {Boston, MA},\ \bibinfo {year} {2000})\BibitemShut
  {NoStop}%
\bibitem [{\citenamefont {Balan}\ \emph {et~al.}(2017)\citenamefont {Balan},
  \citenamefont {Eshet}, \citenamefont {Olshansky}, \citenamefont {Lee},
  \citenamefont {Rabani},\ and\ \citenamefont {Alivisatos}}]{Balan2017}%
  \BibitemOpen
  \bibfield  {author} {\bibinfo {author} {\bibfnamefont {A.~D.}\ \bibnamefont
  {Balan}}, \bibinfo {author} {\bibfnamefont {H.}~\bibnamefont {Eshet}},
  \bibinfo {author} {\bibfnamefont {J.~H.}\ \bibnamefont {Olshansky}}, \bibinfo
  {author} {\bibfnamefont {Y.~V.}\ \bibnamefont {Lee}}, \bibinfo {author}
  {\bibfnamefont {E.}~\bibnamefont {Rabani}}, \ and\ \bibinfo {author}
  {\bibfnamefont {A.~P.}\ \bibnamefont {Alivisatos}},\ }\href {\doibase
  10.1021/acs.nanolett.6b04816} {\bibfield  {journal} {\bibinfo  {journal}
  {Nano Lett.}\ }\textbf {\bibinfo {volume} {17}},\ \bibinfo {pages} {1629}
  (\bibinfo {year} {2017})}\BibitemShut {NoStop}%
\bibitem [{\citenamefont {Cui}\ \emph {et~al.}(2016)\citenamefont {Cui},
  \citenamefont {Beyler}, \citenamefont {Coropceanu}, \citenamefont {Cleary},
  \citenamefont {Avila}, \citenamefont {Chen}, \citenamefont {Cordero},
  \citenamefont {Heathcote}, \citenamefont {Harris}, \citenamefont {Chen},
  \citenamefont {Cao},\ and\ \citenamefont {Bawendi}}]{Bawendi2016}%
  \BibitemOpen
  \bibfield  {author} {\bibinfo {author} {\bibfnamefont {J.}~\bibnamefont
  {Cui}}, \bibinfo {author} {\bibfnamefont {A.~P.}\ \bibnamefont {Beyler}},
  \bibinfo {author} {\bibfnamefont {I.}~\bibnamefont {Coropceanu}}, \bibinfo
  {author} {\bibfnamefont {L.}~\bibnamefont {Cleary}}, \bibinfo {author}
  {\bibfnamefont {T.~R.}\ \bibnamefont {Avila}}, \bibinfo {author}
  {\bibfnamefont {Y.}~\bibnamefont {Chen}}, \bibinfo {author} {\bibfnamefont
  {J.~M.}\ \bibnamefont {Cordero}}, \bibinfo {author} {\bibfnamefont {S.~L.}\
  \bibnamefont {Heathcote}}, \bibinfo {author} {\bibfnamefont {D.~K.}\
  \bibnamefont {Harris}}, \bibinfo {author} {\bibfnamefont {O.}~\bibnamefont
  {Chen}}, \bibinfo {author} {\bibfnamefont {J.}~\bibnamefont {Cao}}, \ and\
  \bibinfo {author} {\bibfnamefont {M.~G.}\ \bibnamefont {Bawendi}},\ }\href
  {\doibase 10.1021/acs.nanolett.5b03790} {\bibfield  {journal} {\bibinfo
  {journal} {Nano Lett.}\ }\textbf {\bibinfo {volume} {16}},\ \bibinfo {pages}
  {289} (\bibinfo {year} {2016})}\BibitemShut {NoStop}%
\bibitem [{\citenamefont {Mack}, \citenamefont {Jethi},\ and\ \citenamefont
  {Kambhampati}(2017)}]{Mack2017}%
  \BibitemOpen
  \bibfield  {author} {\bibinfo {author} {\bibfnamefont {T.~G.}\ \bibnamefont
  {Mack}}, \bibinfo {author} {\bibfnamefont {L.}~\bibnamefont {Jethi}}, \ and\
  \bibinfo {author} {\bibfnamefont {P.}~\bibnamefont {Kambhampati}},\ }\href
  {\doibase 10.1021/acs.jpcc.7b09903} {\bibfield  {journal} {\bibinfo
  {journal} {J. Phys. Chem. C}\ }\textbf {\bibinfo {volume} {121}},\ \bibinfo
  {pages} {28537} (\bibinfo {year} {2017})}\BibitemShut {NoStop}%
\bibitem [{\citenamefont {Peterson}\ \emph {et~al.}(2014)\citenamefont
  {Peterson}, \citenamefont {Cass}, \citenamefont {Harris}, \citenamefont
  {Edme}, \citenamefont {Sung},\ and\ \citenamefont {Weiss}}]{Peterson2014}%
  \BibitemOpen
  \bibfield  {author} {\bibinfo {author} {\bibfnamefont {M.~D.}\ \bibnamefont
  {Peterson}}, \bibinfo {author} {\bibfnamefont {L.~C.}\ \bibnamefont {Cass}},
  \bibinfo {author} {\bibfnamefont {R.~D.}\ \bibnamefont {Harris}}, \bibinfo
  {author} {\bibfnamefont {K.}~\bibnamefont {Edme}}, \bibinfo {author}
  {\bibfnamefont {K.}~\bibnamefont {Sung}}, \ and\ \bibinfo {author}
  {\bibfnamefont {E.~A.}\ \bibnamefont {Weiss}},\ }\href {\doibase
  10.1146/annurev-physchem-040513-103649} {\bibfield  {journal} {\bibinfo
  {journal} {Annu. Rev. Phys. Chem.}\ }\textbf {\bibinfo {volume} {65}},\
  \bibinfo {pages} {317} (\bibinfo {year} {2014})}\BibitemShut {NoStop}%
\bibitem [{\citenamefont {Tvrdy}, \citenamefont {Frantsuzov},\ and\
  \citenamefont {Kamat}(2011)}]{Kamat2011}%
  \BibitemOpen
  \bibfield  {author} {\bibinfo {author} {\bibfnamefont {K.}~\bibnamefont
  {Tvrdy}}, \bibinfo {author} {\bibfnamefont {P.~A.}\ \bibnamefont
  {Frantsuzov}}, \ and\ \bibinfo {author} {\bibfnamefont {P.~V.}\ \bibnamefont
  {Kamat}},\ }\href {\doibase 10.1073/pnas.1011972107} {\bibfield  {journal}
  {\bibinfo  {journal} {Proc. Natl. Acad. Sci. U. S. A.}\ }\textbf {\bibinfo
  {volume} {108}},\ \bibinfo {pages} {29} (\bibinfo {year} {2011})}\BibitemShut
  {NoStop}%
\bibitem [{\citenamefont {Harris}\ \emph {et~al.}(2016)\citenamefont {Harris},
  \citenamefont {Bettis~Homan}, \citenamefont {Kodaimati}, \citenamefont {He},
  \citenamefont {Nepomnyashchii}, \citenamefont {Swenson}, \citenamefont
  {Lian}, \citenamefont {Calzada},\ and\ \citenamefont {Weiss}}]{Harris2016}%
  \BibitemOpen
  \bibfield  {author} {\bibinfo {author} {\bibfnamefont {R.~D.}\ \bibnamefont
  {Harris}}, \bibinfo {author} {\bibfnamefont {S.}~\bibnamefont
  {Bettis~Homan}}, \bibinfo {author} {\bibfnamefont {M.}~\bibnamefont
  {Kodaimati}}, \bibinfo {author} {\bibfnamefont {C.}~\bibnamefont {He}},
  \bibinfo {author} {\bibfnamefont {A.~B.}\ \bibnamefont {Nepomnyashchii}},
  \bibinfo {author} {\bibfnamefont {N.~K.}\ \bibnamefont {Swenson}}, \bibinfo
  {author} {\bibfnamefont {S.}~\bibnamefont {Lian}}, \bibinfo {author}
  {\bibfnamefont {R.}~\bibnamefont {Calzada}}, \ and\ \bibinfo {author}
  {\bibfnamefont {E.~A.}\ \bibnamefont {Weiss}},\ }\href {\doibase
  10.1021/acs.chemrev.6b00102} {\bibfield  {journal} {\bibinfo  {journal}
  {Chem. Rev.}\ }\textbf {\bibinfo {volume} {116}},\ \bibinfo {pages} {12865}
  (\bibinfo {year} {2016})}\BibitemShut {NoStop}%
\bibitem [{\citenamefont {Bozyigit}\ \emph {et~al.}(2016)\citenamefont
  {Bozyigit}, \citenamefont {Yazdani}, \citenamefont {Yarema}, \citenamefont
  {Yarema}, \citenamefont {Lin}, \citenamefont {Volk}, \citenamefont
  {Vuttivorakulchai}, \citenamefont {Luisier}, \citenamefont {Juranyi},\ and\
  \citenamefont {Wood}}]{Wood2016Nature}%
  \BibitemOpen
  \bibfield  {author} {\bibinfo {author} {\bibfnamefont {D.}~\bibnamefont
  {Bozyigit}}, \bibinfo {author} {\bibfnamefont {N.}~\bibnamefont {Yazdani}},
  \bibinfo {author} {\bibfnamefont {M.}~\bibnamefont {Yarema}}, \bibinfo
  {author} {\bibfnamefont {O.}~\bibnamefont {Yarema}}, \bibinfo {author}
  {\bibfnamefont {W.~M.~M.}\ \bibnamefont {Lin}}, \bibinfo {author}
  {\bibfnamefont {S.}~\bibnamefont {Volk}}, \bibinfo {author} {\bibfnamefont
  {K.}~\bibnamefont {Vuttivorakulchai}}, \bibinfo {author} {\bibfnamefont
  {M.}~\bibnamefont {Luisier}}, \bibinfo {author} {\bibfnamefont
  {F.}~\bibnamefont {Juranyi}}, \ and\ \bibinfo {author} {\bibfnamefont
  {V.}~\bibnamefont {Wood}},\ }\href {\doibase 10.1038/nature16977} {\bibfield
  {journal} {\bibinfo  {journal} {Nature}\ }\textbf {\bibinfo {volume} {531}},\
  \bibinfo {pages} {618} (\bibinfo {year} {2016})}\BibitemShut {NoStop}%
\bibitem [{\citenamefont {Mack}\ \emph {et~al.}(2019)\citenamefont {Mack},
  \citenamefont {Jethi}, \citenamefont {Andrews},\ and\ \citenamefont
  {Kambhampati}}]{Mack2019}%
  \BibitemOpen
  \bibfield  {author} {\bibinfo {author} {\bibfnamefont {T.~G.}\ \bibnamefont
  {Mack}}, \bibinfo {author} {\bibfnamefont {L.}~\bibnamefont {Jethi}},
  \bibinfo {author} {\bibfnamefont {M.}~\bibnamefont {Andrews}}, \ and\
  \bibinfo {author} {\bibfnamefont {P.}~\bibnamefont {Kambhampati}},\ }\href
  {\doibase 10.1021/acs.jpcc.8b11098} {\bibfield  {journal} {\bibinfo
  {journal} {J. Phys. Chem. C}\ }\textbf {\bibinfo {volume} {123}},\ \bibinfo
  {pages} {5084} (\bibinfo {year} {2019})}\BibitemShut {NoStop}%
\bibitem [{\citenamefont {Wei}\ and\ \citenamefont {Chou}(1992)}]{Chou1992}%
  \BibitemOpen
  \bibfield  {author} {\bibinfo {author} {\bibfnamefont {S.}~\bibnamefont
  {Wei}}\ and\ \bibinfo {author} {\bibfnamefont {M.~Y.}\ \bibnamefont {Chou}},\
  }\href@noop {} {\bibfield  {journal} {\bibinfo  {journal} {Phys. Rev. Lett.}\
  }\textbf {\bibinfo {volume} {69}},\ \bibinfo {pages} {2799} (\bibinfo {year}
  {1992})}\BibitemShut {NoStop}%
\bibitem [{\citenamefont {Han}\ and\ \citenamefont
  {Bester}(2012)}]{Bester2012}%
  \BibitemOpen
  \bibfield  {author} {\bibinfo {author} {\bibfnamefont {P.}~\bibnamefont
  {Han}}\ and\ \bibinfo {author} {\bibfnamefont {G.}~\bibnamefont {Bester}},\
  }\href {\doibase 10.1103/PhysRevB.85.235422} {\bibfield  {journal} {\bibinfo
  {journal} {Phys. Rev. B}\ }\textbf {\bibinfo {volume} {85}},\ \bibinfo
  {pages} {1} (\bibinfo {year} {2012})}\BibitemShut {NoStop}%
\bibitem [{\citenamefont {Han}\ and\ \citenamefont
  {Bester}(2019)}]{Bester2019}%
  \BibitemOpen
  \bibfield  {author} {\bibinfo {author} {\bibfnamefont {P.}~\bibnamefont
  {Han}}\ and\ \bibinfo {author} {\bibfnamefont {G.}~\bibnamefont {Bester}},\
  }\href {\doibase 10.1103/PhysRevB.99.100302} {\bibfield  {journal} {\bibinfo
  {journal} {Phys. Rev. B}\ }\textbf {\bibinfo {volume} {99}},\ \bibinfo
  {pages} {1} (\bibinfo {year} {2019})}\BibitemShut {NoStop}%
\bibitem [{\citenamefont {Kong}(2011)}]{Kong2011}%
  \BibitemOpen
  \bibfield  {author} {\bibinfo {author} {\bibfnamefont {L.~T.}\ \bibnamefont
  {Kong}},\ }\href {\doibase 10.1016/j.cpc.2011.04.019} {\bibfield  {journal}
  {\bibinfo  {journal} {Comput. Phys. Commun.}\ }\textbf {\bibinfo {volume}
  {182}},\ \bibinfo {pages} {2201} (\bibinfo {year} {2011})}\BibitemShut
  {NoStop}%
\bibitem [{\citenamefont {Liptay}\ \emph {et~al.}(2007)\citenamefont {Liptay},
  \citenamefont {Marshall}, \citenamefont {Rao}, \citenamefont {Ram},\ and\
  \citenamefont {Bawendi}}]{Bawendi2007}%
  \BibitemOpen
  \bibfield  {author} {\bibinfo {author} {\bibfnamefont {T.~J.}\ \bibnamefont
  {Liptay}}, \bibinfo {author} {\bibfnamefont {L.~F.}\ \bibnamefont
  {Marshall}}, \bibinfo {author} {\bibfnamefont {P.~S.}\ \bibnamefont {Rao}},
  \bibinfo {author} {\bibfnamefont {R.~J.}\ \bibnamefont {Ram}}, \ and\
  \bibinfo {author} {\bibfnamefont {M.~G.}\ \bibnamefont {Bawendi}},\ }\href
  {\doibase 10.1103/PhysRevB.76.155314} {\bibfield  {journal} {\bibinfo
  {journal} {Phys. Rev. B: Condens. Matter Mater. Phys.}\ }\textbf {\bibinfo
  {volume} {76}},\ \bibinfo {pages} {1} (\bibinfo {year} {2007})}\BibitemShut
  {NoStop}%
\bibitem [{\citenamefont {Salvador}, \citenamefont {Graham},\ and\
  \citenamefont {Scholes}(2006)}]{Scholes2006JCP}%
  \BibitemOpen
  \bibfield  {author} {\bibinfo {author} {\bibfnamefont {M.~R.}\ \bibnamefont
  {Salvador}}, \bibinfo {author} {\bibfnamefont {M.~W.}\ \bibnamefont
  {Graham}}, \ and\ \bibinfo {author} {\bibfnamefont {G.~D.}\ \bibnamefont
  {Scholes}},\ }\href {\doibase 10.1063/1.2363190} {\bibfield  {journal}
  {\bibinfo  {journal} {J. Chem. Phys.}\ }\textbf {\bibinfo {volume} {125}}
  (\bibinfo {year} {2006}),\ 10.1063/1.2363190}\BibitemShut {NoStop}%
\bibitem [{\citenamefont {Kelley}(2011)}]{Kelley2011}%
  \BibitemOpen
  \bibfield  {author} {\bibinfo {author} {\bibfnamefont {A.~M.}\ \bibnamefont
  {Kelley}},\ }\href {\doibase 10.1021/nn201475d} {\bibfield  {journal}
  {\bibinfo  {journal} {ACS Nano}\ }\textbf {\bibinfo {volume} {5}},\ \bibinfo
  {pages} {5254} (\bibinfo {year} {2011})}\BibitemShut {NoStop}%
\bibitem [{\citenamefont {Gr{\"{u}}nwald}\ \emph {et~al.}(2012)\citenamefont
  {Gr{\"{u}}nwald}, \citenamefont {Zayak}, \citenamefont {Neaton},
  \citenamefont {Geissler},\ and\ \citenamefont {Rabani}}]{Grunwald2012}%
  \BibitemOpen
  \bibfield  {author} {\bibinfo {author} {\bibfnamefont {M.}~\bibnamefont
  {Gr{\"{u}}nwald}}, \bibinfo {author} {\bibfnamefont {A.}~\bibnamefont
  {Zayak}}, \bibinfo {author} {\bibfnamefont {J.~B.}\ \bibnamefont {Neaton}},
  \bibinfo {author} {\bibfnamefont {P.~L.}\ \bibnamefont {Geissler}}, \ and\
  \bibinfo {author} {\bibfnamefont {E.}~\bibnamefont {Rabani}},\ }\href
  {\doibase 10.1063/1.4729468} {\bibfield  {journal} {\bibinfo  {journal} {J.
  Chem. Phys.}\ }\textbf {\bibinfo {volume} {136}},\ \bibinfo {pages} {234111}
  (\bibinfo {year} {2012})}\BibitemShut {NoStop}%
\bibitem [{\citenamefont {Chilla}\ \emph {et~al.}(2008)\citenamefont {Chilla},
  \citenamefont {Kipp}, \citenamefont {Menke}, \citenamefont {Heitmann},
  \citenamefont {Nikolic}, \citenamefont {Fr\"omsdorf}, \citenamefont
  {Kornowski}, \citenamefont {F\"orster},\ and\ \citenamefont
  {Weller}}]{Weller2008}%
  \BibitemOpen
  \bibfield  {author} {\bibinfo {author} {\bibfnamefont {G.}~\bibnamefont
  {Chilla}}, \bibinfo {author} {\bibfnamefont {T.}~\bibnamefont {Kipp}},
  \bibinfo {author} {\bibfnamefont {T.}~\bibnamefont {Menke}}, \bibinfo
  {author} {\bibfnamefont {D.}~\bibnamefont {Heitmann}}, \bibinfo {author}
  {\bibfnamefont {M.}~\bibnamefont {Nikolic}}, \bibinfo {author} {\bibfnamefont
  {A.}~\bibnamefont {Fr\"omsdorf}}, \bibinfo {author} {\bibfnamefont
  {A.}~\bibnamefont {Kornowski}}, \bibinfo {author} {\bibfnamefont
  {S.}~\bibnamefont {F\"orster}}, \ and\ \bibinfo {author} {\bibfnamefont
  {H.}~\bibnamefont {Weller}},\ }\href {\doibase
  10.1103/PhysRevLett.100.057403} {\bibfield  {journal} {\bibinfo  {journal}
  {Phys. Rev. Lett.}\ }\textbf {\bibinfo {volume} {100}},\ \bibinfo {pages}
  {057403} (\bibinfo {year} {2008})}\BibitemShut {NoStop}%
\bibitem [{\citenamefont {Mork}, \citenamefont {Lee},\ and\ \citenamefont
  {Tisdale}(2016)}]{Tisdale2016}%
  \BibitemOpen
  \bibfield  {author} {\bibinfo {author} {\bibfnamefont {A.~J.}\ \bibnamefont
  {Mork}}, \bibinfo {author} {\bibfnamefont {E.~M.~Y.}\ \bibnamefont {Lee}}, \
  and\ \bibinfo {author} {\bibfnamefont {W.~A.}\ \bibnamefont {Tisdale}},\
  }\href {\doibase 10.1039/C6CP05683K} {\bibfield  {journal} {\bibinfo
  {journal} {Phys. Chem. Chem. Phys.}\ }\textbf {\bibinfo {volume} {18}},\
  \bibinfo {pages} {28797} (\bibinfo {year} {2016})}\BibitemShut {NoStop}%
\bibitem [{\citenamefont {Leo}, \citenamefont {R\"uhle},\ and\ \citenamefont
  {Ploog}(1988)}]{Ploog1988}%
  \BibitemOpen
  \bibfield  {author} {\bibinfo {author} {\bibfnamefont {K.}~\bibnamefont
  {Leo}}, \bibinfo {author} {\bibfnamefont {W.~W.}\ \bibnamefont {R\"uhle}}, \
  and\ \bibinfo {author} {\bibfnamefont {K.}~\bibnamefont {Ploog}},\ }\href
  {\doibase 10.1103/PhysRevB.38.1947} {\bibfield  {journal} {\bibinfo
  {journal} {Phys. Rev. B}\ }\textbf {\bibinfo {volume} {38}},\ \bibinfo
  {pages} {1947} (\bibinfo {year} {1988})}\BibitemShut {NoStop}%
\bibitem [{\citenamefont {Klimov}, \citenamefont {Haring~Bolivar},\ and\
  \citenamefont {Kurz}(1995)}]{Klimov1995}%
  \BibitemOpen
  \bibfield  {author} {\bibinfo {author} {\bibfnamefont {V.}~\bibnamefont
  {Klimov}}, \bibinfo {author} {\bibfnamefont {P.}~\bibnamefont
  {Haring~Bolivar}}, \ and\ \bibinfo {author} {\bibfnamefont {H.}~\bibnamefont
  {Kurz}},\ }\href {\doibase 10.1103/PhysRevB.52.4728} {\bibfield  {journal}
  {\bibinfo  {journal} {Phys. Rev. B}\ }\textbf {\bibinfo {volume} {52}},\
  \bibinfo {pages} {4728} (\bibinfo {year} {1995})}\BibitemShut {NoStop}%
\bibitem [{\citenamefont {Li}\ and\ \citenamefont {Lian}(2017)}]{Li2017}%
  \BibitemOpen
  \bibfield  {author} {\bibinfo {author} {\bibfnamefont {Q.}~\bibnamefont
  {Li}}\ and\ \bibinfo {author} {\bibfnamefont {T.}~\bibnamefont {Lian}},\
  }\href@noop {} {\bibfield  {journal} {\bibinfo  {journal} {Nano Lett.}\
  }\textbf {\bibinfo {volume} {17}},\ \bibinfo {pages} {3152} (\bibinfo {year}
  {2017})}\BibitemShut {NoStop}%
\bibitem [{\citenamefont {Han}\ and\ \citenamefont
  {Bester}(2017)}]{Bester2017}%
  \BibitemOpen
  \bibfield  {author} {\bibinfo {author} {\bibfnamefont {P.}~\bibnamefont
  {Han}}\ and\ \bibinfo {author} {\bibfnamefont {G.}~\bibnamefont {Bester}},\
  }\href {\doibase 10.1103/PhysRevB.96.195436} {\bibfield  {journal} {\bibinfo
  {journal} {Phys. Rev. B}\ }\textbf {\bibinfo {volume} {96}},\ \bibinfo
  {pages} {195436} (\bibinfo {year} {2017})}\BibitemShut {NoStop}%
\bibitem [{\citenamefont {Fernée}\ \emph {et~al.}(2014)\citenamefont
  {Fernée}, \citenamefont {Sinito}, \citenamefont {Mulvaney}, \citenamefont
  {Tamarat},\ and\ \citenamefont {Lounis}}]{Lounis2014}%
  \BibitemOpen
  \bibfield  {author} {\bibinfo {author} {\bibfnamefont {M.~J.}\ \bibnamefont
  {Fernée}}, \bibinfo {author} {\bibfnamefont {C.}~\bibnamefont {Sinito}},
  \bibinfo {author} {\bibfnamefont {P.}~\bibnamefont {Mulvaney}}, \bibinfo
  {author} {\bibfnamefont {P.}~\bibnamefont {Tamarat}}, \ and\ \bibinfo
  {author} {\bibfnamefont {B.}~\bibnamefont {Lounis}},\ }\href {\doibase
  10.1039/C4CP02022G} {\bibfield  {journal} {\bibinfo  {journal} {Phys. Chem.
  Chem. Phys.}\ }\textbf {\bibinfo {volume} {16}},\ \bibinfo {pages} {16957}
  (\bibinfo {year} {2014})}\BibitemShut {NoStop}%
\bibitem [{\citenamefont {Szilagyi}\ \emph {et~al.}(2015)\citenamefont
  {Szilagyi}, \citenamefont {Wittenberg}, \citenamefont {Miller}, \citenamefont
  {Lutker}, \citenamefont {Quirin}, \citenamefont {Lemke}, \citenamefont {Zhu},
  \citenamefont {Chollet}, \citenamefont {Robinson}, \citenamefont {Wen},
  \citenamefont {Sokolowski-Tinten},\ and\ \citenamefont
  {Lindenberg}}]{Lindenberg2015}%
  \BibitemOpen
  \bibfield  {author} {\bibinfo {author} {\bibfnamefont {E.}~\bibnamefont
  {Szilagyi}}, \bibinfo {author} {\bibfnamefont {J.~S.}\ \bibnamefont
  {Wittenberg}}, \bibinfo {author} {\bibfnamefont {T.~A.}\ \bibnamefont
  {Miller}}, \bibinfo {author} {\bibfnamefont {K.}~\bibnamefont {Lutker}},
  \bibinfo {author} {\bibfnamefont {F.}~\bibnamefont {Quirin}}, \bibinfo
  {author} {\bibfnamefont {H.}~\bibnamefont {Lemke}}, \bibinfo {author}
  {\bibfnamefont {D.}~\bibnamefont {Zhu}}, \bibinfo {author} {\bibfnamefont
  {M.}~\bibnamefont {Chollet}}, \bibinfo {author} {\bibfnamefont
  {J.}~\bibnamefont {Robinson}}, \bibinfo {author} {\bibfnamefont
  {H.}~\bibnamefont {Wen}}, \bibinfo {author} {\bibfnamefont {K.}~\bibnamefont
  {Sokolowski-Tinten}}, \ and\ \bibinfo {author} {\bibfnamefont {A.~M.}\
  \bibnamefont {Lindenberg}},\ }\href {\doibase 10.1038/ncomms7577} {\bibfield
  {journal} {\bibinfo  {journal} {Nat. Commun.}\ }\textbf {\bibinfo {volume}
  {6}},\ \bibinfo {pages} {6577} (\bibinfo {year} {2015})}\BibitemShut
  {NoStop}%
\bibitem [{\citenamefont {Klein}\ \emph {et~al.}(1990)\citenamefont {Klein},
  \citenamefont {Hache}, \citenamefont {Ricard},\ and\ \citenamefont
  {Flytzanis}}]{Flytzanis1990}%
  \BibitemOpen
  \bibfield  {author} {\bibinfo {author} {\bibfnamefont {M.~C.}\ \bibnamefont
  {Klein}}, \bibinfo {author} {\bibfnamefont {F.}~\bibnamefont {Hache}},
  \bibinfo {author} {\bibfnamefont {D.}~\bibnamefont {Ricard}}, \ and\ \bibinfo
  {author} {\bibfnamefont {C.}~\bibnamefont {Flytzanis}},\ }\href {\doibase
  10.1103/PhysRevB.42.11123} {\bibfield  {journal} {\bibinfo  {journal} {Phys.
  Rev. B}\ }\textbf {\bibinfo {volume} {42}},\ \bibinfo {pages} {11123}
  (\bibinfo {year} {1990})}\BibitemShut {NoStop}%
\bibitem [{\citenamefont {Nomura}\ and\ \citenamefont
  {Kobayashi}(1992)}]{Kobayashi1992}%
  \BibitemOpen
  \bibfield  {author} {\bibinfo {author} {\bibfnamefont {S.}~\bibnamefont
  {Nomura}}\ and\ \bibinfo {author} {\bibfnamefont {T.}~\bibnamefont
  {Kobayashi}},\ }\href@noop {} {\bibfield  {journal} {\bibinfo  {journal}
  {Phys. Rev. B}\ }\textbf {\bibinfo {volume} {45}},\ \bibinfo {pages} {1305}
  (\bibinfo {year} {1992})}\BibitemShut {NoStop}%
\bibitem [{\citenamefont {Marini}, \citenamefont {Stebe},\ and\ \citenamefont
  {Kartheuser}(1994)}]{Kartheuser1994}%
  \BibitemOpen
  \bibfield  {author} {\bibinfo {author} {\bibfnamefont {J.~C.}\ \bibnamefont
  {Marini}}, \bibinfo {author} {\bibfnamefont {B.}~\bibnamefont {Stebe}}, \
  and\ \bibinfo {author} {\bibfnamefont {E.}~\bibnamefont {Kartheuser}},\
  }\href {\doibase 10.1103/PhysRevB.50.14302} {\bibfield  {journal} {\bibinfo
  {journal} {Phys. Rev. B: Condens. Matter Mater. Phys.}\ }\textbf {\bibinfo
  {volume} {50}},\ \bibinfo {pages} {14302} (\bibinfo {year}
  {1994})}\BibitemShut {NoStop}%
\bibitem [{\citenamefont {Takagahara}(1996)}]{Takagahara1996}%
  \BibitemOpen
  \bibfield  {author} {\bibinfo {author} {\bibfnamefont {T.}~\bibnamefont
  {Takagahara}},\ }\href {\doibase 10.1016/0022-2313(96)00050-6} {\bibfield
  {journal} {\bibinfo  {journal} {J. Lumin.}\ }\textbf {\bibinfo {volume}
  {70}},\ \bibinfo {pages} {129} (\bibinfo {year} {1996})}\BibitemShut
  {NoStop}%
\bibitem [{\citenamefont {Hamma}\ \emph {et~al.}(2007)\citenamefont {Hamma},
  \citenamefont {Miranda}, \citenamefont {Vasilevskiy},\ and\ \citenamefont
  {Zorkani}}]{Zorkani2007}%
  \BibitemOpen
  \bibfield  {author} {\bibinfo {author} {\bibfnamefont {M.}~\bibnamefont
  {Hamma}}, \bibinfo {author} {\bibfnamefont {R.~P.}\ \bibnamefont {Miranda}},
  \bibinfo {author} {\bibfnamefont {M.~I.}\ \bibnamefont {Vasilevskiy}}, \ and\
  \bibinfo {author} {\bibfnamefont {I.}~\bibnamefont {Zorkani}},\ }\href@noop
  {} {\bibfield  {journal} {\bibinfo  {journal} {J. Phys. Condens. Matter}\
  }\textbf {\bibinfo {volume} {19}} (\bibinfo {year} {2007})}\BibitemShut
  {NoStop}%
\bibitem [{\citenamefont {Craig}, \citenamefont {Duncan},\ and\ \citenamefont
  {Prezhdo}(2005)}]{Prezhdo2005PRL}%
  \BibitemOpen
  \bibfield  {author} {\bibinfo {author} {\bibfnamefont {C.~F.}\ \bibnamefont
  {Craig}}, \bibinfo {author} {\bibfnamefont {W.~R.}\ \bibnamefont {Duncan}}, \
  and\ \bibinfo {author} {\bibfnamefont {O.~V.}\ \bibnamefont {Prezhdo}},\
  }\href {\doibase 10.1103/PhysRevLett.95.163001} {\bibfield  {journal}
  {\bibinfo  {journal} {Phys. Rev. Lett.}\ }\textbf {\bibinfo {volume} {95}},\
  \bibinfo {pages} {163001} (\bibinfo {year} {2005})}\BibitemShut {NoStop}%
\bibitem [{\citenamefont {Akimov}\ and\ \citenamefont
  {Prezhdo}(2013)}]{Prezhdo2013JCTC}%
  \BibitemOpen
  \bibfield  {author} {\bibinfo {author} {\bibfnamefont {A.~V.}\ \bibnamefont
  {Akimov}}\ and\ \bibinfo {author} {\bibfnamefont {O.~V.}\ \bibnamefont
  {Prezhdo}},\ }\href {\doibase 10.1021/ct400641n} {\bibfield  {journal}
  {\bibinfo  {journal} {J. Chem. Theory Comput.}\ }\textbf {\bibinfo {volume}
  {9}},\ \bibinfo {pages} {4959} (\bibinfo {year} {2013})}\BibitemShut
  {NoStop}%
\bibitem [{\citenamefont {Yazdani}\ \emph {et~al.}(2018)\citenamefont
  {Yazdani}, \citenamefont {Bozyigit}, \citenamefont {Vuttivorakulchai},
  \citenamefont {Luisier}, \citenamefont {Infante},\ and\ \citenamefont
  {Wood}}]{Wood2018NL}%
  \BibitemOpen
  \bibfield  {author} {\bibinfo {author} {\bibfnamefont {N.}~\bibnamefont
  {Yazdani}}, \bibinfo {author} {\bibfnamefont {D.}~\bibnamefont {Bozyigit}},
  \bibinfo {author} {\bibfnamefont {K.}~\bibnamefont {Vuttivorakulchai}},
  \bibinfo {author} {\bibfnamefont {M.}~\bibnamefont {Luisier}}, \bibinfo
  {author} {\bibfnamefont {I.}~\bibnamefont {Infante}}, \ and\ \bibinfo
  {author} {\bibfnamefont {V.}~\bibnamefont {Wood}},\ }\href {\doibase
  10.1021/acs.nanolett.7b04729} {\bibfield  {journal} {\bibinfo  {journal}
  {Nano Lett.}\ }\textbf {\bibinfo {volume} {18}},\ \bibinfo {pages} {2233}
  (\bibinfo {year} {2018})}\BibitemShut {NoStop}%
\bibitem [{\citenamefont {Yazdani}\ \emph {et~al.}(2020)\citenamefont
  {Yazdani}, \citenamefont {Volk}, \citenamefont {Yarema}, \citenamefont
  {Yarema},\ and\ \citenamefont {Wood}}]{Wood2021}%
  \BibitemOpen
  \bibfield  {author} {\bibinfo {author} {\bibfnamefont {N.}~\bibnamefont
  {Yazdani}}, \bibinfo {author} {\bibfnamefont {S.}~\bibnamefont {Volk}},
  \bibinfo {author} {\bibfnamefont {O.}~\bibnamefont {Yarema}}, \bibinfo
  {author} {\bibfnamefont {M.}~\bibnamefont {Yarema}}, \ and\ \bibinfo {author}
  {\bibfnamefont {V.}~\bibnamefont {Wood}},\ }\href {\doibase
  10.1021/acsphotonics.0c00034} {\bibfield  {journal} {\bibinfo  {journal} {ACS
  Photonics}\ }\textbf {\bibinfo {volume} {7}},\ \bibinfo {pages} {1088}
  (\bibinfo {year} {2020})}\BibitemShut {NoStop}%
\bibitem [{\citenamefont {Zeng}\ and\ \citenamefont {He}(2021)}]{Zeng2021}%
  \BibitemOpen
  \bibfield  {author} {\bibinfo {author} {\bibfnamefont {T.}~\bibnamefont
  {Zeng}}\ and\ \bibinfo {author} {\bibfnamefont {Y.}~\bibnamefont {He}},\
  }\href {\doibase 10.1103/PhysRevB.103.035428} {\bibfield  {journal} {\bibinfo
   {journal} {Phys. Rev. B}\ }\textbf {\bibinfo {volume} {103}},\ \bibinfo
  {pages} {1} (\bibinfo {year} {2021})}\BibitemShut {NoStop}%
\bibitem [{\citenamefont {Morello}\ \emph {et~al.}(2007)\citenamefont
  {Morello}, \citenamefont {{De Giorgi}}, \citenamefont {Kudera}, \citenamefont
  {Manna}, \citenamefont {Cingolani},\ and\ \citenamefont {Anni}}]{Anni2007}%
  \BibitemOpen
  \bibfield  {author} {\bibinfo {author} {\bibfnamefont {G.}~\bibnamefont
  {Morello}}, \bibinfo {author} {\bibfnamefont {M.}~\bibnamefont {{De
  Giorgi}}}, \bibinfo {author} {\bibfnamefont {S.}~\bibnamefont {Kudera}},
  \bibinfo {author} {\bibfnamefont {L.}~\bibnamefont {Manna}}, \bibinfo
  {author} {\bibfnamefont {R.}~\bibnamefont {Cingolani}}, \ and\ \bibinfo
  {author} {\bibfnamefont {M.}~\bibnamefont {Anni}},\ }\href {\doibase
  10.1021/jp068307t} {\bibfield  {journal} {\bibinfo  {journal} {J. Phys. Chem.
  C}\ }\textbf {\bibinfo {volume} {111}},\ \bibinfo {pages} {5846} (\bibinfo
  {year} {2007})}\BibitemShut {NoStop}%
\bibitem [{\citenamefont {Sagar}\ \emph
  {et~al.}(2008{\natexlab{a}})\citenamefont {Sagar}, \citenamefont {Cooney},
  \citenamefont {Sewall}, \citenamefont {Dias}, \citenamefont {Barsan},
  \citenamefont {Butler},\ and\ \citenamefont
  {Kambhampati}}]{Kambhampati2008PRB}%
  \BibitemOpen
  \bibfield  {author} {\bibinfo {author} {\bibfnamefont {D.~M.}\ \bibnamefont
  {Sagar}}, \bibinfo {author} {\bibfnamefont {R.~R.}\ \bibnamefont {Cooney}},
  \bibinfo {author} {\bibfnamefont {S.~L.}\ \bibnamefont {Sewall}}, \bibinfo
  {author} {\bibfnamefont {E.~A.}\ \bibnamefont {Dias}}, \bibinfo {author}
  {\bibfnamefont {M.~M.}\ \bibnamefont {Barsan}}, \bibinfo {author}
  {\bibfnamefont {I.~S.}\ \bibnamefont {Butler}}, \ and\ \bibinfo {author}
  {\bibfnamefont {P.}~\bibnamefont {Kambhampati}},\ }\href {\doibase
  10.1103/PhysRevB.77.235321} {\bibfield  {journal} {\bibinfo  {journal} {Phys.
  Rev. B}\ }\textbf {\bibinfo {volume} {77}},\ \bibinfo {pages} {1} (\bibinfo
  {year} {2008}{\natexlab{a}})}\BibitemShut {NoStop}%
\bibitem [{\citenamefont {Sagar}\ \emph
  {et~al.}(2008{\natexlab{b}})\citenamefont {Sagar}, \citenamefont {Cooney},
  \citenamefont {Sewall},\ and\ \citenamefont
  {Kambhampati}}]{Kambhampati2008JPCC}%
  \BibitemOpen
  \bibfield  {author} {\bibinfo {author} {\bibfnamefont {D.~M.}\ \bibnamefont
  {Sagar}}, \bibinfo {author} {\bibfnamefont {R.~R.}\ \bibnamefont {Cooney}},
  \bibinfo {author} {\bibfnamefont {S.~L.}\ \bibnamefont {Sewall}}, \ and\
  \bibinfo {author} {\bibfnamefont {P.}~\bibnamefont {Kambhampati}},\ }\href
  {\doibase 10.1021/jp803386g} {\bibfield  {journal} {\bibinfo  {journal} {J.
  Phys. Chem. C}\ }\textbf {\bibinfo {volume} {112}},\ \bibinfo {pages} {9124}
  (\bibinfo {year} {2008}{\natexlab{b}})}\BibitemShut {NoStop}%
\bibitem [{\citenamefont {Scamarcio}\ \emph {et~al.}(1996)\citenamefont
  {Scamarcio}, \citenamefont {Spagnolo}, \citenamefont {Ventruti},
  \citenamefont {Lugar{\'{a}}},\ and\ \citenamefont {Righini}}]{Righini1996}%
  \BibitemOpen
  \bibfield  {author} {\bibinfo {author} {\bibfnamefont {G.}~\bibnamefont
  {Scamarcio}}, \bibinfo {author} {\bibfnamefont {V.}~\bibnamefont {Spagnolo}},
  \bibinfo {author} {\bibfnamefont {G.}~\bibnamefont {Ventruti}}, \bibinfo
  {author} {\bibfnamefont {M.}~\bibnamefont {Lugar{\'{a}}}}, \ and\ \bibinfo
  {author} {\bibfnamefont {G.}~\bibnamefont {Righini}},\ }\href {\doibase
  10.1103/PhysRevB.53.R10489} {\bibfield  {journal} {\bibinfo  {journal} {Phys.
  Rev. B}\ }\textbf {\bibinfo {volume} {53}},\ \bibinfo {pages} {R10489}
  (\bibinfo {year} {1996})}\BibitemShut {NoStop}%
\bibitem [{\citenamefont {Heitz}\ \emph {et~al.}(1999)\citenamefont {Heitz},
  \citenamefont {Mukhametzhanov}, \citenamefont {Stier}, \citenamefont
  {Madhukar},\ and\ \citenamefont {Bimberg}}]{Bimberg1999}%
  \BibitemOpen
  \bibfield  {author} {\bibinfo {author} {\bibfnamefont {R.}~\bibnamefont
  {Heitz}}, \bibinfo {author} {\bibfnamefont {I.}~\bibnamefont
  {Mukhametzhanov}}, \bibinfo {author} {\bibfnamefont {O.}~\bibnamefont
  {Stier}}, \bibinfo {author} {\bibfnamefont {A.}~\bibnamefont {Madhukar}}, \
  and\ \bibinfo {author} {\bibfnamefont {D.}~\bibnamefont {Bimberg}},\ }\href
  {\doibase 10.1103/PhysRevLett.83.4654} {\bibfield  {journal} {\bibinfo
  {journal} {Phys. Rev. Lett.}\ }\textbf {\bibinfo {volume} {83}},\ \bibinfo
  {pages} {4654} (\bibinfo {year} {1999})}\BibitemShut {NoStop}%
\bibitem [{\citenamefont {Lin}\ \emph {et~al.}(2015{\natexlab{a}})\citenamefont
  {Lin}, \citenamefont {Gong}, \citenamefont {Kelley},\ and\ \citenamefont
  {Kelley}}]{Kelley2015ACSNano}%
  \BibitemOpen
  \bibfield  {author} {\bibinfo {author} {\bibfnamefont {C.}~\bibnamefont
  {Lin}}, \bibinfo {author} {\bibfnamefont {K.}~\bibnamefont {Gong}}, \bibinfo
  {author} {\bibfnamefont {D.~F.}\ \bibnamefont {Kelley}}, \ and\ \bibinfo
  {author} {\bibfnamefont {A.~M.}\ \bibnamefont {Kelley}},\ }\href {\doibase
  10.1021/acsnano.5b02230} {\bibfield  {journal} {\bibinfo  {journal} {ACS
  Nano}\ }\textbf {\bibinfo {volume} {9}},\ \bibinfo {pages} {8131} (\bibinfo
  {year} {2015}{\natexlab{a}})}\BibitemShut {NoStop}%
\bibitem [{\citenamefont {Lin}\ \emph {et~al.}(2015{\natexlab{b}})\citenamefont
  {Lin}, \citenamefont {Gong}, \citenamefont {Kelley},\ and\ \citenamefont
  {Kelley}}]{Lin2015}%
  \BibitemOpen
  \bibfield  {author} {\bibinfo {author} {\bibfnamefont {C.}~\bibnamefont
  {Lin}}, \bibinfo {author} {\bibfnamefont {K.}~\bibnamefont {Gong}}, \bibinfo
  {author} {\bibfnamefont {D.~F.}\ \bibnamefont {Kelley}}, \ and\ \bibinfo
  {author} {\bibfnamefont {A.~M.}\ \bibnamefont {Kelley}},\ }\href@noop {}
  {\bibfield  {journal} {\bibinfo  {journal} {J. Phys. Chem. C}\ } (\bibinfo
  {year} {2015}{\natexlab{b}})}\BibitemShut {NoStop}%
\bibitem [{\citenamefont {Achermann}\ \emph {et~al.}(2006)\citenamefont
  {Achermann}, \citenamefont {Bartko}, \citenamefont {Hollingsworth},\ and\
  \citenamefont {Klimov}}]{Achermann2006}%
  \BibitemOpen
  \bibfield  {author} {\bibinfo {author} {\bibfnamefont {M.}~\bibnamefont
  {Achermann}}, \bibinfo {author} {\bibfnamefont {A.~P.}\ \bibnamefont
  {Bartko}}, \bibinfo {author} {\bibfnamefont {J.~A.}\ \bibnamefont
  {Hollingsworth}}, \ and\ \bibinfo {author} {\bibfnamefont {V.~I.}\
  \bibnamefont {Klimov}},\ }\href {\doibase 10.1038/nphys363} {\bibfield
  {journal} {\bibinfo  {journal} {Nat. Phys.}\ }\textbf {\bibinfo {volume}
  {2}},\ \bibinfo {pages} {557} (\bibinfo {year} {2006})}\BibitemShut {NoStop}%
\bibitem [{\citenamefont {Harvey}\ \emph {et~al.}(2018)\citenamefont {Harvey},
  \citenamefont {Phelan}, \citenamefont {Hannah}, \citenamefont {Brown},
  \citenamefont {Young}, \citenamefont {Kirschner}, \citenamefont
  {Wasielewski},\ and\ \citenamefont {Schaller}}]{Harvey2018}%
  \BibitemOpen
  \bibfield  {author} {\bibinfo {author} {\bibfnamefont {S.~M.}\ \bibnamefont
  {Harvey}}, \bibinfo {author} {\bibfnamefont {B.~T.}\ \bibnamefont {Phelan}},
  \bibinfo {author} {\bibfnamefont {D.~C.}\ \bibnamefont {Hannah}}, \bibinfo
  {author} {\bibfnamefont {K.~E.}\ \bibnamefont {Brown}}, \bibinfo {author}
  {\bibfnamefont {R.~M.}\ \bibnamefont {Young}}, \bibinfo {author}
  {\bibfnamefont {M.~S.}\ \bibnamefont {Kirschner}}, \bibinfo {author}
  {\bibfnamefont {M.~R.}\ \bibnamefont {Wasielewski}}, \ and\ \bibinfo {author}
  {\bibfnamefont {R.~D.}\ \bibnamefont {Schaller}},\ }\href {\doibase
  10.1021/acs.jpclett.8b01504} {\bibfield  {journal} {\bibinfo  {journal} {J.
  Phys. Chem. Lett.}\ }\textbf {\bibinfo {volume} {9}},\ \bibinfo {pages}
  {4481} (\bibinfo {year} {2018})}\BibitemShut {NoStop}%
\bibitem [{\citenamefont {Correa}\ \emph {et~al.}(2012)\citenamefont {Correa},
  \citenamefont {Dauler}, \citenamefont {Nair}, \citenamefont {Pan},
  \citenamefont {Rosenberg}, \citenamefont {Kerman}, \citenamefont {Molnar},
  \citenamefont {Hu}, \citenamefont {Marsili}, \citenamefont {Anant},
  \citenamefont {Berggren},\ and\ \citenamefont {Bawendi}}]{Correa2012}%
  \BibitemOpen
  \bibfield  {author} {\bibinfo {author} {\bibfnamefont {R.~E.}\ \bibnamefont
  {Correa}}, \bibinfo {author} {\bibfnamefont {E.~A.}\ \bibnamefont {Dauler}},
  \bibinfo {author} {\bibfnamefont {G.}~\bibnamefont {Nair}}, \bibinfo {author}
  {\bibfnamefont {S.~H.}\ \bibnamefont {Pan}}, \bibinfo {author} {\bibfnamefont
  {D.}~\bibnamefont {Rosenberg}}, \bibinfo {author} {\bibfnamefont {A.~J.}\
  \bibnamefont {Kerman}}, \bibinfo {author} {\bibfnamefont {R.~J.}\
  \bibnamefont {Molnar}}, \bibinfo {author} {\bibfnamefont {X.}~\bibnamefont
  {Hu}}, \bibinfo {author} {\bibfnamefont {F.}~\bibnamefont {Marsili}},
  \bibinfo {author} {\bibfnamefont {V.}~\bibnamefont {Anant}}, \bibinfo
  {author} {\bibfnamefont {K.~K.}\ \bibnamefont {Berggren}}, \ and\ \bibinfo
  {author} {\bibfnamefont {M.~G.}\ \bibnamefont {Bawendi}},\ }\href {\doibase
  10.1021/nl300642k} {\bibfield  {journal} {\bibinfo  {journal} {Nano Lett.}\
  }\textbf {\bibinfo {volume} {12}},\ \bibinfo {pages} {2953} (\bibinfo {year}
  {2012})}\BibitemShut {NoStop}%
\bibitem [{\citenamefont {Utzat}\ \emph {et~al.}(2019)\citenamefont {Utzat},
  \citenamefont {Sun}, \citenamefont {Kaplan}, \citenamefont {Krieg},
  \citenamefont {Ginterseder}, \citenamefont {Spokoyny}, \citenamefont {Klein},
  \citenamefont {Shulenberger}, \citenamefont {Perkinson}, \citenamefont
  {Kovalenko},\ and\ \citenamefont {Bawendi}}]{Utzat2019}%
  \BibitemOpen
  \bibfield  {author} {\bibinfo {author} {\bibfnamefont {H.}~\bibnamefont
  {Utzat}}, \bibinfo {author} {\bibfnamefont {W.}~\bibnamefont {Sun}}, \bibinfo
  {author} {\bibfnamefont {A.~E.~K.}\ \bibnamefont {Kaplan}}, \bibinfo {author}
  {\bibfnamefont {F.}~\bibnamefont {Krieg}}, \bibinfo {author} {\bibfnamefont
  {M.}~\bibnamefont {Ginterseder}}, \bibinfo {author} {\bibfnamefont
  {B.}~\bibnamefont {Spokoyny}}, \bibinfo {author} {\bibfnamefont {N.~D.}\
  \bibnamefont {Klein}}, \bibinfo {author} {\bibfnamefont {K.~E.}\ \bibnamefont
  {Shulenberger}}, \bibinfo {author} {\bibfnamefont {C.~F.}\ \bibnamefont
  {Perkinson}}, \bibinfo {author} {\bibfnamefont {M.~V.}\ \bibnamefont
  {Kovalenko}}, \ and\ \bibinfo {author} {\bibfnamefont {M.~G.}\ \bibnamefont
  {Bawendi}},\ }\href {\doibase 10.1126/science.aau7392} {\bibfield  {journal}
  {\bibinfo  {journal} {Science}\ }\textbf {\bibinfo {volume} {363}},\ \bibinfo
  {pages} {1068} (\bibinfo {year} {2019})}\BibitemShut {NoStop}%
\bibitem [{\citenamefont {Klimov}\ \emph {et~al.}(2008)\citenamefont {Klimov},
  \citenamefont {McGuire}, \citenamefont {Schaller},\ and\ \citenamefont
  {Rupasov}}]{Klimov2008}%
  \BibitemOpen
  \bibfield  {author} {\bibinfo {author} {\bibfnamefont {V.~I.}\ \bibnamefont
  {Klimov}}, \bibinfo {author} {\bibfnamefont {J.~A.}\ \bibnamefont {McGuire}},
  \bibinfo {author} {\bibfnamefont {R.~D.}\ \bibnamefont {Schaller}}, \ and\
  \bibinfo {author} {\bibfnamefont {V.~I.}\ \bibnamefont {Rupasov}},\
  }\href@noop {} {\bibfield  {journal} {\bibinfo  {journal} {Phys. Rev. B}\
  }\textbf {\bibinfo {volume} {77}},\ \bibinfo {pages} {195324} (\bibinfo
  {year} {2008})}\BibitemShut {NoStop}%
\bibitem [{\citenamefont {Baghani}\ \emph {et~al.}(2015)\citenamefont
  {Baghani}, \citenamefont {Oleary}, \citenamefont {Fedin}, \citenamefont
  {Talapin},\ and\ \citenamefont {Pelton}}]{Baghani2015}%
  \BibitemOpen
  \bibfield  {author} {\bibinfo {author} {\bibfnamefont {E.}~\bibnamefont
  {Baghani}}, \bibinfo {author} {\bibfnamefont {S.~K.}\ \bibnamefont {Oleary}},
  \bibinfo {author} {\bibfnamefont {I.}~\bibnamefont {Fedin}}, \bibinfo
  {author} {\bibfnamefont {D.~V.}\ \bibnamefont {Talapin}}, \ and\ \bibinfo
  {author} {\bibfnamefont {M.}~\bibnamefont {Pelton}},\ }\href {\doibase
  10.1021/acs.jpclett.5b00143} {\bibfield  {journal} {\bibinfo  {journal} {J.
  Phys. Chem. Lett.}\ }\textbf {\bibinfo {volume} {6}},\ \bibinfo {pages}
  {1032} (\bibinfo {year} {2015})}\BibitemShut {NoStop}%
\bibitem [{\citenamefont {Ben-Shahar}\ \emph {et~al.}(2016)\citenamefont
  {Ben-Shahar}, \citenamefont {Scotognella}, \citenamefont {Kriegel},
  \citenamefont {Moretti}, \citenamefont {Cerullo}, \citenamefont {Rabani},\
  and\ \citenamefont {Banin}}]{Ben-Shahar2016}%
  \BibitemOpen
  \bibfield  {author} {\bibinfo {author} {\bibfnamefont {Y.}~\bibnamefont
  {Ben-Shahar}}, \bibinfo {author} {\bibfnamefont {F.}~\bibnamefont
  {Scotognella}}, \bibinfo {author} {\bibfnamefont {I.}~\bibnamefont
  {Kriegel}}, \bibinfo {author} {\bibfnamefont {L.}~\bibnamefont {Moretti}},
  \bibinfo {author} {\bibfnamefont {G.}~\bibnamefont {Cerullo}}, \bibinfo
  {author} {\bibfnamefont {E.}~\bibnamefont {Rabani}}, \ and\ \bibinfo {author}
  {\bibfnamefont {U.}~\bibnamefont {Banin}},\ }\href {\doibase
  10.1038/ncomms10413} {\bibfield  {journal} {\bibinfo  {journal} {Nat.
  Commun.}\ }\textbf {\bibinfo {volume} {7}},\ \bibinfo {pages} {10416}
  (\bibinfo {year} {2016})}\BibitemShut {NoStop}%
\bibitem [{\citenamefont {Li}\ \emph {et~al.}(2016)\citenamefont {Li},
  \citenamefont {Wu}, \citenamefont {Chen}, \citenamefont {Chen}, \citenamefont
  {McBride},\ and\ \citenamefont {Lian}}]{Li2016}%
  \BibitemOpen
  \bibfield  {author} {\bibinfo {author} {\bibfnamefont {Q.}~\bibnamefont
  {Li}}, \bibinfo {author} {\bibfnamefont {K.}~\bibnamefont {Wu}}, \bibinfo
  {author} {\bibfnamefont {J.}~\bibnamefont {Chen}}, \bibinfo {author}
  {\bibfnamefont {Z.}~\bibnamefont {Chen}}, \bibinfo {author} {\bibfnamefont
  {J.~R.}\ \bibnamefont {McBride}}, \ and\ \bibinfo {author} {\bibfnamefont
  {T.}~\bibnamefont {Lian}},\ }\href {\doibase 10.1021/acsnano.6b00787}
  {\bibfield  {journal} {\bibinfo  {journal} {ACS Nano}\ }\textbf {\bibinfo
  {volume} {10}},\ \bibinfo {pages} {3843} (\bibinfo {year}
  {2016})}\BibitemShut {NoStop}%
\bibitem [{\citenamefont {Pelton}\ \emph {et~al.}(2017)\citenamefont {Pelton},
  \citenamefont {Andrews}, \citenamefont {Fedin}, \citenamefont {Talapin},
  \citenamefont {Leng},\ and\ \citenamefont {O'Leary}}]{Pelton2017}%
  \BibitemOpen
  \bibfield  {author} {\bibinfo {author} {\bibfnamefont {M.}~\bibnamefont
  {Pelton}}, \bibinfo {author} {\bibfnamefont {J.~J.}\ \bibnamefont {Andrews}},
  \bibinfo {author} {\bibfnamefont {I.}~\bibnamefont {Fedin}}, \bibinfo
  {author} {\bibfnamefont {D.~V.}\ \bibnamefont {Talapin}}, \bibinfo {author}
  {\bibfnamefont {H.}~\bibnamefont {Leng}}, \ and\ \bibinfo {author}
  {\bibfnamefont {S.~K.}\ \bibnamefont {O'Leary}},\ }\href {\doibase
  10.1021/acs.nanolett.7b03294} {\bibfield  {journal} {\bibinfo  {journal}
  {Nano Lett.}\ }\textbf {\bibinfo {volume} {17}},\ \bibinfo {pages} {6900}
  (\bibinfo {year} {2017})}\BibitemShut {NoStop}%
\bibitem [{\citenamefont {Garc{\'{i}}a-Santamar{\'{i}}a}\ \emph
  {et~al.}(2011)\citenamefont {Garc{\'{i}}a-Santamar{\'{i}}a}, \citenamefont
  {Brovelli}, \citenamefont {Viswanatha}, \citenamefont {Hollingsworth},
  \citenamefont {Htoon}, \citenamefont {Crooker},\ and\ \citenamefont
  {Klimov}}]{Garcia-Santamaria2011}%
  \BibitemOpen
  \bibfield  {author} {\bibinfo {author} {\bibfnamefont {F.}~\bibnamefont
  {Garc{\'{i}}a-Santamar{\'{i}}a}}, \bibinfo {author} {\bibfnamefont
  {S.}~\bibnamefont {Brovelli}}, \bibinfo {author} {\bibfnamefont
  {R.}~\bibnamefont {Viswanatha}}, \bibinfo {author} {\bibfnamefont {J.~A.}\
  \bibnamefont {Hollingsworth}}, \bibinfo {author} {\bibfnamefont
  {H.}~\bibnamefont {Htoon}}, \bibinfo {author} {\bibfnamefont {S.~A.}\
  \bibnamefont {Crooker}}, \ and\ \bibinfo {author} {\bibfnamefont {V.~I.}\
  \bibnamefont {Klimov}},\ }\href {\doibase 10.1021/nl103801e} {\bibfield
  {journal} {\bibinfo  {journal} {Nano Lett.}\ }\textbf {\bibinfo {volume}
  {11}},\ \bibinfo {pages} {687} (\bibinfo {year} {2011})}\BibitemShut
  {NoStop}%
\bibitem [{\citenamefont {Stolle}\ \emph {et~al.}(2017)\citenamefont {Stolle},
  \citenamefont {Lu}, \citenamefont {Yu}, \citenamefont {Schaller},\ and\
  \citenamefont {Korgel}}]{Stolle2017}%
  \BibitemOpen
  \bibfield  {author} {\bibinfo {author} {\bibfnamefont {C.~J.}\ \bibnamefont
  {Stolle}}, \bibinfo {author} {\bibfnamefont {X.}~\bibnamefont {Lu}}, \bibinfo
  {author} {\bibfnamefont {Y.}~\bibnamefont {Yu}}, \bibinfo {author}
  {\bibfnamefont {R.~D.}\ \bibnamefont {Schaller}}, \ and\ \bibinfo {author}
  {\bibfnamefont {B.~A.}\ \bibnamefont {Korgel}},\ }\href {\doibase
  10.1021/acs.nanolett.7b02386} {\bibfield  {journal} {\bibinfo  {journal}
  {Nano Lett.}\ }\textbf {\bibinfo {volume} {17}},\ \bibinfo {pages} {5580}
  (\bibinfo {year} {2017})}\BibitemShut {NoStop}%
\bibitem [{\citenamefont {She}\ \emph {et~al.}(2015)\citenamefont {She},
  \citenamefont {Fedin}, \citenamefont {Dolzhnikov}, \citenamefont {Dahlberg},
  \citenamefont {Engel}, \citenamefont {Schaller},\ and\ \citenamefont
  {Talapin}}]{She2015}%
  \BibitemOpen
  \bibfield  {author} {\bibinfo {author} {\bibfnamefont {C.}~\bibnamefont
  {She}}, \bibinfo {author} {\bibfnamefont {I.}~\bibnamefont {Fedin}}, \bibinfo
  {author} {\bibfnamefont {D.~S.}\ \bibnamefont {Dolzhnikov}}, \bibinfo
  {author} {\bibfnamefont {P.~D.}\ \bibnamefont {Dahlberg}}, \bibinfo {author}
  {\bibfnamefont {G.~S.}\ \bibnamefont {Engel}}, \bibinfo {author}
  {\bibfnamefont {R.~D.}\ \bibnamefont {Schaller}}, \ and\ \bibinfo {author}
  {\bibfnamefont {D.~V.}\ \bibnamefont {Talapin}},\ }\href {\doibase
  10.1021/acsnano.5b02509} {\bibfield  {journal} {\bibinfo  {journal} {ACS
  Nano}\ }\textbf {\bibinfo {volume} {9}},\ \bibinfo {pages} {9475} (\bibinfo
  {year} {2015})}\BibitemShut {NoStop}%
\bibitem [{\citenamefont {Franceschetti}\ and\ \citenamefont
  {Zunger}(1997{\natexlab{b}})}]{Franceschetti1997}%
  \BibitemOpen
  \bibfield  {author} {\bibinfo {author} {\bibfnamefont {A.}~\bibnamefont
  {Franceschetti}}\ and\ \bibinfo {author} {\bibfnamefont {A.}~\bibnamefont
  {Zunger}},\ }\href {\doibase 10.1103/PhysRevLett.78.915} {\bibfield
  {journal} {\bibinfo  {journal} {Phys. Rev. Lett.}\ }\textbf {\bibinfo
  {volume} {78}},\ \bibinfo {pages} {915} (\bibinfo {year}
  {1997}{\natexlab{b}})}\BibitemShut {NoStop}%
\bibitem [{\citenamefont {Baskoutas}(2005)}]{Baskoutas2005}%
  \BibitemOpen
  \bibfield  {author} {\bibinfo {author} {\bibfnamefont {S.}~\bibnamefont
  {Baskoutas}},\ }\href {\doibase https://doi.org/10.1016/j.cplett.2005.01.075}
  {\bibfield  {journal} {\bibinfo  {journal} {Chem. Phys. Lett.}\ }\textbf
  {\bibinfo {volume} {404}},\ \bibinfo {pages} {107} (\bibinfo {year}
  {2005})}\BibitemShut {NoStop}%
\bibitem [{\citenamefont {Rajadell}\ \emph {et~al.}(2009)\citenamefont
  {Rajadell}, \citenamefont {Climente}, \citenamefont {Planelles},\ and\
  \citenamefont {Bertoni}}]{Rajadell2009}%
  \BibitemOpen
  \bibfield  {author} {\bibinfo {author} {\bibfnamefont {F.}~\bibnamefont
  {Rajadell}}, \bibinfo {author} {\bibfnamefont {J.~I.}\ \bibnamefont
  {Climente}}, \bibinfo {author} {\bibfnamefont {J.}~\bibnamefont {Planelles}},
  \ and\ \bibinfo {author} {\bibfnamefont {A.}~\bibnamefont {Bertoni}},\ }\href
  {\doibase 10.1021/jp902652z} {\bibfield  {journal} {\bibinfo  {journal} {J.
  Phys. Chem. C}\ }\textbf {\bibinfo {volume} {113}},\ \bibinfo {pages} {11268}
  (\bibinfo {year} {2009})}\BibitemShut {NoStop}%
\bibitem [{\citenamefont {Royo}\ \emph {et~al.}(2010)\citenamefont {Royo},
  \citenamefont {Climente}, \citenamefont {Movilla},\ and\ \citenamefont
  {Planelles}}]{Royo2010}%
  \BibitemOpen
  \bibfield  {author} {\bibinfo {author} {\bibfnamefont {M.}~\bibnamefont
  {Royo}}, \bibinfo {author} {\bibfnamefont {J.~I.}\ \bibnamefont {Climente}},
  \bibinfo {author} {\bibfnamefont {J.~L.}\ \bibnamefont {Movilla}}, \ and\
  \bibinfo {author} {\bibfnamefont {J.}~\bibnamefont {Planelles}},\ }\href
  {\doibase 10.1088/0953-8984/23/1/015301} {\bibfield  {journal} {\bibinfo
  {journal} {J. Phys. Condens. Matter}\ }\textbf {\bibinfo {volume} {23}},\
  \bibinfo {pages} {015301} (\bibinfo {year} {2010})}\BibitemShut {NoStop}%
\bibitem [{\citenamefont {Scott}\ \emph {et~al.}(2016)\citenamefont {Scott},
  \citenamefont {Achtstein}, \citenamefont {Prudnikau}, \citenamefont
  {Antanovich}, \citenamefont {Siebbeles}, \citenamefont {Artemyev},\ and\
  \citenamefont {Woggon}}]{Scott2016}%
  \BibitemOpen
  \bibfield  {author} {\bibinfo {author} {\bibfnamefont {R.}~\bibnamefont
  {Scott}}, \bibinfo {author} {\bibfnamefont {A.~W.}\ \bibnamefont
  {Achtstein}}, \bibinfo {author} {\bibfnamefont {A.~V.}\ \bibnamefont
  {Prudnikau}}, \bibinfo {author} {\bibfnamefont {A.}~\bibnamefont
  {Antanovich}}, \bibinfo {author} {\bibfnamefont {L.~D.~A.}\ \bibnamefont
  {Siebbeles}}, \bibinfo {author} {\bibfnamefont {M.}~\bibnamefont {Artemyev}},
  \ and\ \bibinfo {author} {\bibfnamefont {U.}~\bibnamefont {Woggon}},\ }\href
  {\doibase 10.1021/acs.nanolett.6b03244} {\bibfield  {journal} {\bibinfo
  {journal} {Nano Lett.}\ }\textbf {\bibinfo {volume} {16}},\ \bibinfo {pages}
  {6576} (\bibinfo {year} {2016})}\BibitemShut {NoStop}%
\bibitem [{\citenamefont {Rajadell}, \citenamefont {Climente},\ and\
  \citenamefont {Planelles}(2017)}]{Rajadell2017}%
  \BibitemOpen
  \bibfield  {author} {\bibinfo {author} {\bibfnamefont {F.}~\bibnamefont
  {Rajadell}}, \bibinfo {author} {\bibfnamefont {J.~I.}\ \bibnamefont
  {Climente}}, \ and\ \bibinfo {author} {\bibfnamefont {J.}~\bibnamefont
  {Planelles}},\ }\href {\doibase 10.1103/PhysRevB.96.035307} {\bibfield
  {journal} {\bibinfo  {journal} {Phys. Rev. B}\ }\textbf {\bibinfo {volume}
  {96}},\ \bibinfo {pages} {035307} (\bibinfo {year} {2017})}\BibitemShut
  {NoStop}%
\bibitem [{\citenamefont {Cragg}\ and\ \citenamefont
  {Efros}(2010)}]{Cragg2010}%
  \BibitemOpen
  \bibfield  {author} {\bibinfo {author} {\bibfnamefont {G.~E.}\ \bibnamefont
  {Cragg}}\ and\ \bibinfo {author} {\bibfnamefont {A.~L.}\ \bibnamefont
  {Efros}},\ }\href {\doibase 10.1021/nl903592h} {\bibfield  {journal}
  {\bibinfo  {journal} {Nano Lett.}\ }\textbf {\bibinfo {volume} {10}},\
  \bibinfo {pages} {313} (\bibinfo {year} {2010})}\BibitemShut {NoStop}%
\bibitem [{\citenamefont {Taguchi}\ \emph {et~al.}(2011)\citenamefont
  {Taguchi}, \citenamefont {Saruyama}, \citenamefont {Teranishi},\ and\
  \citenamefont {Kanemitsu}}]{Taguchi2011}%
  \BibitemOpen
  \bibfield  {author} {\bibinfo {author} {\bibfnamefont {S.}~\bibnamefont
  {Taguchi}}, \bibinfo {author} {\bibfnamefont {M.}~\bibnamefont {Saruyama}},
  \bibinfo {author} {\bibfnamefont {T.}~\bibnamefont {Teranishi}}, \ and\
  \bibinfo {author} {\bibfnamefont {Y.}~\bibnamefont {Kanemitsu}},\ }\href
  {\doibase 10.1103/PhysRevB.83.155324} {\bibfield  {journal} {\bibinfo
  {journal} {Phys. Rev. B}\ }\textbf {\bibinfo {volume} {83}},\ \bibinfo
  {pages} {155324} (\bibinfo {year} {2011})}\BibitemShut {NoStop}%
\bibitem [{\citenamefont {Htoon}\ \emph {et~al.}(2003)\citenamefont {Htoon},
  \citenamefont {Hollingsworth}, \citenamefont {Dickerson},\ and\ \citenamefont
  {Klimov}}]{Htoon2003}%
  \BibitemOpen
  \bibfield  {author} {\bibinfo {author} {\bibfnamefont {H.}~\bibnamefont
  {Htoon}}, \bibinfo {author} {\bibfnamefont {J.~A.}\ \bibnamefont
  {Hollingsworth}}, \bibinfo {author} {\bibfnamefont {R.}~\bibnamefont
  {Dickerson}}, \ and\ \bibinfo {author} {\bibfnamefont {V.~I.}\ \bibnamefont
  {Klimov}},\ }\href {\doibase 10.1103/PhysRevLett.91.227401} {\bibfield
  {journal} {\bibinfo  {journal} {Phys. Rev. Lett.}\ }\textbf {\bibinfo
  {volume} {91}},\ \bibinfo {pages} {227401} (\bibinfo {year}
  {2003})}\BibitemShut {NoStop}%
\bibitem [{\citenamefont {Zhu}\ and\ \citenamefont {Lian}(2012)}]{Zhu2012}%
  \BibitemOpen
  \bibfield  {author} {\bibinfo {author} {\bibfnamefont {H.}~\bibnamefont
  {Zhu}}\ and\ \bibinfo {author} {\bibfnamefont {T.}~\bibnamefont {Lian}},\
  }\href {\doibase 10.1021/ja304724u} {\bibfield  {journal} {\bibinfo
  {journal} {J. Am. Chem. Soc.}\ }\textbf {\bibinfo {volume} {134}},\ \bibinfo
  {pages} {11289} (\bibinfo {year} {2012})}\BibitemShut {NoStop}%
\bibitem [{\citenamefont {Refaely-Abramson}\ \emph {et~al.}(2017)\citenamefont
  {Refaely-Abramson}, \citenamefont {{Da Jornada}}, \citenamefont {Louie},\
  and\ \citenamefont {Neaton}}]{Refaely-Abramson2017}%
  \BibitemOpen
  \bibfield  {author} {\bibinfo {author} {\bibfnamefont {S.}~\bibnamefont
  {Refaely-Abramson}}, \bibinfo {author} {\bibfnamefont {F.~H.}\ \bibnamefont
  {{Da Jornada}}}, \bibinfo {author} {\bibfnamefont {S.~G.}\ \bibnamefont
  {Louie}}, \ and\ \bibinfo {author} {\bibfnamefont {J.~B.}\ \bibnamefont
  {Neaton}},\ }\href {\doibase 10.1103/PhysRevLett.119.267401} {\bibfield
  {journal} {\bibinfo  {journal} {Phys. Rev. Lett.}\ }\textbf {\bibinfo
  {volume} {119}},\ \bibinfo {pages} {267401} (\bibinfo {year}
  {2017})}\BibitemShut {NoStop}%
\bibitem [{\citenamefont {Neuhauser}\ \emph {et~al.}(2016)\citenamefont
  {Neuhauser}, \citenamefont {Rabani}, \citenamefont {Cytter},\ and\
  \citenamefont {Baer}}]{Neuhauser2016}%
  \BibitemOpen
  \bibfield  {author} {\bibinfo {author} {\bibfnamefont {D.}~\bibnamefont
  {Neuhauser}}, \bibinfo {author} {\bibfnamefont {E.}~\bibnamefont {Rabani}},
  \bibinfo {author} {\bibfnamefont {Y.}~\bibnamefont {Cytter}}, \ and\ \bibinfo
  {author} {\bibfnamefont {R.}~\bibnamefont {Baer}},\ }\href {\doibase
  10.1021/acs.jpca.5b10573} {\bibfield  {journal} {\bibinfo  {journal} {J.
  Phys. Chem. A}\ }\textbf {\bibinfo {volume} {120}},\ \bibinfo {pages} {3071}
  (\bibinfo {year} {2016})}\BibitemShut {NoStop}%
\bibitem [{\citenamefont {Koley}\ \emph {et~al.}(2021)\citenamefont {Koley},
  \citenamefont {Cui}, \citenamefont {Panfil},\ and\ \citenamefont
  {Banin}}]{Koley2021}%
  \BibitemOpen
  \bibfield  {author} {\bibinfo {author} {\bibfnamefont {S.}~\bibnamefont
  {Koley}}, \bibinfo {author} {\bibfnamefont {J.}~\bibnamefont {Cui}}, \bibinfo
  {author} {\bibfnamefont {Y.~E.}\ \bibnamefont {Panfil}}, \ and\ \bibinfo
  {author} {\bibfnamefont {U.}~\bibnamefont {Banin}},\ }\href {\doibase
  10.1021/acs.accounts.0c00691} {\bibfield  {journal} {\bibinfo  {journal}
  {Acc. Chem. Res.}\ }\textbf {\bibinfo {volume} {54}},\ \bibinfo {pages}
  {1178} (\bibinfo {year} {2021})}\BibitemShut {NoStop}%
\bibitem [{\citenamefont {Philbin}\ \emph {et~al.}(2021)\citenamefont
  {Philbin}, \citenamefont {Kelly}, \citenamefont {Peng}, \citenamefont
  {Coropceanu}, \citenamefont {Hazarika}, \citenamefont {Talapin},
  \citenamefont {Rabani}, \citenamefont {Ma},\ and\ \citenamefont
  {Narang}}]{Philbin2021arxiv}%
  \BibitemOpen
  \bibfield  {author} {\bibinfo {author} {\bibfnamefont {J.~P.}\ \bibnamefont
  {Philbin}}, \bibinfo {author} {\bibfnamefont {J.}~\bibnamefont {Kelly}},
  \bibinfo {author} {\bibfnamefont {L.}~\bibnamefont {Peng}}, \bibinfo {author}
  {\bibfnamefont {I.}~\bibnamefont {Coropceanu}}, \bibinfo {author}
  {\bibfnamefont {A.}~\bibnamefont {Hazarika}}, \bibinfo {author}
  {\bibfnamefont {D.~V.}\ \bibnamefont {Talapin}}, \bibinfo {author}
  {\bibfnamefont {E.}~\bibnamefont {Rabani}}, \bibinfo {author} {\bibfnamefont
  {X.}~\bibnamefont {Ma}}, \ and\ \bibinfo {author} {\bibfnamefont
  {P.}~\bibnamefont {Narang}},\ }\href {\doibase 10.48550/ARXIV.2104.06452}
  {\enquote {\bibinfo {title} {Room temperature single-photon superfluorescence
  from a single epitaxial cuboid nano-heterostructure},}\ } (\bibinfo {year}
  {2021})\BibitemShut {NoStop}%
\bibitem [{\citenamefont {Baer}\ and\ \citenamefont {Rabani}(2012)}]{Baer2012}%
  \BibitemOpen
  \bibfield  {author} {\bibinfo {author} {\bibfnamefont {R.}~\bibnamefont
  {Baer}}\ and\ \bibinfo {author} {\bibfnamefont {E.}~\bibnamefont {Rabani}},\
  }\href {\doibase 10.1021/nl300452c} {\bibfield  {journal} {\bibinfo
  {journal} {Nano Lett.}\ }\textbf {\bibinfo {volume} {12}},\ \bibinfo {pages}
  {2123} (\bibinfo {year} {2012})}\BibitemShut {NoStop}%
\bibitem [{\citenamefont {Ben-Shahar}\ \emph {et~al.}(2018)\citenamefont
  {Ben-Shahar}, \citenamefont {Philbin}, \citenamefont {Scotognella},
  \citenamefont {Ganzar}, \citenamefont {Cerullo}, \citenamefont {Rabani},\
  and\ \citenamefont {Banin}}]{Ben-Shahar2018}%
  \BibitemOpen
  \bibfield  {author} {\bibinfo {author} {\bibfnamefont {Y.}~\bibnamefont
  {Ben-Shahar}}, \bibinfo {author} {\bibfnamefont {J.~P.}\ \bibnamefont
  {Philbin}}, \bibinfo {author} {\bibfnamefont {F.}~\bibnamefont
  {Scotognella}}, \bibinfo {author} {\bibfnamefont {L.}~\bibnamefont {Ganzar}},
  \bibinfo {author} {\bibfnamefont {G.}~\bibnamefont {Cerullo}}, \bibinfo
  {author} {\bibfnamefont {E.}~\bibnamefont {Rabani}}, \ and\ \bibinfo {author}
  {\bibfnamefont {U.}~\bibnamefont {Banin}},\ }\href@noop {} {\bibfield
  {journal} {\bibinfo  {journal} {Nano Lett.}\ }\textbf {\bibinfo {volume}
  {18}},\ \bibinfo {pages} {5211} (\bibinfo {year} {2018})}\BibitemShut
  {NoStop}%
\bibitem [{\citenamefont {Yan}\ \emph {et~al.}(2021)\citenamefont {Yan},
  \citenamefont {Weinberg}, \citenamefont {Jasrasaria}, \citenamefont
  {Kolaczkowski}, \citenamefont {Liu}, \citenamefont {Philbin}, \citenamefont
  {Balan}, \citenamefont {Liu}, \citenamefont {Schwartzberg}, \citenamefont
  {Rabani},\ and\ \citenamefont {Alivisatos}}]{Yan2021}%
  \BibitemOpen
  \bibfield  {author} {\bibinfo {author} {\bibfnamefont {C.}~\bibnamefont
  {Yan}}, \bibinfo {author} {\bibfnamefont {D.}~\bibnamefont {Weinberg}},
  \bibinfo {author} {\bibfnamefont {D.}~\bibnamefont {Jasrasaria}}, \bibinfo
  {author} {\bibfnamefont {M.~A.}\ \bibnamefont {Kolaczkowski}}, \bibinfo
  {author} {\bibfnamefont {Z.-j.}\ \bibnamefont {Liu}}, \bibinfo {author}
  {\bibfnamefont {J.~P.}\ \bibnamefont {Philbin}}, \bibinfo {author}
  {\bibfnamefont {A.~D.}\ \bibnamefont {Balan}}, \bibinfo {author}
  {\bibfnamefont {Y.}~\bibnamefont {Liu}}, \bibinfo {author} {\bibfnamefont
  {A.~M.}\ \bibnamefont {Schwartzberg}}, \bibinfo {author} {\bibfnamefont
  {E.}~\bibnamefont {Rabani}}, \ and\ \bibinfo {author} {\bibfnamefont {A.~P.}\
  \bibnamefont {Alivisatos}},\ }\href {\doibase 10.1021/acsnano.0c08158}
  {\bibfield  {journal} {\bibinfo  {journal} {ACS Nano}\ }\textbf {\bibinfo
  {volume} {15}},\ \bibinfo {pages} {2281} (\bibinfo {year}
  {2021})}\BibitemShut {NoStop}%
\bibitem [{\citenamefont {Frederick}\ and\ \citenamefont
  {Weiss}(2010)}]{Frederick2010}%
  \BibitemOpen
  \bibfield  {author} {\bibinfo {author} {\bibfnamefont {M.~T.}\ \bibnamefont
  {Frederick}}\ and\ \bibinfo {author} {\bibfnamefont {E.~A.}\ \bibnamefont
  {Weiss}},\ }\href {\doibase 10.1021/nn1007435} {\bibfield  {journal}
  {\bibinfo  {journal} {ACS Nano}\ }\textbf {\bibinfo {volume} {4}},\ \bibinfo
  {pages} {3195} (\bibinfo {year} {2010})}\BibitemShut {NoStop}%
\bibitem [{\citenamefont {Jain}\ \emph {et~al.}(2016)\citenamefont {Jain},
  \citenamefont {Voznyy}, \citenamefont {Hoogland}, \citenamefont
  {Korkusinski}, \citenamefont {Hawrylak},\ and\ \citenamefont
  {Sargent}}]{Jain2016}%
  \BibitemOpen
  \bibfield  {author} {\bibinfo {author} {\bibfnamefont {A.}~\bibnamefont
  {Jain}}, \bibinfo {author} {\bibfnamefont {O.}~\bibnamefont {Voznyy}},
  \bibinfo {author} {\bibfnamefont {S.}~\bibnamefont {Hoogland}}, \bibinfo
  {author} {\bibfnamefont {M.}~\bibnamefont {Korkusinski}}, \bibinfo {author}
  {\bibfnamefont {P.}~\bibnamefont {Hawrylak}}, \ and\ \bibinfo {author}
  {\bibfnamefont {E.~H.}\ \bibnamefont {Sargent}},\ }\href {\doibase
  10.1021/acs.nanolett.6b03059} {\bibfield  {journal} {\bibinfo  {journal}
  {Nano Lett.}\ }\textbf {\bibinfo {volume} {16}},\ \bibinfo {pages} {6491}
  (\bibinfo {year} {2016})}\BibitemShut {NoStop}%
\bibitem [{\citenamefont {Sagar}\ \emph {et~al.}(2020)\citenamefont {Sagar},
  \citenamefont {Bappi}, \citenamefont {Johnston}, \citenamefont {Chen},
  \citenamefont {Todorović}, \citenamefont {Levina}, \citenamefont
  {Saidaminov}, \citenamefont {García~de Arquer}, \citenamefont {Nam},
  \citenamefont {Choi}, \citenamefont {Hoogland}, \citenamefont {Voznyy},\ and\
  \citenamefont {Sargent}}]{Sagar2020}%
  \BibitemOpen
  \bibfield  {author} {\bibinfo {author} {\bibfnamefont {L.~K.}\ \bibnamefont
  {Sagar}}, \bibinfo {author} {\bibfnamefont {G.}~\bibnamefont {Bappi}},
  \bibinfo {author} {\bibfnamefont {A.}~\bibnamefont {Johnston}}, \bibinfo
  {author} {\bibfnamefont {B.}~\bibnamefont {Chen}}, \bibinfo {author}
  {\bibfnamefont {P.}~\bibnamefont {Todorović}}, \bibinfo {author}
  {\bibfnamefont {L.}~\bibnamefont {Levina}}, \bibinfo {author} {\bibfnamefont
  {M.~I.}\ \bibnamefont {Saidaminov}}, \bibinfo {author} {\bibfnamefont
  {F.~P.}\ \bibnamefont {García~de Arquer}}, \bibinfo {author} {\bibfnamefont
  {D.-H.}\ \bibnamefont {Nam}}, \bibinfo {author} {\bibfnamefont {M.-J.}\
  \bibnamefont {Choi}}, \bibinfo {author} {\bibfnamefont {S.}~\bibnamefont
  {Hoogland}}, \bibinfo {author} {\bibfnamefont {O.}~\bibnamefont {Voznyy}}, \
  and\ \bibinfo {author} {\bibfnamefont {E.~H.}\ \bibnamefont {Sargent}},\
  }\href {\doibase 10.1021/acs.chemmater.0c01788} {\bibfield  {journal}
  {\bibinfo  {journal} {Chem. Mater.}\ }\textbf {\bibinfo {volume} {32}},\
  \bibinfo {pages} {7703} (\bibinfo {year} {2020})}\BibitemShut {NoStop}%
\bibitem [{\citenamefont {Kong}\ \emph {et~al.}(2018)\citenamefont {Kong},
  \citenamefont {Jia}, \citenamefont {Ren}, \citenamefont {Xie}, \citenamefont
  {Wu},\ and\ \citenamefont {Lian}}]{Kong2018}%
  \BibitemOpen
  \bibfield  {author} {\bibinfo {author} {\bibfnamefont {D.}~\bibnamefont
  {Kong}}, \bibinfo {author} {\bibfnamefont {Y.}~\bibnamefont {Jia}}, \bibinfo
  {author} {\bibfnamefont {Y.}~\bibnamefont {Ren}}, \bibinfo {author}
  {\bibfnamefont {Z.}~\bibnamefont {Xie}}, \bibinfo {author} {\bibfnamefont
  {K.}~\bibnamefont {Wu}}, \ and\ \bibinfo {author} {\bibfnamefont
  {T.}~\bibnamefont {Lian}},\ }\href {\doibase 10.1021/acs.jpcc.8b01234}
  {\bibfield  {journal} {\bibinfo  {journal} {J. Phys. Chem. C}\ }\textbf
  {\bibinfo {volume} {122}},\ \bibinfo {pages} {14091} (\bibinfo {year}
  {2018})}\BibitemShut {NoStop}%
\bibitem [{\citenamefont {Zhu}, \citenamefont {Song},\ and\ \citenamefont
  {Lian}(2010)}]{Zhu2010}%
  \BibitemOpen
  \bibfield  {author} {\bibinfo {author} {\bibfnamefont {H.}~\bibnamefont
  {Zhu}}, \bibinfo {author} {\bibfnamefont {N.}~\bibnamefont {Song}}, \ and\
  \bibinfo {author} {\bibfnamefont {T.}~\bibnamefont {Lian}},\ }\href {\doibase
  10.1021/ja106710m} {\bibfield  {journal} {\bibinfo  {journal} {J. Am. Chem.
  Soc.}\ }\textbf {\bibinfo {volume} {132}},\ \bibinfo {pages} {15038}
  (\bibinfo {year} {2010})}\BibitemShut {NoStop}%
\bibitem [{\citenamefont {Williams}\ \emph {et~al.}(2009)\citenamefont
  {Williams}, \citenamefont {Tisdale}, \citenamefont {Leschkies}, \citenamefont
  {Haugstad}, \citenamefont {Norris}, \citenamefont {Aydil},\ and\
  \citenamefont {Zhu}}]{Williams2009}%
  \BibitemOpen
  \bibfield  {author} {\bibinfo {author} {\bibfnamefont {K.~J.}\ \bibnamefont
  {Williams}}, \bibinfo {author} {\bibfnamefont {W.~A.}\ \bibnamefont
  {Tisdale}}, \bibinfo {author} {\bibfnamefont {K.~S.}\ \bibnamefont
  {Leschkies}}, \bibinfo {author} {\bibfnamefont {G.}~\bibnamefont {Haugstad}},
  \bibinfo {author} {\bibfnamefont {D.~J.}\ \bibnamefont {Norris}}, \bibinfo
  {author} {\bibfnamefont {E.~S.}\ \bibnamefont {Aydil}}, \ and\ \bibinfo
  {author} {\bibfnamefont {X.-Y.}\ \bibnamefont {Zhu}},\ }\href {\doibase
  10.1021/nn9001819} {\bibfield  {journal} {\bibinfo  {journal} {ACS Nano}\
  }\textbf {\bibinfo {volume} {3}},\ \bibinfo {pages} {1532} (\bibinfo {year}
  {2009})}\BibitemShut {NoStop}%
\bibitem [{\citenamefont {Evers}\ \emph {et~al.}(2015)\citenamefont {Evers},
  \citenamefont {Schins}, \citenamefont {Aerts}, \citenamefont {Kulkarni},
  \citenamefont {Capiod}, \citenamefont {Berthe}, \citenamefont {Grandidier},
  \citenamefont {Delerue}, \citenamefont {van~der Zant}, \citenamefont {van
  Overbeek}, \citenamefont {Peters}, \citenamefont {Vanmaekelbergh},\ and\
  \citenamefont {Siebbeles}}]{Evers2015}%
  \BibitemOpen
  \bibfield  {author} {\bibinfo {author} {\bibfnamefont {W.~H.}\ \bibnamefont
  {Evers}}, \bibinfo {author} {\bibfnamefont {J.~M.}\ \bibnamefont {Schins}},
  \bibinfo {author} {\bibfnamefont {M.}~\bibnamefont {Aerts}}, \bibinfo
  {author} {\bibfnamefont {A.}~\bibnamefont {Kulkarni}}, \bibinfo {author}
  {\bibfnamefont {P.}~\bibnamefont {Capiod}}, \bibinfo {author} {\bibfnamefont
  {M.}~\bibnamefont {Berthe}}, \bibinfo {author} {\bibfnamefont
  {B.}~\bibnamefont {Grandidier}}, \bibinfo {author} {\bibfnamefont
  {C.}~\bibnamefont {Delerue}}, \bibinfo {author} {\bibfnamefont {H.~S.~J.}\
  \bibnamefont {van~der Zant}}, \bibinfo {author} {\bibfnamefont
  {C.}~\bibnamefont {van Overbeek}}, \bibinfo {author} {\bibfnamefont {J.~L.}\
  \bibnamefont {Peters}}, \bibinfo {author} {\bibfnamefont {D.}~\bibnamefont
  {Vanmaekelbergh}}, \ and\ \bibinfo {author} {\bibfnamefont {L.~D.~A.}\
  \bibnamefont {Siebbeles}},\ }\href {\doibase 10.1038/ncomms9195} {\bibfield
  {journal} {\bibinfo  {journal} {Nat. Commun.}\ }\textbf {\bibinfo {volume}
  {6}},\ \bibinfo {pages} {8195} (\bibinfo {year} {2015})}\BibitemShut
  {NoStop}%
\bibitem [{\citenamefont {Lee}, \citenamefont {Tisdale},\ and\ \citenamefont
  {Willard}(2018)}]{Lee2018}%
  \BibitemOpen
  \bibfield  {author} {\bibinfo {author} {\bibfnamefont {E.~M.~Y.}\
  \bibnamefont {Lee}}, \bibinfo {author} {\bibfnamefont {W.~A.}\ \bibnamefont
  {Tisdale}}, \ and\ \bibinfo {author} {\bibfnamefont {A.~P.}\ \bibnamefont
  {Willard}},\ }\href {\doibase 10.1116/1.5046694} {\bibfield  {journal}
  {\bibinfo  {journal} {J. Vac. Sci. Technol. A}\ }\textbf {\bibinfo {volume}
  {36}},\ \bibinfo {pages} {068501} (\bibinfo {year} {2018})}\BibitemShut
  {NoStop}%
\bibitem [{\citenamefont {Ondry}\ \emph {et~al.}(2021)\citenamefont {Ondry},
  \citenamefont {Philbin}, \citenamefont {Lostica}, \citenamefont {Rabani},\
  and\ \citenamefont {Alivisatos}}]{Ondry2021}%
  \BibitemOpen
  \bibfield  {author} {\bibinfo {author} {\bibfnamefont {J.~C.}\ \bibnamefont
  {Ondry}}, \bibinfo {author} {\bibfnamefont {J.~P.}\ \bibnamefont {Philbin}},
  \bibinfo {author} {\bibfnamefont {M.}~\bibnamefont {Lostica}}, \bibinfo
  {author} {\bibfnamefont {E.}~\bibnamefont {Rabani}}, \ and\ \bibinfo {author}
  {\bibfnamefont {A.~P.}\ \bibnamefont {Alivisatos}},\ }\href {\doibase
  10.1021/acsnano.0c07202} {\bibfield  {journal} {\bibinfo  {journal} {ACS
  Nano}\ }\textbf {\bibinfo {volume} {15}},\ \bibinfo {pages} {2251} (\bibinfo
  {year} {2021})}\BibitemShut {NoStop}%
\bibitem [{\citenamefont {Notot}\ \emph {et~al.}(2022)\citenamefont {Notot},
  \citenamefont {Walravens}, \citenamefont {Berthe}, \citenamefont {Peric},
  \citenamefont {Addad}, \citenamefont {Wallart}, \citenamefont {Delerue},
  \citenamefont {Hens}, \citenamefont {Grandidier},\ and\ \citenamefont
  {Biadala}}]{Notot2022}%
  \BibitemOpen
  \bibfield  {author} {\bibinfo {author} {\bibfnamefont {V.}~\bibnamefont
  {Notot}}, \bibinfo {author} {\bibfnamefont {W.}~\bibnamefont {Walravens}},
  \bibinfo {author} {\bibfnamefont {M.}~\bibnamefont {Berthe}}, \bibinfo
  {author} {\bibfnamefont {N.}~\bibnamefont {Peric}}, \bibinfo {author}
  {\bibfnamefont {A.}~\bibnamefont {Addad}}, \bibinfo {author} {\bibfnamefont
  {X.}~\bibnamefont {Wallart}}, \bibinfo {author} {\bibfnamefont
  {C.}~\bibnamefont {Delerue}}, \bibinfo {author} {\bibfnamefont
  {Z.}~\bibnamefont {Hens}}, \bibinfo {author} {\bibfnamefont {B.}~\bibnamefont
  {Grandidier}}, \ and\ \bibinfo {author} {\bibfnamefont {L.}~\bibnamefont
  {Biadala}},\ }\href {\doibase 10.1021/acsnano.1c10596} {\bibfield  {journal}
  {\bibinfo  {journal} {ACS Nano}\ }\textbf {\bibinfo {volume} {16}},\ \bibinfo
  {pages} {3081} (\bibinfo {year} {2022})}\BibitemShut {NoStop}%
\end{thebibliography}
%

\end{document}